\documentclass[11pt]{article}
\usepackage{amsmath}
\usepackage{amssymb}
\usepackage{mathptmx}
\usepackage{amsfonts}
\usepackage[mathscr]{eucal}
\usepackage[dvips]{graphicx}
\usepackage{color}

\newcommand{\QED}{ \textrm{QED} }

\newcommand{\D}{ \textrm{D} }
\newcommand{\Dir}{ \textrm{Dir} }
\newcommand{\Dirac}{\textrm{Dirac}}
\newcommand{\rad}{\textrm{rad}}

\newcommand{\I}{\textrm{I}}
\newcommand{\II}{\textrm{II}}

\newcommand{\bos}{\textrm{b}}
\newcommand{\fer}{\textrm{f}}
\newcommand{\fin}{\textrm{fin}}

\newcommand{\ess}{ \textrm{ess} }

\newcommand{\Fb}{\mathscr{F}_{\textrm{b}}  } 
\newcommand{\Ff}{\mathscr{F}_{\textrm{f}}  }

\newcommand{\Gammab}{\Gamma_{\textrm{b}}}
\newcommand{\Gammaf}{\Gamma_{\textrm{f}}}

\newcommand{\dGammab}{d \Gamma_{\textrm{b}}}  
\newcommand{\dGammaf}{d \Gamma_{\textrm{f}}} 

\newcommand{\Frad}{\mathscr{F}_{\textrm{rad}}}
\newcommand{\FDirac}{\mathscr{F}_{\textrm{Dirac}}}
\newcommand{\FDir}{\mathscr{F}_{\textrm{Dir}}}
\newcommand{\FQED}{\mathscr{F}_{\textrm{QED}}}

\newcommand{\Hrad}{H_{\textrm{rad}}}
\newcommand{\HQED}{H_{\textrm{QED}}}

\newcommand{\Prad}{P_{\textrm{rad}}}

\newcommand{\HI}{H_{\textrm{I}}}
\newcommand{\HII}{H_{\textrm{II}}}

\newcommand{\kappaI}{\kappa_{\textrm{I}}}
\newcommand{\kappaII}{\kappa_{\textrm{II}}}

\newcommand{\chiIx}{\chi_{\textrm{I}} (\mathbf{x}) }

\newcommand{\chiIIx}{\chi_{\textrm{II}} (\mathbf{x}) }
\newcommand{\chiIIy}{\chi_{\textrm{II}}  (\mathbf{y}) }

\newcommand{\Hm}{H_{m}}
\newcommand{\Hradm}{H_{\textrm{rad}, \, m}}

\newcommand{\mbf}[1]{\ensuremath{\mathbf{#1}}}
\newcommand{\ms}[1]{\ensuremath{\mathscr{#1}}}
\newcommand{\sqz}[1]{\ensuremath{d\Gamma({#1}) }}
\newcommand{\sqzb}[1]{\ensuremath{d\Gamma_{\textrm{b}}({#1}) }}
\newcommand{\sqzf}[1]{\ensuremath{d\Gamma_{\textrm{f}}({#1}) }}

\newcommand{\tens}{\otimes}
\newcommand{\ntens}{\otimes^{n}}
\newcommand{\nstens}{\otimes^{n}_{\textrm{s}}}
\newcommand{\natens}{\otimes^{n}_{\textrm{a}}}

\newcommand{\Rthree}{\mathbf{R}^{3} }
\newcommand{\dx}{d \mathbf{x} }
\newcommand{\dy}{d \mathbf{y} }

\newcommand{\dk}{d \mathbf{k} }

\newcommand{\restr}{\upharpoonright}

\newcommand{\intRthree}{\int_{\mathbf{R}^{3}}}

\newcommand{\psix}{ \psi (\mbf{x})  }
\newcommand{\psiy}{ \psi (\mbf{y})  }

\newcommand{\Omegarad}{\Omega_{\textrm{rad}}}

\newcommand{\POmegaD}{P_{\Omega_{\textrm{D}}}}

\newcommand{\psidaggerx}{\psi^{\dagger} (\mathbf{x}) }
\newcommand{\psidaggery}{\psi^{\dagger} (\mathbf{y}) }

\newcommand{\1}{{\small \text{1}}\hspace{-0.32em}1}

\newtheorem{theorem}{Theorem}[section]
\newtheorem{proposition}[theorem]{Proposition}
\newtheorem{lemma}[theorem]{Lemma}
\newtheorem{corollary}[theorem]{Corollary}
\newtheorem{remark}{Remark}[section]

\begin{document}
\begin{center}
{\LARGE Ground States of Quantum  
 Electrodynamics with  Cutoffs } \\
 $\;$ \\
 {\large Toshimitsu Takaesu }  \\
 $\;$ \\
\textit{Faculty of Science and Technology, Gunma University,\\ Gunma, 371-8510, Japan }
\end{center}

\begin{quote}
\textbf{Abstract} In this paper, we investigate a  system of quantum electrodynamics with cutoffs. 
The total Hamiltonian is defined on   a tensor product of a fermion Fock  space and a boson Fock. It is shown that, under spatially localized conditions and momentum regularity conditions, the total Hamiltonian has a  ground state for all values of coupling constants. In particular,  its multiplicity is finite. \\
$\; $ \\
{\small
MSC 2010 : 47A10, 81Q10.  $\; $ \\
key words : Fock spaces, Spectral analysis, Quantum Electrodynamics}.
\end{quote}

\section{Introduction } 
$\;$ This articles is concerned with a system of quantum electrodynamics with cutoffs. In quantum field theory,
the interactions of charged particles and photons are described by quantum electrodynamics.
  We consider the system of  a massive Dirac field coupled to a  radiation field. The radiation filed is quantized in the Coulomb gauge.  In this system,  the process of electron-positron pair production and annihilation occurs. We  mathematically investigate  the spectrum of  the total Hamiltonian for the system. The  Hilbert  space for the system is defined by a tensor product of a fermion Fock space and boson Fock space, which is called a boson-fermion Fock space. The total Hamiltonian is given by 
\begin{align}
 \HQED   =   H_{\D}\tens \1 + \1  \tens \Hrad 
 & + \kappaI \sum_{j=1}^{3} \intRthree  
  \chi_{\I} (\mbf{x}) ( \psidaggerx \alpha^j \psix  \tens A_j (\mbf{x} ) ) d\mbf{x} \notag  \\
  & \; \; + \kappaII \int_{\Rthree \times \Rthree} 
     \frac{\chiIIx \chiIIy}{|\mbf{x} -\mbf{y}|} 
\left(  \psidaggerx \psix  \psidaggery  \psiy \tens \1 \right)
 d \mbf{x} \, d \mbf{y}  \notag 
\end{align}
on the Hilbert space.  Here $H_{\D}$ and $H_{\rad} $ denote the energy Hamiltonians of the Dirac field and radiation field, respectively, $\psi (\mbf{x})$  the Dirac field operator,
$\mbf{A}(\mbf{x})=(A_j(\mbf{x}))_{j=1}^3$ the radiation field operator,
 $\mbf{\alpha}=(\alpha_j)_{j=1}^3$  $4\times 4$ Dirac matrices,  and  $\chiIx$ and $\chiIIx$ the spatial cutoffs. The constants $\kappaI \in \mbf{R}$ and $\kappaII \in \mbf{R}$ are called coupling constants. Ultraviolet cutoffs  are imposed on $\psix$ and $\mbf{A}(\mbf{x})$, respectively. \\

By making use of the  spacial  cutoffs and ultraviolet cutoffs,   $\HQED$ is self-adjoint operator on the Hilbert space, and the spectrum of $\HQED$ is  bounded from below.
The main  interest in this paper is   the  lower bound of the spectrum of $\HQED$. If the infimum of the spectrum of a  self-adjoint operator is eigenvalue,  the eigenvector is called  ground state. 
The infimum of the spectrum of  $H_{0}= H_{\D} \tens \1 + \1 \tens \Hrad$ is eigenvalue, but  it is embedded in continuous spectrum. This is because   the radiation field is a massless field. It is not clear that $\HQED$ has a ground state since the embedded eigenvalue is not stable when interactions are turned on. \\

The  ground state of $\HQED$   for sufficiently small values of coupling constants was proven in \cite{Ta09}. The aim of this paper is to prove  that  $\HQED$ has  a ground state for all values of coupling constants. In particular,   its multiplicity is finite.
For the  ground states of  other QED models,
 Dimassi-Guillot \cite{DiGu03} and Barbaroux-Dimassi-Guillot \cite{BDG04} investigated the system of the Dirac field in external potential coupled to the radiation field. They 
 proved   the existence of the ground state  of the total Hamiltonian  with generalized interactions for sufficiently small values of coupling constants.  As far as we know,   the existence of the ground states  for  the  systems of a  fermionic field coupled to a massless bosonic field, which include QED models,   has not been proven for all values of coupling constants until now. \\

To prove the existence of the ground state of  $\HQED$ for all values of coupling constants, we apply the methods for   systems of   particles coupled bosonic fields. 
The spectral analysis and scattering theory for   these systems, which include the non-relativistic QED models, have been progressed since the middle of '90s. 
The  existence  of the  ground states   was established by Arai-Hirokawa \cite{AH97}, Bach-Fr{\"o}hlich-Sigal \cite{BFS98, BFS99}, G{\'e}rard \cite{Ge00}, Griesemer-Lieb-Loss \cite{GLL01}, Lieb-Loss \cite{LiLo03}, Spohn \cite{Sp98} and many researchers. The strategy  is as follows.

$\; $\\
\textbf{[1st Step]} 
 We introduce approximating  Hamiltonians $H_{m}$, $m>0$. Physically,  $m>0$ denotes the  artificial mass of photon, and we call $H_{m}$ a  massive Hamiltonian.  To  prove the existence of  ground states of $H_{m}$,  
  we use   partition of unity on Fock space, which was developed by 
 Derezi\'{n}ski-G\'{e}rard \cite{DeGe99}. We especially need the partitions of unity for  both Dirac field and radiation field. By  the partitions of unity and the Weyl sequence method,  we  prove that 
 a positive spectral gap above the infimum of the spectrum  exists  for all values of coupling constants. From this, the
existence of the ground states of $H_m$ for all values of coupling constants follows.

$\; $ \\
\textbf{[2nd Step]} 
Let $\Psi_m $ be the ground state of   $H_{m}$, $m>0$. 
Without loss of generality, we may assume that  the  $\Psi_m$ is normalized.
Then, there exists a subsequence of $\{ \Psi_{m_j} \}_{j=1}^\infty$ with $m_{j+1} < m_{j}$, $j\in \mbf{N}$, such that the weak limit of $ \{ \Psi_{m_j}\}_{j=1}^{\infty}$ exists. 
The key point   is to show that the the weak limit is non-zero vector. To prove this, 
we consider a   combined method of  Gerard \cite{Ge00} and  Griesemer-Lieb-Loss in \cite{GLL01}. We use the electron positron derivative bounds and photon derivative bounds. To derive these bounds,  the argument of   the spatially localization   is needed.  For the spatially localized conditions, we suppose
\[
\int_{\Rthree}  |\mbf{x}| \, | \chi_{\I}(\mbf{x})  | dx\, < \, \infty  , \qquad 
 \int_{\Rthree \times \Rthree}  
\frac{| \chi_{\II}(\mbf{x}) \chi_{\II}(\mbf{y}) |}{|\mbf{x}- \mbf{y}|} |\mbf{x}| \dx \dy <  \infty.
\]
I addition, We  imposed  momentum regularity conditions on the Dirac field and radiation field, which include  the infrared regularity   condition
\[
  \int_{\Rthree} \frac{ | \chi_{\rad}(\mbf{k}) |^2}{| \mbf{k}  |^{5} } \dk 
< \infty .
  \]
 $\;$\\

$\; $\\ 
We briefly  review the results for the systems of  fermionic fields coupled bosonic fields. For  QED models, the Gell-Mann - Low formula of $\HQED$ was obtained by Futakuchi-Usui \cite{FuUs14}.
 For  the Yukawa model, which is the system for a massive Dirac field interacting with a massive Klein-Gordon field, the existence of the ground state was proven  in \cite{Ta11}. The spectaral analysis  for the  the weak interaction models has been analyzed, and   refer to Barbaroux-Faupin-Guillot \cite{BFG14}, Guillot \cite{Gu15} and the reference therein.

$\;$ \\

This paper is organized as follows. In section 2, full  Fock spaces, fermion Fock spaces and boson Fock spaces are introduced, and    Dirac field  operators and radiation field  operators are defined on a Fermion Fock space  and boson Fock space, respectively.  The total Hamiltonian is defined on a boson-fermion Fock space and  the main theorem is  stated.  In Section 3, partitions of unity for the  Dirac field and  radiation field are investigated. Then  the existence of the ground state of $H_{m}$ is proven. In section 4, the derivative bounds for electrons-positrons and photons are derived. In Section 5, we give the proof of  the main theorem. \\

 \section{Notations and Main Results} 

\subsection{Fock Spaces}
\textbf{(i) Full Fock Space}\\
The full Fock space over a complex Hilbert space $\ms{Z}$ is defined by $ \ms{F}( \ms{Z})= \oplus_{n=0}^\infty ( \ntens \ms{Z})$ where $\ntens \ms{Z}$ is the $n$ fold tensor product of $Z$.  The Fock vacuum is defined by $\Omega = \{1, 0,0, \cdots  \} \in \ms{F}( \ms{Z})$. 
Let  $ \ms{L} (\ms{Z})$ be the set which consists  of all linear operators on  $\ms{Z}$. The functor  of $Q\in \ms{L}(\ms{Z})$ is defined by $\Gamma (Q) = \oplus_{n=0}^{\infty} \left( \ntens Q \right)$ and the second quantization  of $T\in\ms{L}(\ms{Z})$ is given by 
$ \sqz{T} = \oplus_{n=0}^{\infty}  \tilde{T}^{(n)} $ with 
$ \tilde{T}^{(n)} =\sum\limits_{j=1}^{n} ( (\tens^{j-1} \1 ) \tens T \tens ( \tens^{n-j} \1 )  )$.  The number operator  is defined by $N= \sqz{\1} $. \\

$\;$ \\
\textbf{(ii) Fermion Fock Space} \\
The fermion Fock space over a complex Hilbert space $\ms{X}$ is defined by $ \Ff ( \ms{X})= \oplus_{n=0}^\infty ( \natens \ms{X})$ where $ \natens \ms{X}$ denotes the $n$-fold anti-symmetric tensor product of $\ms{X}$. The Fock vacuum is defined by $\Omega_{\fer} = \{1, 0, 0, \cdots  \} \in \ms{F}( \ms{X})$. 
Let $T_{\fer}$ and $Q_{\fer}$ be  linear operators on  $\ms{X}$. We set $\sqzf{T_{\fer}}= \sqz{T_{\fer}}_{\upharpoonright \Ff ( \ms{X})}$ and $\Gamma_{\fer}( Q_{\fer })= \Gamma (Q_{\fer})_{\upharpoonright \Ff ( \ms{X})}$ where   $ X_{\upharpoonright \ms{M}} $ is the restriction of the operator $X$  to  the subspace  $\ms{M}  $.  The number operator is defined by $N_{\fer}= \sqzf{\1} $.  The creation operator $C^{\,  \dagger}(f)$, $f \in \ms{X}$, is defined by 
$(C^{\, \dagger}(f) \Psi )^{(n)} = \sqrt{n} \, U_{\textrm{a}}^{ n}(f \tens \Psi^{(n-1)})$, $n \geq 1$,
and $(C^{\, \dagger }(f) \Psi )^{(0)} = 0 $ where $ U_{\textrm{a}}^{ n} $ is the projection from 
$ \ntens \ms{X} $ to $\natens \ms{X} $. The annihilation operator $C(f)$ is defined by  $C(f)=(C^{\, \dagger}(f))^{\ast}$ where $X^{\ast} $ denotes the adjoint of the operator $X$. 
For each subspace $\ms{M} \subset \ms{X}$, the finite particle space  $ \ms{F}_{\fer}^{\, \fin} (\ms{M}) $ is defined by the linear hull  of $ \Omega_{\fer} $  and  $C^{\, \dagger}(f_{1}) , \cdots C^{\, \dagger}(f_{n}) \Omega_{\fer}  $, $j=1 , \cdots , n$, $n \in \mbf{N}$. 
The creation  and  annihilation operators satisfy the canonical anti-commutation relations on 
$\Ff (\ms{X})$ : 
\[
\qquad 
\{ C(f) , \, C^{ \, \dagger}(f')  \} = (f , f' ) , \quad   \{ C^{ \, \dagger }(f) ,
 C^{ \, \dagger }(f')  \} 
= \{ C(f) , C (f')  \} =0 , \quad f, f' \in \ms{X},
 \]
 where $\{ X ,Y \} = XY +YX $.

$\;$ \\
\textbf{(ii) Boson Fock Space}\\
 The boson Fock space over a complex Hilbert space $\ms{Y}$ is defined by $ \Fb ( \ms{Y})= \oplus_{n=0}^\infty ( \nstens \ms{Y})$ where $ \nstens \ms{Y}$ denotes the $ n$-fold symmetric tensor product of $\ms{Y}$.  The Fock vacuum is given by $\Omega_{\bos} = \{1, 0, 0,   \cdots  \} \in \ms{F}( \ms{Y})$
Let $T_{\bos}$ and $Q_{\bos}$ be  linear operators on   $\ms{Y}$. Then we define $\sqzb{T_{\bos}}= \sqz{T_{\bos} }_{\upharpoonright \Fb ( \ms{Y})}$ and  $\Gamma_{\bos}( Q)= \Gamma (Q_{\bos})_{\upharpoonright \Fb ( \ms{Y})}$.   The number operator is defined by $N_{\fer}= \sqzf{\1} $.  The creation operator $A^{\dagger}(g)$, $g \in \ms{Y}$, is defined by 
$(A^{\dagger}(g) \Phi )^{(n)} = \sqrt{n} \, U_{\textrm{s}}^{ n}(f \tens \Phi^{(n-1)})$, $n \geq 1$,
and $(A^{\dagger}(g) \Phi )^{(0)} = 0 $ where $ U_{\textrm{s}}^{ n} $ is the projection from 
$ \ntens  \ms{Y} $ to $ \nstens \ms{Y}  $. The annihilation operator $A(f)$ is defined by  $A(g)= ( A^{\dagger}(g) )^{\ast}  $. The finite particle space  $ \ms{F}_{\bos}^{\, \fin} (\ms{N}) $ on  the  subspace $\ms{N} \subset \ms{Y}$ defined by the  linear hull  of  $ \Omega_{\fer} $  and  $A^{\, \dagger}(g_{1}) , \cdots A^{\, \dagger}(g_{n}) \Omega_{\fer}  $, $j=1 , \cdots , n$, $n \in \mbf{N}$. 
The creation  and  annihilation operators satisfy the canonical commutation relations on  $ \ms{F}_{\bos}^{\, \fin} (\ms{N}) $:
\[
\qquad
[  A(g),A^{ \dagger }(g ')   ] = (g , g' ) , \quad   [ A^{  \dagger }(g) , A^{ \dagger }(g')  ] 
= [ A(g) , A (g') ] =0 , \quad g, g' \in \ms{Y} ,
 \]
where $[X ,Y]=XY-YX $. \\

\subsection{Dirac field}
Let $\ms{F}_{\Dir}  = \Ff (L^2( \Rthree ;\mbf{C}^4))$.  The energy Hamiltonian of the Dirac field is defined by 
\[
 H_{\D} = \sqzf{\omega_{\,M} } 
\]
 where  $\omega_M (\mbf{p}) = \sqrt{| \mbf{p}|^2 + M^2 }$, $\mbf{p} \in \Rthree $, and $M>0$.  
Let $C^{\, \dagger } (^{t}(f_{1} , \cdots , f_{4}))$, $f_{l} \in L^2( \Rthree )$, $l=1, \cdots ,4$, be the creation operator on $\ms{F}_{\Dirac}$. For each $f \in L^2( \Rthree  )$, we set 
\begin{align*}
&b_{1/2}^{\dagger}(f) = C^{\, \dagger}({}^{t} (f,0,0,0) ), \quad \quad  
b_{-1/2}^{\dagger}(f) = C^{\, \dagger}({}^{t}(0,f,0,0) ), \\
&d_{1/2}^{\, \dagger}(f) = C^{\, \dagger}({}^{t}(0,0,f,0) ), \quad \quad  
d_{-1/2}^{\, \dagger}(f) = C^{\, \dagger}({}^{t}(0,0,0,f) ) . 
\end{align*}
 We define  $b_s ( f)$ and $d_{s} ( g)$ by the conjugate of $b_s^{\dagger} ( f)$ and $d_{s}^{ \, \dagger } ( g)$, respectively. Then the  canonical  anti-commutation relations 
\begin{align}
& \{ b_s (f) ,b_{s'}^{\dagger} (g ) \} = \{ d_s (f) ,d_{s'}^{\, \dagger } (g ) \}= \delta_{s, s'}
 (f,g  ),  \quad  \label{ACR1} \\
& \{ b_s (f) ,b_{s'} (g) \} =\{ d_s (f) ,d_{s'} (g) \}=  \{ b_s (g) ,d_{s'}^{\, \dagger} (g) \} = 0 ,  \quad  \label{ACR2}
\end{align}
are satisfied and it holds  that 
\begin{equation}
 \|b_s (f)  \| = \|b_s^{\dagger }(f)  \| =  \| f \| , \qquad \|d_s(g)  \| = \|d_s^{\, \dagger }(g)  \|=  \|  g \|  . \label{bdNorm}
 \end{equation}
Let $h_{\D}(\mbf{p})=\mbf{\alpha} \cdot \mbf{p}+ M\beta $ be the Fourier transformed Dirac operator with $4 \times 4$ Dirac matrices $\boldsymbol{\alpha} = (\alpha^j)_{j=1}^3$ and $\beta $.   
 Let  $\mbf{S}(\mbf{p})= \mbf{S} \cdot \mbf{p}$, $\mbf{p} \in \Rthree$, where $\mbf{S}= -\frac{i}{4} \boldsymbol{\alpha} \wedge \boldsymbol{\alpha} $ is the spin angular momentum. The spinors  $ u_{s}(\mbf{p} ) = 
(u_{s}^{l} (\mbf{p}) )_{l=1}^{4} \; $ and  
 $v_{s}(\mbf{p} ) = (v_{s}^{\, l} (\mbf{p}) )_{l=1}^{4}  \; $ are function which
satisfy the following :  
\begin{align*}
\textbf{(D.1)}\;  \; \;  &h_{D} (\mbf{p}) u_{s} (\mbf{p}) = E_{M} (\mbf{p}) u_{s} (\mbf{p}), 
\quad h_{D} (\mbf{p}) v_{s} (\mbf{p}) = 
-E_{M} (\mbf{p}) v_{s} (\mbf{p}),    \\ 
\textbf{(D.2)}  \; \;  \; & S(\mbf{p} ) u_{s} (\mbf{p}) = s | \mbf{p} | u_{s} (\mbf{p}),
\quad  S(\mbf{p} ) v_{s} (\mbf{p}) = s | \mbf{p} | v_{s} (\mbf{p}) , \\
 \textbf{(D.3)} \; \; \; & \sum_{l=1}^4 u_{s}^{\, l} (\mbf{p} )^{\ast}    u_{s'}^{\,l} (\mbf{p}' ) 
=   \sum_{l=1}^4 v_{s}^{\, l} (\mbf{p} )^{\ast}    v_{s'}^{\, l} (\mbf{p}' ) 
=\delta_{s,s'}   ,\quad  
\sum_{l=1}^4 u_{s}^{\, l} (\mbf{p} )^{\ast}    v_{s'}^{\, l} (\mbf{p}' )  = 0 . 
\end{align*}

\begin{remark} \label{exa1} \normalfont
We review the example  of   spinors   in the standard representation  (see  \cite{Tha} ; Section 1). The Pauli matrices are defined by $ \sigma_{1} = \left( 
\begin{array}{cc}
0 & 1 \\
1  & 0
\end{array}
\right)  $,  
 $ \sigma_{2} = \left( 
\begin{array}{cc}
0 & -i \\
i  & 0
\end{array}
\right)
$ and 
$ \sigma_{3} = \left( 
\begin{array}{cc}
1 & 0 \\
0  & -1
\end{array}
\right)  $. Then, the Dirac marices are $
  \alpha =  \left( 
\begin{array}{cc}
O & \sigma \\
\sigma  & O
\end{array}
\right) $, $
 \beta = 
 \left( 
\begin{array}{cc}
\1 & O \\
O  & -\1
\end{array}
\right) $, and
 the spin angular momentum is $\mbf{S} = \frac{1}{2} \left( \begin{array}{cc}
\ \sigma & O \\
O  & \sigma
\end{array} \right)
 $.  Let  $ O_{\textrm{SR}} = \{  \mbf{p}  = (p^{1} , p^{2}, p^{3} )  \in \Rthree \; \left| \frac{}{} 
 \right. |\mbf{p}| -p^3 = 0 \}  $. We see that the Lebesgue measure of $O_{\textrm{SR} }$ is zero. We set  
 \[ 
\eta_{+} (\mbf{p} ) = \left\{
\begin{array}{c}
\frac{1}{\sqrt{ 2 | \mbf{p} | (  | \mbf{p} | -p^{3} ) } }
 \begin{pmatrix}
  p^{1} - i p^{2}  \\  | \mbf{p} |  - p^{3}
\end{pmatrix} , \;  \mbf{p}  \notin O_{\textrm{SR} }    ,  \\ 
 \qquad \qquad
 \begin{pmatrix} 
  1 \\ 0   \end{pmatrix} ,  \quad  \quad \qquad 
    \mbf{p}  \in O_{\textrm{SR} }   ,
 \end{array} 
 \right.
\; \;   
 \eta_{-} (\mbf{p} ) = 
 \left\{
 \begin{array}{c}
\frac{1}{\sqrt{ 2 | \mbf{p} | ( | \mbf{p} | - p^{3}   ) } }
    \begin{pmatrix}
p^{3} - | \mbf{p} |  \\  p^{1} + i p^{2} 
\end{pmatrix}  , \,
    \mbf{p}  \notin O_{\textrm{SR} }    ,  \\ 
\qquad \quad   \begin{pmatrix}   0  \\  1 \end{pmatrix} ,
 \quad  \qquad \qquad \mbf{p}  \in O_{\textrm{SR} } .
  \end{array}  
\right. 
\]
 Let
  \[
 u_{\pm1/2} (\mbf{p}) = 
 \begin{pmatrix} \lambda_{+} (\mbf{p} ) \eta_{\pm} (\mbf{p}) 
 \\  \pm \lambda_{-} (\mbf{p} ) \eta_{\pm} (\mbf{p})   \end{pmatrix} ,  \qquad
 v_{\pm 1/2} (\mbf{p}) = 
 \begin{pmatrix} \mp \lambda_{-} (\mbf{p} ) \phi_{\pm} (\mbf{p}) 
 \\  \pm \lambda_{+} (\mbf{p} ) \eta_{\pm} (\mbf{p})   \end{pmatrix} ,
 \]
with  $ \lambda_{\pm} (\mbf{p}) = 
\frac{1}{\sqrt{2}} \sqrt{  1 \pm  M \, E_{M} (\mbf{p})^{-1 }}$. Here note that   $ u_{s} $ and
 $  v_{s} $ satisfy   $  u_{s} , v_s  \in 
 \oplus^4 (  C^{1} ( \Rthree \backslash   O_{\textrm{SR}} ) ) $. \\
 $\;$ 
\end{remark}

$\;$ \\
The Dirac field operator $  \psi(\mbf{x}) =  {}^{t} (\psi_{1}(\mbf{x}) ,  \cdots ,  \psi_{4}(\mbf{x}))$  is  defined by 
\[
\qquad \qquad \quad \psi_{l}(\mbf{x}) = \sum_{s=\pm 1/2} \left( b_s ( f_{s , \mbf{x}}^{\, l} ) +  
 d^{\, \dagger}_s ( g_{s , \mbf{x}}^l) \frac{}{} \right) , \quad \qquad l=1, \cdots , 4 ,
\]
where $f_{s , \mbf{x}}^{\, l} (\mbf{p})= f_{s }^{\, l} (\mbf{p})e^{-\mbf{p}\cdot \mbf{x} }$ with 
$f_{s }^{\, l} (\mbf{p})= \frac{1}{\sqrt{ (2 \pi )^3 }} \chi_{\D}(\mbf{p}) u_s^{l}(\mbf{p})$ and $g_{s , \mbf{x}}^l (\mbf{p})= g_{s }^l (\mbf{p})e^{-\mbf{p}\cdot \mbf{x} }$ with 
$g_{s }^l (\mbf{p})= \frac{1}{\sqrt{ (2 \pi )^3 }}\chi_{\D}(\mbf{p}) \tilde{v}_s^{l}(\mbf{p})$ and $\tilde{v}_s^{\, l}(\mbf{p})= v_{s}^{\, l} (-\mbf{p})$. Here 
$\chi_{\D} $ satisfy  the following condition.\\

\begin{quote}
\textbf{(A.1 ; Ultraviolet Cutoff for Dirac Field)}
\[
\int_{\Rthree} | \chi_{\D}(\mbf{p})  |^2 d \mbf{p} 
< \infty  .
\]
\end{quote}
Then  it holds that 
\begin{equation}
\qquad 
\|\psi_l (\mbf{x}) \| \leq  c_{\, \D }^{ \, l } ,  \label{psiBound} 
\end{equation}
where $ c_{\, \D }^{\, l } = \frac{1}{\sqrt{(2\pi)^3}} \sum\limits_{s= \pm 1/2} 
\left( \| \chi_{\D} \, u_{s}^{\,l}  \| +  \| \chi_{\D} \,  \tilde{v}_{s}^{\, l} \|  \right)$, 
$l=1 , \cdots , 4.$\\

\subsection{Radiation Field in the Coulomb Gauge}
Let $\Frad  = \Fb ( \oplus_{r= 1,2} L^2( \Rthree   ))$.  The free Hamiltonian is defined by 
\[
 \Hrad = \sqzb{\omega } 
\]
 where $\omega (\mbf{k}) = |\mbf{k}|$, $\mbf{k} \in \Rthree $.
Let $A^{\ast }(h_1 , h_2 )$, $h_{r} \in L^2( \Rthree   )$, $r=1,2$, be the creation operators on $\Frad $.  Let 
\[
\qquad 
a^{\dagger }_{1} (h ) = A (( h, 0)) , \quad a^{\dagger }_{2} (h )= A ((  0, h )) , \qquad h \in L^2( \Rthree   ) ,
\]
and   $a_{r}(h')=(a^{\dagger} (h'))^{\ast}$, $h' \in L^2( \Rthree   ) $, $r=1,2$.  
The creation operators and annihilation operators satisfy the canonical commutation relations 
\begin{align}
& [a_r (h) ,a_{r'}^{\dagger} (h') ] = \delta_{r, r'} (h, \, h' ), \; \; 
 \label{radCCR1}  \\
& [a_r (h) ,a_{r'} ( h' ) ] = [a_r^{\dagger } (h' ) ,a_{r'}^{\dagger} (h') ] = 0 , 
 \label{radCCR2}
\end{align}
on  $\Frad^{\, \fin}(\ms{M})$ where  $\ms{M} $ is a subspace of $\oplus_{r= 1,2} L^2( \Rthree   )	$. For all  $h \in \ms{D}(\omega^{-1/2}) $, it follows that 
\begin{equation}
\quad
\| a_r (h )  ( \Hrad   +1 )^{-1/2} \| \leq \| \frac{h}{\sqrt{\omega }} \|  , \quad
 \| a^{\dagger }_r (h )  ( \Hrad   +1 )^{-1/2} \| \leq  \| \frac{h}{\sqrt{\omega }}    \|  + \| h \|  . \label{radfiedBound}
\end{equation}
 The polarization vectors $\mbf{e}_{r} (\mbf{k}) = (e_r^{j} (\mbf{k}))$, $r=1,2$, satisfy the 
 following relations.
\begin{equation}
 \textbf{(R.1)} \; \qquad  \mbf{e}_{r} (\mbf{k}) \cdot \mbf{e}_{r'} (\mbf{k}) =0 , \quad 
\mbf{e}_{r} (\mbf{k})\cdot \mbf{k} = 0 , \quad \quad \mbf{k} \in \Rthree \backslash \{ \mbf{0} \} . \notag
\end{equation}
$\;$ \
\begin{remark} \label{exa2} \normalfont
We check  the example of the polarization vectors. For all $\mbf{k} \in \Rthree \backslash    \{ \mbf{0} \} $, we set
\[ \qquad 
\mbf{e}_1(\mbf{k}) = \frac{1}{\sqrt{(k^1)^2 + (k^2) ^2}} \left( \begin{array}{c} - k^{2} \\ k^{1} \\ 0 \end{array} \right)  , \; \; \;  \mbf{e}_2(\mbf{k})   =\frac{1}{|\mbf{k}|\sqrt{k_1^2 +k_2^2}}
 \left( \begin{array}{c}  k^{1} k^3 \\ k^{2} k^3 \\ - (k^1)^2 - (k^2)^2 \end{array} \right)  . \quad 
 \]
Then  \textbf{(R.1)} is satisfied. Here it is noted that  $\mbf{e}_{r} \in \oplus^3 \, ( C^{1}(\Rthree \backslash \{ \mbf{0} \} ))$, $r=1,2$. 
\end{remark}

$\;$ \\
The radiation field operator $\mbf{A}(\mbf{x})= ( A_j (\mbf{x}))_{j=1}^3$ is defined  by 
\[
A_j(\mbf{x}) = \sum_{r=1,2} \left( a_r (h_{r , \mbf{x}}^j) +   a^{\dagger}_r (h_{r , \mbf{x}}^j) \frac{}{} \right)
\]
where $h_{r , \mbf{x}}^j (\mbf{k})= h_{r }^j (\mbf{k})e^{-\mbf{k}\cdot \mbf{x} }$ with 
$h_{r }^j (\mbf{k})= \frac{1}{\sqrt{(2 \pi )^3}}  \frac{\chi_{\rad}(\mbf{k}) e_r^{j}(\mbf{k})}{\sqrt{2  \omega(\mbf{k})}}$, and $\chi_{\rad}$ satisfy the following condition.  \\
\begin{quote}
\textbf{(A.2 : Ultraviolet Cutoff for Radiation Field)}
\[
\qquad 
\int_{\Rthree} \frac{ | \chi_{\rad}(\mbf{k}) |^2}{ \omega(\mbf{k})^{l}} \dk 
< \infty , \quad l= 1,2 .
\]
\end{quote}
Then 
\begin{equation}
 \|A_j (\mbf{x}) ( \Hrad +1 )^{-1/2}  \| \leq  c^{\, j}_{\rad }  
\end{equation}
where  $c^{\,j}_{\rad }=  \frac{1}{\sqrt{(2\pi )^3}}\sum\limits_{r=1,2}\left( \sqrt{2} \| \frac{\chi_{\rad} e_r^j }{ \omega} \|  +   \|  \frac{\chi_{\rad} e_r^j }{\sqrt{2 \omega}}  \| \right) $.\\

\subsection{Total Hamiltonian and Main Theorem}
We define the system of the Dirac field  interacting with the radiation field. The Hilbert space for the system is defined by $\FQED = \FDirac \tens \Frad $. The free Hamiltonian is defined by 
\begin{equation}
H_0= H_{\D} \tens \1_{\rad} + \1_{\D} \tens \Hrad  \notag
\end{equation}
on  the domain $\ms{D}(H_{0}) = \ms{D}( H_{\D} \tens \1_{\rad} )  \cap \ms{D}(\1_{\D} \tens \Hrad )$. To  define the interactions, we introduce spatial cutoff $\chi_{\I}$ and $\chi_{\II}$, which satisfy the  condition below.\\
\begin{quote}
\textbf{(A.3 : Spatial Cutoff )}
\[
\int_{\Rthree} | \chiIx | \dx < \infty , \qquad   \int_{\Rthree \times \Rthree} 
     \frac{|\chiIIx \chiIIy|}{|\mbf{x} -\mbf{y}|} \dx \dy < \infty . 
\]
\end{quote}
First we define a functional on  $  \ms{F}_{\QED} \times \ms{F}_{\QED}  $   by
\[
\qquad \qquad 
\ell_{\I}(\Phi , \Psi ) = \sum_{j=1}^3   \int_{\Rthree} \chiIx 
\left( \Phi , (\psidaggerx \alpha^j \psix  \tens A_j (\mbf{x}) ) \Psi \right) ,
\; \; \Phi \in \ms{F}_{\QED} , \, \Psi \in \ms{D}(H_0 ) ,
\]
where $ \psidaggerx = ( \psi_{1}(\mbf{x})^{\ast} , \cdots , \psi_{4}(\mbf{x})^{\ast} ) $. We see that 
\begin{equation}
| \ell_{\I}(\Phi , \Psi ) |   \leq  \left( \int_{\Rthree} | \chiIx | \dx  \right) \,
\sum\limits_{j=1}^3 \, \sum_{l,l'=1}^4 |\alpha_{l,l'}^j | c_{\, \D }^{\, l } c_{\, \D }^{ \, l' } c_{\rad}^{\,  j } \,
 \|  \Phi \| \,  \| \1_{\D} \tens (\Hrad +1)^{1/2} \Psi \| .  \notag
\end{equation}
 By the 
Riesz representation theorem,  we can define  the operator $\HI$ which satisfy 
$  (\Phi , \HI \Psi ) 
= \ell_{\I}(\Phi , \Psi )   $ and
\begin{equation}
 \|  \HI \Psi \| \leq c_{\, \I} \,   \| \1_{\D} \tens (\Hrad +1)^{1/2} \Psi \|   , 
  \label{HIbound}
\end{equation} 
where $ c_{\, \I} = \| \chi_{\I} \|_{L^1} \sum\limits_{j=1}^3 \, \sum\limits_{l,l'=1}^4 |\alpha_{l,l'}^j | c_{\, \D }^{\, l } c_{\, \D }^{ \, l' } c_{\rad}^{\,  j }$. 
By the spectral decomposition theorem, it is proven  that for all $\epsilon > 0$, 
\begin{equation}
\|\HI \Psi \| \leq c_{\I} \epsilon \|H_{0} \Psi \| +c_\I \left( \frac{1 }{2 \epsilon}  +1 \right)  \| \Psi \|  .
 \label{HIbound'}
\end{equation}
 Next we  define a functional on  $ \ms{F}_{\QED} \tens  \ms{F}_{\QED} $ by 
\[
 \ell_{\II}(\Phi , \Psi ) =    \int_{\Rthree \times \Rthree} 
     \frac{\chiIIx \chiIIy}{|\mbf{x} -\mbf{y}|} 
\left( \Phi , \left(  \psidaggerx \psix  \psidaggery \psiy \tens \1_{\rad}  \right) \Psi \right) 
 d \mbf{x} \, d \mbf{y}  , \; \; \Phi , \, \Psi \in \ms{F}_{\QED}  . \\
\]
We see that 
\begin{equation}
|\ell_{\II}(\Phi , \Psi ) |   \leq 
 \left( \int_{\Rthree \times \Rthree} \left| \frac{\chiIIx \chiIIy}{ |\mbf{x}-\mbf{y}|} \right| \dx \dy \right)
 \,  \sum_{l , l' =1}^4
 (c_{\, \D}^{ \, l} c_{\, \D }^{\, l'})^2 \| \Phi  \| \, \| \Psi \| .  \notag
\end{equation}
Then, by the Riesz representation theorem,  we can define  an operator $\HII$ satisfying  $  (\Phi , \HII \Psi ) = \ell_{\II}(\Phi , \Psi )$ and 
\begin{equation} 
\|\HII \| \leq  c_{\II} ,   \label{HIIbound}
\end{equation}
where $ c_{\II} = \left\| \frac{\chiIIx \chiIIy}{ |\mbf{x}-\mbf{y}|} \right\|_{L^1} \,  \sum\limits_{l , l' =1}^4  (c_{\, \D}^{ \, l} c_{\, \D }^{\, l'})^2$. 
 By (\ref{HIbound'}) and (\ref{HIIbound}), it holds that
\begin{equation}
\| (\kappaI \HI  + \kappaII \HII ) \Psi  \| 
\leq    c_{\I} \kappaI   \epsilon  \| H_0 \Psi \| + 
\left(  c_{\I} \kappaI \left(  \frac{1 }{2 \epsilon}  +1 \right)  +  c_{\II} \kappaII \right) \| \Psi \|
. \notag 
\end{equation}
Then the  Kato-Rellich theorem  yields  that that   $\HQED$ is self-adjoint on $\ms{D}(H_{0})$ and essentially self-djoint on any core of $H_{0}$. Hence, in particular, $\HQED$ is essentially self-adjoin on
\[
\ms{D}_{0} = \FDir^{\, \fin}{\ms{D}(\omega_{\, M})} \hat{\tens }\Frad^{\, \fin} (\ms{D}(\omega ))
\]
where $\hat{\tens }$ denotes the algebraic tensor product.

$\; $ \\
To prove the existence of the ground  state of $\HQED$, we suppose additional conditions below.

\begin{quote}
\textbf{(A.4 : Spatial Localization)}
\[
 \int_{\Rthree} |\mbf{x}|| \chiIx | \dx < \infty ,  \quad  \int_{\Rthree \times \Rthree} 
     \frac{|\chiIIx \chiIIy|}{|\mbf{x} -\mbf{y}|}|\mbf{x}| \dx \dy < \infty       .
\]
\end{quote}

\begin{quote}
\textbf{(A.5 : Momentum Regularity Condition for Dirac Field)} \\
 There exists a subset  $O_{\D} \subset \Rthree $ with Lebesgue measure zero such that $u_{s}, v_{s} \in \oplus^4 \, ( C^1 (\Rthree \backslash O_{\D})) $, $s= \pm 1/2$.   $\chi_{\D} \in  C^{1}(\Rthree) $, and it satisfies that
\[
   \int_{\Rthree}  | \partial_{p^{\nu}}\chi_{\D}(\mbf{p}) |^2 d \mbf{p}
< \infty ,
 \;   \int_{\Rthree} | \chi_{\D }(\mbf{p }) \partial_{p^{\nu }} u_{s}^{\, l} (\mbf{p}) |^2 d \mbf{p}
< \infty , \;  \int_{\Rthree} | \chi_{\D}(\mbf{p }) \partial_{p^{\nu }} v_{s}^{\, l} (-\mbf{p}) |^2 d \mbf{p}
< \infty ,
\]
for all   $\nu =1, 2,3$, $ l = 1 , \cdots ,4$, $  s= \pm 1/2 $.
\end{quote}

\begin{quote}
\textbf{(A.6  : Momentum Regularity Condition for Radiation Field )} \\
There exists a subset  $O_{\rad} \subset \Rthree $ with Lebesgue measure  zero such that $\mbf{e}_{r} \in \oplus^3 \, ( C^{1}(\Rthree \backslash O_{\rad} ))$, $r=1,2$, where $O_{\rad} $. $\chi_{\rad} \in C^{1}(\Rthree)$ and it satisfies that
\[
\int_{\Rthree} \frac{ | \chi_{\rad}(\mbf{k}) |^2}{ | \mbf{k}|^5} \dk 
< \infty , \quad 
\int_{\Rthree} \frac{ | \partial_{k^\nu }\chi_{\rad}(\mbf{k}) |^2}{ | \mbf{k} |^3} \dk 
< \infty , \quad  \int_{\Rthree} \frac{ | \chi_{\rad}(\mbf{k}) \partial_{k^\nu } e_{r}^{j} (\mbf{k}) |^2}{ |\mbf{k}|^3} \dk 
< \infty , 
\]
for all  $\nu =1, 2, 3$, $j=1, 2, 3$,   $r=1,2$.
\end{quote}

\begin{remark} \label{exa3} \normalfont
Examples of $O_{\D}$ and $O_{\rad}$ in \textbf{(A.5)} and \textbf{(A.6)} are as follows. 
In the case of  the standard representation,    
$O_{\D} = O_{\textrm{SR}}$ where $ O_{\textrm{SR}} $ is defined in  Remark \ref{exa1}.  
For the  polarization vectors considered in Remark 
\ref{exa2},  $O_{\rad} = \{ \mbf{0} \}$. 
\end{remark}

$\; $ \\ 
The main theorem in this paper is as follows.

\begin{theorem} \normalfont  \label{Main-Theorem} 
\textbf{(Existence of a Ground State)} \\
Suppose \textbf{(A.1)} - \textbf{(A.6)}.   
   Then    $ \HQED$ has a ground state for  all values of coupling constants.
   In particular, its multiplicity is finite. 
\end{theorem}


\section{Ground States of Massive case}
In this section, we consider a massive Hamiltonian defined by 
\begin{equation}
H_m = H_{\D} \tens \1 + \1 \tens \Hradm + \kappaI \HI + \kappaII \HII ,  \notag 
\end{equation}
where $\Hradm = \sqzb{\omega_{m}}$ with $\omega_m (\mbf{k})=\sqrt{\mbf{k}^2 + m^2} $, $m>0$. 
\\

\subsection{Fock Spaces on Direct Sum of Hilbert Spaces} $\, $
We review  basic properties of   Fock spaces on direct sum of Hilbert spaces. These are useful for constructing partitions of unity on Fock spaces (see, Derezi\'{n}ski-G\'{e}rard \cite{DeGe99}).

$\, $ \\
\textbf{(i) Full Fock Space on } $\ms{Z} \oplus \ms{Z}$  \\
Let   $ Z = \left[ \begin{array}{c} Z_0 \\ Z_\infty \end{array} \right] $,  
$ Z_0 , Z_\infty \in   \ms{L} (\ms{Z}) $, where $\ms{Z}$ is a complex Hilbert space. We consider 
 $ Z = \left[ \begin{array}{c} Z_0 \\ Z_\infty \end{array} \right] $ is an operator  $ \ms{Z} \to \ms{Z} \oplus \ms{Z}$ which acts for 
\[
\qquad 
 \qquad Z h = \left[ \begin{array}{c} Z_0 \,  h \\ Z_\infty \, h \end{array} \right] , \quad \; h \in  \ms{D} (Z_{0}) \cap \ms{D} (Z_{\infty}) .
\]
Let $J = \left[ \begin{array}{c} J_0 \,   \\ J_\infty \,  \end{array} \right]$,
 $ J_0 , J_\infty \in   \ms{L} (\ms{Z}) $ and $ B = \left[ \begin{array}{c} B_0 \,   \\ B_\infty \,  \end{array} \right] $, $ B_0 , B_\infty \in   \ms{L} (\ms{Z}) $. We define
$d \Gamma ( J , B) : \ms{F}(\ms{Z}) \to \ms{F} (\ms{Z}) \oplus  \ms{F}(\ms{Z})$ by
\[
d \Gamma (J , B ) = 
\oplus_{n=0}^{\infty} \left( \sum\limits_{j=1}^n 
 ( \tens^{j-1} J )   \tens  B
  \tens ( \tens^{n-j} J ) \right) .
\]
If $B_{0}$ and $B_{\infty}$ are bounded, and    $J_{0}^{\ast}J_0 + J_{\infty}^{\ast}J_{\infty} \leq 1 $,  it holds that 
\begin{equation}
\|   d \Gamma ( J, B ) (N +1)^{-1} \| \leq  \sqrt{ \|B_{0} \|^2 + \|  B_{\infty}  \|^2 } .  
    \label{3.1.1}
\end{equation}
Let  $T \in \ms{L}(\ms{Z})$.   
 Then it holds that 
\begin{equation}
\Gamma (J ) \sqz{T} =  d \Gamma \left( \left[ 
\begin{array}{cc} T & 0 \\ 
 0 & T  \end{array} \right]   \right)
\Gamma ( J ) + d \Gamma ( J ,\tilde{\text{ad}}_{T}( J )  ) ,
  \label{3.1.2}
\end{equation}
where  $ \tilde{\text{ad}}_{T}(J)  :  \ms{Z} \to \ms{Z} \oplus \ms{Z}$ is defined by 
\[
\qquad  \qquad \quad 
\tilde{\text{ad}}_{T}(J) h=\left[ 
\begin{array}{c} [T ,J_0 ] h \\ 
 {[} T , J_\infty {]} h  \end{array} \right] , \quad h \in \ms{D}([T ,J_0 ]) 
\cap \ms{D}([T ,J_\infty ] ) .
\]

$\;$ \\
\textbf{(ii) Fermion Fock Space on }$\ms{X} \oplus \ms{X}$ \\
Let  $\ms{X}$ be a complex Hilbert space. 
 Let $ J_{\fer}  = \left[ \begin{array}{c} J_{\fer}^{\, 0 } \,   \\ J_{\fer }^{\, \infty} \,  \end{array} \right] $, $ J_{\fer}^{\, 0 } ,  J_{\fer }^{\, \infty} \in   \ms{L} (\ms{X}) $ and $  B_{\fer} = \left[ \begin{array}{c}
 B_{\fer}^{\, 0 } \,   \\   B_{\fer }^{\, \infty}\,  \end{array} \right] $, $ B_{\fer}^{\, 0 }, B^{\, \infty}_{\fer } \in   \ms{L} (\ms{X}) $. We set $\sqzf{ J_{\fer} , B_{\fer} } = 
\sqz{ J_{\fer} , B_{\fer}  }_{\restr \Ff (\ms{X})} $. 
Suppose that $B_{\fer}^{\, 0 } $ and $B^{\, \infty}_{\fer } $ are bounded, and   
 $ (J_{\fer}^{\, 0 })^\ast J_{\fer}^{\, 0 } +   (J_{\fer }^{\, \infty})^{\ast} J_{\fer }^{\, \infty}  \leq 1 $. By (\ref{3.1.1}), it holds that 
\begin{equation}
\|   d \Gammaf ( J_{\fer} , B_{\fer} ) (N_{\fer} +1)^{-1} \| \leq 
 \sqrt{ \| B_{\fer}^{\, 0 } \|^2 + \|   B_{\fer }^{\, \infty}  \|^2 } . 
\label{3.1.3}
\end{equation}
Let  $T_{\fer}  \in \ms{L}(\ms{X})$.   
 From (\ref{3.1.2}), it holds that 
\begin{equation}
\Gammaf ( J_{\fer} ) \sqzf{T_{\fer} } =  d \Gammaf \left( \left[ 
\begin{array}{cc} T_{\fer}  & 0 \\ 
 0 & T_{\fer}  \end{array} \right]   \right)
\Gammaf ( J_{\fer} ) + d \Gammaf ( J_{\fer}  ,\tilde{\text{ad}}_{T_{\fer} }(
 J_{\fer} )  ).  \label{3.1.4}
\end{equation}
Let $ C(f)$ and $C^{ \, \dagger} (f)$, $f \in \ms{X}$,  be the annihilation  and creation  operators on $\Ff (\ms{X})$, respectively. Then it follows that
\begin{align}
& \Gamma_{\fer} ( J_{\fer} ) C (f) = C \left( 
\left[  \begin{array}{c} f  \\   0  \end{array} \right]  
 \right)  \Gamma_{\fer } ( J_{\fer} ) + \Gamma_{\fer} 
( J_{\fer} ) \,  C \left( (1-(  J_{\fer}^{\, 0 })^{\ast})f \frac{}{}\right) ,
 \label{3.1.5}
 \\
 & \Gamma ( J_{\fer} ) C^{\, \dagger} (f) = C^{\, \dagger} \left( 
\left[  \begin{array}{c} f  \\   0  \end{array} \right]  
 \right)  \Gamma_{\fer} ( J_{\fer} ) +  C^{\, \dagger} \left( 
\left[  \begin{array}{c} J_{\fer}^{\, 0 } -1  \\      J_{\fer }^{\, \infty}   \end{array} \right]  
 f \right) \Gamma_{\fer } ( J_{\fer} )  . 
\label{3.1.6}
\end{align}
$\;$ \\

$\; $ \\
\textbf{(iii) Boson Fock Space on }$\ms{Y} \oplus \ms{Y}$ \\
Let  $\ms{Y}$ be a complex Hilbert space.  
Let $ J_{\bos}  = \left[ \begin{array}{c} J_{\bos}^{\, 0 } \,   \\ J_{\bos }^{\, \infty} \,  \end{array} \right] $, $ J_{\bos}^{\, 0 } ,  J_{\bos }^{\, \infty} \in   \ms{L} (\ms{Y}) $ and $  B_{\bos} = \left[ \begin{array}{c}
 B_{\bos}^{\, 0 } \,   \\   B_{\bos }^{\, \infty} \,  \end{array} \right] $, $ B_{\bos}^{\, 0 },  B_{\bos }^{\, \infty} \in   \ms{L} (\ms{Y}) $. We define $\sqzb{ J_{\bos} , B_{\bos} } = 
\sqz{ J_{\bos} , B_{\bos}  }_{\restr \Fb (\ms{Y})} $. 
Assume that $B_{\bos}^{\, 0 } $ and $ B_{\bos }^{\, \infty} $ are bounded, and   
 $ (J_{\bos }^{\, 0 })^\ast J_{\bos}^{\, 0 } +   (J_{\bos }^{\, \infty})^{\ast} J_{\bos }^{\, \infty}  \leq 1 $. By (\ref{3.1.1}), it follows that
\begin{equation}
\|   d \Gammab ( J_{\bos} , B_{\bos} ) (N_{\bos} +1)^{-1} \| \leq 
 \sqrt{ \| B_{\bos}^{\, 0 } \|^2 + \|   B_{\bos }^{\, \infty}  \|^2 } . 
\label{3.1.7}
\end{equation}
Let   $ T_{\bos } \in \ms{L}(\ms{Y})$. Then    
 (\ref{3.1.2}) yields that 
 \begin{equation}
\Gammab ( J_{\bos} ) \sqzb{T_{\bos } } =  d \Gammab \left( \left[ 
\begin{array}{cc} T_{\bos}  & 0 \\ 
 0 & T_{\bos}  \end{array} \right]   \right)
\Gammab ( J_{\fer} ) + d \Gammab ( J_{\bos}  ,\tilde{\text{ad}}_{T_{\bos} }(
 J_{\bos} )  ).   \label{3.1.8}
\end{equation}
Let $ A(g)$   and $A^{\dagger} (g)$, $g \in \ms{Y}$,   be the annihilation  and creation  operators on $\Fb (\ms{Y})$, respectively. Then it follows that
\begin{align}
& \Gamma_{\bos} ( J_{\bos} ) A (g) = A \left( 
\left[  \begin{array}{c} g  \\   0  \end{array} \right]  
 \right)  \Gammab ( J_{\bos } ) + \Gammab 
( J_{\bos} ) \, A \left( (1-( J_{\bos}^{\, 0 } )^{\ast}) g \frac{}{}\right) , 
 \label{3.1.9} \\
 & \Gammab ( J_{\bos} ) A^{\dagger } (g) = A^{\dagger} \left( 
\left[  \begin{array}{c} g  \\   0  \end{array} \right]  
 \right)  \Gamma_{\bos} ( J_{\bos} ) +  A^{\dagger } \left( 
\left[  \begin{array}{c}   J_{\bos}^{\, 0 } -1 \\    J_{\bos}^{\, \infty }  \end{array} \right]   g \right) \Gammab ( J_{\bos} )  .
\label{3.1.10}
\end{align}

\newpage
\subsection{Partition of Unity for the Dirac Field}
We construct a partition of unity for the Dirac field. For general properties of  partition of unity for fermionic fields, refer to  Ammari \cite{Am04}. 

$\;$ \\
Let
\[ \quad 
c_{ \tau , s} (f) = \left\{ \begin{array}{c} b_{s}(f), \; \; \tau = + , \\
 d_{s}(f) , \; \;\tau = - . \end{array}  \right.  
\]
 Let  $U_{\fer}:   \Ff \left( L^2( \Rthree_{\bf{p}} ; \mbf{C}^4  ) \oplus L^2( \Rthree_{\bf{p}} ; \mbf{C}^4  ) \right)  \to \ms{F}_{\Dirac} \tens \ms{F}_{\Dirac} $ be  an isometric operator which satisfy  $U_{\fer} \,  \Omega_{\D} =\Omega_{\D} \tens \Omega_{\D} $
 and  
\begin{align}
&U_{\fer }  \;  c^{\dagger }_{\tau_1, s_1} \left(\left[ \begin{array}{c} f_1 \\ g_1 \end{array} \right]\right) \cdots  c^{\dagger }_{\tau_1 , s_1} \left(\left[ \begin{array}{c} f_1 \\ g_n \end{array} \right]\right) \Omega_{\D}   \notag \\
&  =  \left( c^{\dagger}_{\tau_1 , s_1} (f_1) \tens \1  + (-1)^{N_{\D}}
 \tens c^{\dagger}_{\tau_1 ,  s_1}(g_1) \frac{}{}\right)
\cdots  \left( c^{\dagger }_{\tau_n, s_n} (f_n) \tens \1  + (-1)^{N_{\D}} \tens c^{\dagger}_{\tau_n , r_n}(g_n) \frac{}{} \right)
 \Omega_{\D} \tens \Omega_{\D} . \notag
\end{align}
Here note  that  $(-1)^{N_{\D}}\Psi = (-1)^n \Psi  $ for  the vector  of the form $ \Psi = c^\dagger_{\tau_1 ,  s_1}(f_1)  \cdots   c^\dagger_{\tau_n, s_n} (f_n) \Omega_{\D} $, $f_{j} \in L^2 (\Rthree) $, $j= 1, \cdots ,n$, $n \in \mbf{N}$. 
Let $j_{0} , j_{\infty} \in C^{\, \infty} (\mbf{R})$. We assume  that 
$j_0 \geq 0 $, $j_{\infty} \geq 0$,  $j_{0} (\mbf{x})^2 +  j_{\infty} (\mbf{x})^2 =1$, 
$j_{0}(\mbf{x})=1 $ for $|\mbf{x}| \leq 1 $ and $j_{0}(\mbf{x})=0 $ for $|\mbf{x}| \geq 2 $. 
Let $j_{\fer , R } = \left[ \begin{array}{c} j_{\fer , R }^{\, 0}   \\ 
  j_{\fer , R }^{\, \infty}  \end{array} \right]$ where 
$ j_{\fer , R }^{\, 0}= j_{0} (\frac{-i \, \mbf{\nabla_{\mbf{p}}}}{R})$ and $ j_{\fer , R }^{\, \infty}=  j_{\infty}
 (\frac{-i \, \mbf{ \nabla_{\mbf{p}}}}{R})    $ with $\nabla_{\mbf{p}}= (\partial_{p^1}, \partial_{p^2},\partial_{p^3} ) $. \\
 
 $\;$ \\
Let $X_{\fer , R} : \FDir \to  \ms{F}_{\Dir} \tens \ms{F}_{\Dir}  $  defined by  
\begin{equation}
\quad 
 X_{\fer , R}  = \,    U_{\fer } \, \Gamma_{\fer}(j_{\fer , R}) . \notag 
\end{equation}
From (\ref{3.1.4})-(\ref{3.1.6}), it holds that 
 \begin{align}
 &  X_{\fer , R} \, H_{\D} = 
\left( H_{\D} \tens \1 + \1 \tens  H_{\D} \frac{}{} \right) 
  X_{\fer , R} \, +   U_{\fer } \,  d \Gamma_{\fer} (j_{\fer, R} , \tilde{\text{ad}}_{\omega_{M} }(j_{\fer ,R})) , \label{3.2.1}  \\
& X_{\fer  , R} \,  c_{\tau , s}(f)
=  (c_{\tau, s } (f) \tens \1 )  X_{\fer , R} \, +
 X_{\fer , R} \,
 c_{\tau , s}((1-j_{\fer , R}^{\, 0} ) f) , \label{3.2.2} \\
& X_{\fer , R} \, c^{\dagger}_{\tau , s }(f) 
= ( c^\ast_{\tau , s } (f) \tens \1 ) X_{\fer , R} 
+    \left( c^{\dagger}_{\tau , s } ((j_{\fer , R}^{\, 0} -1 )f ) \tens \1 
+  (-1)^{N_{\D}} \tens c^{\dagger }_{\tau , s }(j_{\fer, R}^{\, \infty} f ) \right)
X_{\fer , R}  . \label{3.2.3}
\end{align}

\begin{lemma} \label{lemma32a} \normalfont  
Assume \textbf{(A.1)}. Then,
\begin{align*}
& \textbf{(i)}
\left\| \left(  X_{\fer , R} \,  H_{\D}
-   (      H_{\D} \tens  \1    + \1 \tens  H_{\D} ) \right) 
  X_{\fer , R}  ( N_{\D} + 1 )^{-1} \right\|  \leq  \frac{c_{ \, \fer }}{R} ,  \\
 & \textbf{(ii)} 
\left\|   X_{\fer , R} \, \psi_{l}(\mbf{x})
-   (       \psi_{l}(\mbf{x}) \tens \1  ) X_{\fer , R}    
 \right\|  \leq \delta^{1,l}_{ \, \fer ,R } (\mbf{x}) ,\quad l= 1, \cdots 4,  \\
&\textbf{(iii)} 
\left\|   X_{\fer , R} \, \psi_{l}(\mbf{x})^{\ast}
-   (       \psi_{l}(\mbf{x})^{\ast} \tens \1  ) X_{\fer , R}     \right\|  \leq
 \delta^{2 , l}_{ \, \fer ,R } (\mbf{x}) , \quad  l= 1, \cdots 4 . 
\end{align*}
Here  $c_{ \, \fer } \geq 0$ is a constant, and $\delta^{i, l}_{\, \fer , R} (\mbf{x}) \geq0  $, $l=1 ,\cdots ,4$, $i= 1, 2$,  are error terms which    satisfy $ \sup\limits_{\mbf{x} \in \Rthree} | \delta^{ i , l }_{\, \fer , R} (\mbf{x}) | < \infty$ and 
 $\lim\limits_{R \to \infty }\delta^{i,l}_{\, \fer , R} (\mbf{x}) =0  $ for all $\mbf{x} \in \Rthree $. 
\end{lemma}
$\;$ \\
\textbf{(Proof)} 
\textbf{(i)} By  (\ref{3.2.1}), we have 
\begin{equation}
 \left\| \left(   X_{\fer  , R}  H_{\D}
-   (     H_{\D} \tens  \1   + \1 \tens  H_{\D}  ) X_{\fer  , R} \right) 
   (N_{\D } +1)^{-1}  \right\| 
   \leq   \|  d \Gamma_{\fer} ( j_{\fer, R} , \tilde{\text{ad}}_{\omega_{\, M}}(j_{\fer ,R})) (N_{\D} +1)^{-1}   \|  , \notag
\end{equation}
and (\ref{3.1.3}) yields that
\begin{equation}
 \|  d \Gamma_{\fer} ( j_{\fer, R} , \tilde{\text{ad}}_{\omega_{\, M}}(j_{\fer ,R})) (N_{\D} +1)^{-1}   \|
\leq  \sqrt{ \| [\omega_{\, M},j^{\, 0}_{\fer , R} ]  \|^2_{B(L^2(\Rthree ))} 
+ \| [\omega_{\, M},j^{\, \infty}_{\fer , R} ]  \|^2_{B(L^2(\Rthree ) )} }  \notag
\end{equation}
By pseudo-differential calculus (e.g., \cite{FGS02} ; Appendix A, \cite{Hida11} ; Section IV), it follows that $\;$ $\| [\omega_{M},j^{\, \sharp}_{\fer , R} ]  \|_{B(L^2(\Rthree ) )} \leq$  $ \frac{ c_{\sharp }}{R}$, $\sharp = 0, \infty $, where $c_{\sharp} \geq 0$ are  constants. Thus \textbf{(i)} is proven.   \\
\textbf{(ii)} By the definition of  $\psi_{l} (\mbf{x})= \sum\limits_{s= \pm 1/2}(b_{s} (f_{s, \mbf{x}}^{\,l}) + d^{\, \dagger }_{s} (g_{s, \mbf{x}}^{\,l}) )$, we have from  (\ref{3.2.2}) and (\ref{3.2.3}) that 
\begin{align*}
& X_{\fer , R} \,  \psi_{l}(\mbf{x}) -  ( \psi_{l}(\mbf{x}) \tens \1  )  X_{\fer , R}   \notag \\
& =\sum_{s= \pm 1/2} \left( X_{\fer , R} \,  b_{s} ((1-j_{\fer, R}^{\, 0}) f_{s, \mbf{x}}^{\,l})
 + \left(     d^{\, \dagger }_{s} ((j_{\fer , R}^{\, 0}-1) g_{s,\mbf{x}}^{\, l})\tens \1    +  (-1)^{N_{\D}} \tens  d_{s}^{\, \dagger } (j_{\fer , R}^{\, \infty} g_{s,\mbf{x}}^{l\, } )\right)   X_{\fer , R} \right) . 
\end{align*}
Then  we have
\begin{align*}
& \left\|  X_{\fer , R} \,  \psi_{l}(\mbf{x})
-   (      \psi_{l}(\mbf{x}) \tens \1  )  X_{\fer , R}      \right\|   \notag \\
 & \leq  \sum_{s= \pm 1/2 }  \left( 
\|  b_{s} ((1-j_{\fer, R}^{\, 0}) f_{s ,\mbf{x}}^{\, l})   \|
+ \| (   d^{\, \dagger}_{s} ((j_{\fer , R}^{\, 0}-1) g_{s ,\mbf{x}}^l)\tens \1 ) X_{\fer , R}  \| + \|  ( \1 \tens  d_{s}^{\, \dagger } (j_{\fer , R}^{\, \infty} g_{s,\mbf{x}}^l )  X_{\fer , R}   \|   \right)   \notag \\
  & \leq \sum_{s= \pm 1/2}  \left( \| ((1-j_{\fer, R}^{\, 0}) f_{s ,\mbf{x}}^{\, l})  \|
+  \| ((j_{\fer, R}^{\, 0}-1) g_{s,\mbf{x}}^l \|
+ \| j_{\fer , R}^{\, \infty} g_{s,\mbf{x}}^l   \| \frac{}{}   \right)  .
\end{align*}
 Let $\delta^{1,l}_{\, \fer , R} (\mbf{x}) = \sum\limits_{s= \pm 1/2 } \left( \| ((1-j_{\fer, R}^{\, 0}) f_{s,\mbf{x}}^{\, l})  \|
+  \| ((1-j_{\fer, R}^{\, 0}) g_{s,\mbf{x}}^l) \|
+ \| j_{\fer , R}^{\, \infty} g_{s,\mbf{x}}^l   \| \frac{}{}   \right) $. 
We see that  $  \sup\limits_{\mbf{x} \in \Rthree} |
\delta^{1, l}_{\, \fer , R} (\mbf{x})| \leq   \sum\limits_{s= \pm 1/2 } \left(   \|f_{s}^{\,l} \| + 2 \|g_{s}^l \|\right) $ and  $\lim\limits_{R \to \infty}
\delta^{1,l}_{\, \fer , R} (\mbf{x}) =0$ for all $\mbf{x} \in \mbf{R}$.  
  Hence  \textbf{(ii)} follows. \\
$\; $ \\  
 \textbf{(iii)} From the definition of  $\psi_{l} (\mbf{x})^{\ast}= \sum\limits_{s= \pm 1/2}(b^{\dagger }_{s} (f_{s, \mbf{x}}^{\,l}) + d_{s} (g_{s, \mbf{x}}^{\,l}) )$,  (\ref{3.2.2}) and (\ref{3.2.3}) yield that   
\begin{align*}
& X_{\fer , R} \,  \psi_{l}(\mbf{x})^{\ast} -  ( \psi_{l} (\mbf{x})^{\ast} \tens \1  )  X_{\fer  , R}   \notag \\
&  = \sum_{s= \pm 1/2} \left( 
  \left(     b^{\dagger }_{s} ((j_{\fer , R}^{\, 0}-1) f_{s,\mbf{x}}^{\, l})\tens \1    +  (-1)^{N_{\D }} \tens  b_{s}^{\dagger } (j_{\fer , R}^{\, \infty} f_{s,\mbf{x}}^{\, l } )
\right)   X_{\fer , R} +
X_{\fer  , R} \,  d_{s} ((1-j_{\fer, R}^{\, 0}) g_{s, \mbf{x}}^{\,l}) 
\right) . 
\end{align*}
Then it follows that 
\begin{equation}
\| X_{\fer , R} \,  \psi_{l}(\mbf{x})^{\ast} -  ( \psi_{l}(\mbf{x})^{\ast} \tens \1  )  X_{\fer , R}   \| \leq 
\delta^{2,l}_{\, \fer , R} (\mbf{x}) ,
\notag 
\end{equation}
where  $\delta^{2,l}_{\, \fer , R} (\mbf{x}) = \sum\limits_{s= \pm 1/2 } \left( 
  \| ((j_{\fer, R}^{\, 0} -1) f_{s,\mbf{x}}^l) \|
+ \| j_{\fer , R}^{\, \infty} f_{s,\mbf{x}}^l   \|
+ \| ((1-j_{\fer, R}^{\, 0}) g_{s,\mbf{x}}^{\, l})  \|
\,   \right) $. 
It is seen that  $  \sup\limits_{\mbf{x} \in \Rthree} |
\delta^{2, l}_{\, \fer , R} (\mbf{x})| \leq   \sum\limits_{s= \pm 1/2 } \left(    2 \|f_{s}^{\,l} \| +  \|g_{s}^l \|\right) $ and  $\lim\limits_{R \to \infty}
\delta^{2,l}_{\, \fer , R} (\mbf{x}) =0$ for all $\mbf{x} \in \mbf{R}$.  
  Thus we obtain \textbf{(iii)}. $\blacksquare $ \\

\begin{corollary}  \label{coro32a} \normalfont  
Assume \textbf{(A.1)}. Then, for all $l ,l' = 1, \cdots 4$,
 \begin{align*}
\textbf{(i)} \, & \left\|   X_{\fer , R} \,  \psi_{l}(\mbf{x})^{\ast} \psi_{l'}(\mbf{x}) 
-   (       \psi_{l}  (\mbf{x})^{\ast} \psi_{l'}(\mbf{x}) \tens \1  ) X_{\fer , R}     \right\|  \leq \delta^{3, l,l'}_{ \, \fer ,R } (\mbf{x}) , 
\\
\textbf{(ii)}  
& \left\|   X_{\fer , R} \,   \psi_{l}(\mbf{x})^{\ast} \psi_{l}(\mbf{x})   \psi_{l'}(\mbf{y})^{\ast} \psi_{l'}(\mbf{y})   
-   (    \psi_{l}(\mbf{x})^{\ast} \psi_{l}(\mbf{x})   \psi_{l'}(\mbf{y})^{\ast} \psi_{l'}(\mbf{y})    \tens \1  ) X_{\fer , R}     \right\|  \leq \delta^{ \, 4, l, l'  }_{ \, \fer ,R } (\mbf{x}, \mbf{y}) .\notag
\end{align*}
 Here  $\delta^{ 3 , l, l' }_{\, \fer , R} (\mbf{x}) \geq 0  $ 
 satisfies  $ \sup\limits_{\mbf{x} \in \Rthree} | \delta^{ 3, l,l'}_{\, \fer , R} (\mbf{x}) | < \infty$ and 
 $\lim\limits_{R \to \infty }\delta^{ 3, l ,l'}_{\, \fer , R} (\mbf{x}) =0  $ for all 
$\mbf{x} \in \Rthree $, and  $\delta^{ 4 , l, l' }_{\, \fer , R} (\mbf{x} ,\mbf{y} ) \geq0 $ satisfies 
 $ \sup\limits_{ (\mbf{x} ,\mbf{y}) \in \Rthree \times \Rthree } | \delta^{ \, 4, l,l'}_{\, \fer , R} (\mbf{x} ,\mbf{y}) | < \infty$ and  
 $\lim\limits_{R \to \infty }\delta^{ \, 4, l ,l'}_{\, \fer , R} (\mbf{x} , \mbf{y}) =0  $ for all $ \mbf{x} , \mbf{y} \in  \Rthree  $. 
\end{corollary}
\textbf{(Proof)}
\textbf{(i)} By Lemma \ref{lemma32a} \textbf{(ii)} and  \textbf{(iii)}, it is seen that 
\begin{align}
& \left\|   X_{\fer , R} \, \psi_{l}(\mbf{x})^{\ast} \psi_{l'}(\mbf{x})
-   (       \psi_{l}(\mbf{x})^{\ast} \psi_{l'}(\mbf{x}) \tens \1  ) X_{\fer , R}  \right\|
\notag \\
&\leq \left\|  \left(  X_{\fer , R} \, \psi_{l} (\mbf{x})^{\ast}  
 -    \left( (       \psi_{l}(\mbf{x})^{\ast}\tens \1  ) X_{\fer , R} \right) \right) \psi_{l'}(\mbf{x}) \right\|  \notag \\
& \qquad \qquad \qquad + \left\| (\psi_{l}(\mbf{x})^{\ast} \tens \1 ) \left(  X_{\fer , R} \,   \psi_{l'}(\mbf{x})   -   (       \psi_{l'}(\mbf{x})\tens \1  ) X_{\fer , R} \right)\right\| \notag \\
& \leq \delta_{\fer , R}^{2,l} \|\psi_{l'}(\mbf{x})  \| 
+ \delta_{\fer , R}^{1,l'} \| \psi_{l}(\mbf{x})^{\ast}  \|  . \notag 
\end{align}
Note that  $  \|   \psi_{\, l'}(\mbf{x})  \| \leq c_{\, \D}^{\, l'}$ and 
 $ \| \psi_{l}(\mbf{x})^{\ast}  \| \leq c_{\, \D}^{l} $. Hence  \textbf{(i)} is obtained. Similarly, 
 we can prove \textbf{(ii)} by using  \textbf{(i)}. $\blacksquare$ \\

\subsection{Partition of Unity for  Radiation Field }
Let  $U_{\bos}:   \ms{F}_{\bos }( L^2( \Rthree_{\mbf{k}} \times \{ 1,2  \}  )  \oplus  L^2( \Rthree_{\mbf{k}} \times \{ 1,2  \}  ) ) \to  \ms{F}_{\rad} \tens \ms{F}_{\rad} $ an isometric operator satisfying   $U_{\bos} \,  \Omegarad = \Omegarad \tens \Omegarad $ and 
\begin{align*}
&U_{\bos } \, a_{r_1}^{\dagger } \left(\left[ \begin{array}{c} f_1 \\ g_1 \end{array} \right]\right) \cdots  a_{r_1}^{\dagger } \left(\left[ \begin{array}{c} f_1 \\ g_n \end{array} \right]\right) \Omegarad     \\
= & \,  \left( a^{\dagger }_{r_1} (f_1) \tens \1  + \1 \tens a^{\dagger }_{r_1}(g_1) \frac{}{}\right)
\cdots  \left( a^{\dagger}_{r_n} (f_n) \tens \1  + \1 \tens a^{\dagger}_{r_n}(g_n) \frac{}{} \right)
 \Omegarad \tens \Omegarad .  
\end{align*}
 Let  $j_{0} , j_{\, \infty} \in C^{\infty} (\mbf{R})$. We suppose that $j_0 \geq 0$, $j_{\infty} \geq0 $,  $j_{0}^2 +  j_{\infty}^2 =1$, 
$j_{0}(\mbf{y})=1 $ if $|\mbf{y}| \leq 1 $ and $j_{0}(\mbf{y})=0 $ if $|\mbf{y}| \geq 2 $.  We set $j_{\bos , R } =\left[ \begin{array}{c} j_{\bos , R }^{\, 0} \\ j_{\bos , R }^{\, \infty}  \end{array} \right] $ where
$ j_{\bos , R }^{\, 0}= j_{0} (\frac{-i \mbf{\nabla_{\mbf{k}}}}{R})$ and $ j_{\bos , R }^{\, \infty}=  j_{\infty}
 (\frac{-i \mbf{ \nabla_{\mbf{k}}}}{R})    $ with $\nabla_{\mbf{k}}= (\partial_{k^1}, \partial_{k^2},\partial_{k^3} ) $.   \\
 
 $\;$ \\
Let $Y_{\bos , R} : \Frad \to \ms{F}_{\rad} \tens \ms{F}_{\rad} $  defined by 
 \begin{equation}
 Y_{\bos  , R} = U_{\bos } \, \Gamma_{\bos}(j_{\bos, R})  .
\notag
 \end{equation}
 From 
(\ref{3.1.8}) - (\ref{3.1.10}), it follows that 
 \begin{align}
& Y_{\bos , R} \, \Hradm = 
\left( \Hradm \tens \1 + \1 \tens \Hradm \frac{}{} \right) 
Y_{\bos , R}
-  U_{\bos } \, d \Gamma_{\bos} ( j_{\bos, R} , \tilde{\text{ad}}_{\omega_{m} }(j_{\bos ,R})) ,  \label{3.3.1} \\
& Y_{\bos , R} \,  a_{r}(h)
= (a_r (h) \tens \1 ) Y_{\bos  , R}
+ Y_{\bos , R} \,
 a_{r}((1-j_{\bos , R}^{\, 0} )h) ,   \label{3.3.2} \\
&Y_{\bos , R} \, a^{\dagger}_{r}(h)
= (a^\ast_r (h) \tens \1 ) Y_{\bos , R}
+ \left(a^{\dagger}_r ((j_{\bos , R}^{\, 0}-1) h ) \tens \1 
+ \1 \tens a_{r}^{\dagger }(j_{\bos, R}^{\, \infty} h) \right) Y_{\bos , R}  .
  \label{3.3.3}
\end{align}

\begin{lemma}   \label{lemma33a} \normalfont 
Assume \textbf{(A.2)}. Then 
\begin{align*}
& \textbf{(i)}
\left\| \left(  Y_{\bos , R} \,  \Hradm 
-   (      \Hradm \tens  \1    + \1 \tens  \Hradm  ) \, Y_{\bos , R} \frac{}{} \right) 
    (N_{\rad} +1)^{-1} \right\|  \leq  \frac{c_{ \, \bos}}{R}  , \\   
 & \textbf{(ii)} 
\left\| \left(  Y_{\bos , R} \, A_{j}(\mbf{x})
-   (      A_{j}(\mbf{x}) \tens \1  ) Y_{\bos , R} \frac{}{} \right) (N_{\rad}+1)^{-1/2} \right\|  \leq \delta^j_{ \, \bos ,R } (\mbf{x}) .
\end{align*}
Here $c_{ \, \bos} \geq 0$ is a constant and $\delta^j_{\, \bos , R} (\mbf{x}) \geq0  $, $j=1, 2,3$, are error terms which satisfy  $ \sup\limits_{\mbf{x} \in \Rthree} | \delta^j_{\, \bos , R} (\mbf{x}) | < \infty$ and 
 $\lim\limits_{R \to \infty }\delta^j_{\, \bos , R} (\mbf{x}) =0  $ for all $ \mbf{x} \in \Rthree$.
\end{lemma}
\textbf{(Proof)}
\textbf{(i)} It is  proven in a similar way to
 Lemma \ref{lemma32a} \textbf{(i)}. \\
\textbf{(ii)} By the definition of  $ A_{j}(\mbf{x}) = \sum\limits_{r=1,2} \left( a_{r}(h^j_{r , \mbf{x}}) +   a^{\dagger}_{r}(h^j_{r , \mbf{x}}) \right)$, it follows from (\ref{3.3.2}) and (\ref{3.3.3}) that 
\begin{align}
&  Y_{\bos , R} \,  A_{j}(\mbf{x})
-   (      A_{j}(\mbf{x}) \tens \1  )  Y_{\bos , R}    \notag \\
&= \sum_{r=1,2} \left( Y_{\bos , R} \,
 a_{r}((1-j_{\bos , R}^{\, 0} )h_{r, \mbf{x}}^j)
+ \left(a^{\dagger}_r ((j_{\bos , R}^{\, 0}-1) h_{r, \mbf{x}}^j ) \tens \1 
+ \1 \tens a_{r}^{\dagger }(j_{\bos, R}^{\, \infty} h_{r, \mbf{x}}^j ) \right) Y_{\bos , R} \right)
 .   \notag 
\end{align}
 Since   $\|a_{r}(h) (N_{\rad } + 1)^{-1/2}  \| \leq \| h \|$ and  $\|a^{\dagger }_{r}(h) (N_{\rad } + 1)^{-1/2}  \| \leq 2 \| h \|$, we have
\begin{align}
& \left\| \left( Y_{\bos , R} \,  A_{j}(\mbf{x})
-   (      A_{j}(\mbf{x}) \tens \1  )  Y_{\bos , R}  \right)  (N_{\rad} +1)^{-1/2}
 \right\|   \notag \\
 & \leq  \sum_{r=1,2}  \left( 
\| ( a_{r} (1-j_{\bos, R}^{\, 0}) h_{r,\mbf{x}}^j)   (N_{\rad} +1)^{-1/2} \|
 \right. \notag  \\
 & \qquad \qquad \left.  
+ \|  (a^{\dagger}_{r} ((j_{\bos, R}^{\, 0}-1) h_{r,\mbf{x}}^j)(N_{\rad} +1)^{-1/2} \tens \1 )  ( (N_{\rad} +1)^{1/2} \tens \1 )Y_{\bos , R}    (N_{\rad} +1)^{-1/2}
 \| \right. \notag  \\
&\qquad \qquad \qquad \left.  + \| ( \1 \tens  a_{r}^{\dagger} (j_{\bos, R}^{\, \infty} h_{r,\mbf{x}}^j ) (N_{\rad} +1)^{-1/2} ) 
( \1 \tens  (N_{\rad} +1)^{1/2}  )
 Y_{\bos , R}    (N_{\rad} +1)^{-1/2}  \|  \frac{}{} \right) \notag \\
& \leq  \sum\limits_{r=1,2}  \left( \| (1-j_{\bos, R}^{\, 0}) h_{r,\mbf{x}}^j  \|
+2  \| (j_{\bos, R}^{\, 0}-1)h_{r,\mbf{x}}^j \|
+ 2 \| j_{\bos, R}^{\, \infty} h_{r,\mbf{x}}^j   \| \frac{}{}   \right) . 
  \notag 
\end{align}
 Let $\delta^j_{\, \bos , R} (\mbf{x}) = \sum\limits_{r=1,2}  \left( 3 \| ((1-j_{\bos, R}^{\, 0}) h_{r,\mbf{x}}^j)  \|
+ 2 \| j_{\bos, R}^{\, \infty} h_{r,\mbf{x}}^j   \| \frac{}{}   \right) $, $j=1,2  ,3$. 
We see that  $  \sup\limits_{\mbf{x} \in \Rthree} |\delta^j_{\, \bos , R} (\mbf{x})| 
\leq 5  \left(   \|h_{1}^j \| + \|h_{2}^j \|\right) $ and   $\lim\limits_{R \to \infty}\delta^j_{\, \bos , R} (\mbf{x}) =0$ for all $\mbf{x} \in \mbf{R}$.  Thus we obtain the proof.   $\blacksquare$ \\

\subsection{Existence of Ground State of $H_{m}$}
We recall that  the massive Hamiltonian is defined by
\begin{equation}
H_{m} = H_{\D} \tens \1_{\rad} + \1_{\D} \tens \Hradm + \kappaI \HI + \kappaII \HII  . \notag 
\end{equation}
 Throughout this subsection, we do not omit the subscripts of the identities $\1_{\D}$ and 
 $\1_{\rad}$.    \\
$\;$ \\
 Since $\frac{1}{\omega_m(\mbf{k})^{\lambda}} \leq \frac{1}{\omega (\mbf{k})^{\lambda}}  $, $\lambda>0$, it holds that  
\begin{equation}
 \|A_j (\mbf{x}) ( \Hradm +1 )^{-1/2}  \| 
\leq \sum_{r=1,2}  \left( 
2 \| \frac{ \chi_{\rad} e_{r}^{\, j}}{ \omega_{m}} \|  +   \| \frac{ \chi_{\rad} e_{r}^{\, j}}{ \sqrt{\omega_m}} \| \right) 
\leq  c_{\rad}^{\, j} . \label{9/1.1}
\end{equation}
Then, we have
\begin{equation}
\| \HI \Psi \| \leq  c_{\,\I } \,   \| \1 \tens (\Hradm +1)^{1/2} \Psi \| , 
  \quad  \label{HImbound} 
\end{equation}
and it holds that for all $\epsilon > 0$, 
\begin{equation}
\|\HI \Psi \| \leq c_{\I} \epsilon \|H_{0 , m} \Psi \| + c_\I \left( \frac{1 }{2 \epsilon}  +1 \right)  \| \Psi \|  .
  . \label{HImbound'} 
\end{equation}
From (\ref{HImbound'}) and $\| H_{\II}\| < \infty $, it is proven that  $H_{m}$ is self-adjoint and  essentially self adjoint on any core of $H_{0,m}$. \\

 \begin{theorem}  \normalfont \label{Massive-Case}
\textbf{(Existence of a Ground State of $H_m $)} \\
Suppose \textbf{(A.1)} - \textbf{(A.3)}. Let $m <M$.  Then 
$H_m $ has purely discrete spectrum in $[ E_{0} (H_m ) , E_{0} (H_m ) + m )$.   
   In particular, $ \Hm $ has a ground state. 
\end{theorem}

$\; $ \\ To prove Theorem \ref{Massive-Case},  we need some preparations.
We define
  $ \tilde{X}_{\fer , R } : \ms{F}_{\QED} \to \ms{F}_{\Dir} \tens \FDir \tens \Frad $ by
\[
 \tilde{X}_{\fer , R } =   X_{\fer , R} \tens \1_{\rad} .
\]
We introduce Hamiltonian $\tilde{H}_{m} : \FDir \tens \FDir \tens \Frad  \to \FDir \tens \FDir \tens \Frad $  defined by 
\begin{equation} 
\tilde{H}_{m} = \tilde{H}_{\D} \tens \1_{\rad} + \1_{\D} \tens \tilde{H}_{\rad}  + \kappaI
\tilde{H}_{\I} + \kappaII \tilde{H}_{\II}  , \notag  
\end{equation}
where $\tilde{H}_{\D} =  H_{\D} \tens  \1_{\D} $, $ \tilde{H}_{\rad} =  \1_{\D}  \tens \Hradm $  and
\begin{align*}
&\tilde{H}_{\I} = \sum_{j=1}^{3} \intRthree  
  \chi_{\I} (\mbf{x}) (\tilde{\psi}^{\dagger}(\mbf{x}) \tilde{\alpha}^j \tilde{\psi}(\mbf{x} )  \tens A_j (\mbf{x} )) d\mbf{x} , \\
& \tilde{H}_{\II} = 
\int_{\Rthree \times \Rthree} 
     \frac{\chiIIx \chiIIy}{|\mbf{x} -\mbf{y}|} 
\left( \tilde{\psi}^{\dagger }  (\mbf{x}) \tilde{\psi}  (\mbf{x}) \tilde{\psi}^{\dagger}   (\mbf{y}) \tilde{\psi} (\mbf{y})\tens  \1_{\rad} \right)
 d \mbf{x} \, d \mbf{y}. 
 \end{align*}
with  $\tilde{\psi}(\mbf{x})=\psi(\mbf{x}) \tens   \1_{\D}$ and $\tilde{\alpha}^{j} = \alpha^j  \tens  \1_{\D}$, $j=1, \cdots 3$. 
$\; $ \\

\begin{proposition} \label{9/9.a} \normalfont 
Assume \textbf{(A.1)} - \textbf{(A.3)}. Let   $\Psi \in \ms{D}(H_{m})$. Then, it holds that 
\begin{align*}
& \textbf{(i)} \; \;
\left\|  \left(   \tilde{X}_{\fer , R }   ( H_{\D} \tens \1_{\rad}  )
-   (    \tilde{H}_{\D}   \tens \1_{\rad}   +  \1_{\D} \tens  H_{\D} \tens  \1_{\rad} ) \tilde{X}_{\fer , R }  \right) \Psi  \right\|    \notag \\
& \quad \;  \; \; \; \leq  \frac{ c_{ \, \fer }}{R} \, 
 \left( \left\|        (   N_{\D} \tens  \1_{\rad}   ) \Psi \right\|  + \left\| \Psi \right\| \right) , \\
 & \textbf{(ii)} \;  \, \,
 \left\|  \left(  \tilde{X}_{\fer , R }  \HI
-         \tilde{\HI}    \tilde{X}_{\fer , R } \right) \Psi      \right\| 
 \leq \delta_{\, \fer , \I } (R )  \,   
 \left( \|  (  \1_{\D} \tens  N_{\rad}^{1/2}  ) \Psi \|
+  \left\|  \Psi \right\| \right), \\
& \textbf{(iii)} \;
\left\|   \left(  \tilde{X}_{\fer , R }  \HII
-         \tilde{\HII}    \tilde{X}_{\fer , R } \right) \Psi   \right\|
  \leq \delta_{\, \fer , \II} (R) \,  \| \Psi \| .
\end{align*}
Here $c_{\fer} \geq 0$ is the constant in Lemma \ref{lemma32a}\textbf{(i)}, and   $\delta_{\, \fer , \I}(R )  \geq 0 $ and  $\delta_{\, \fer , \II} (R)  \geq 0 $ are error terms satisfying  that  $\lim\limits_{R \to \infty }\delta_{\, \fer , \I } (R)  =0  $ and 
$\lim\limits_{R \to \infty }\delta_{\, \fer , \II } (R) =0  $, respectively. 
\end{proposition}
\textbf{(Proof)}\\
\textbf{(i)}  It directly follows from  Lemma \ref{lemma32a} (\textbf{i}). \\
\textbf{(ii)}
 Let $\Psi \in \ms{D}(H_m ) $ and $\tilde{\Phi} \in \ms{F}_{\Dir} \tens  \ms{F}_{\Dir} \tens \Frad $ with $\| \tilde{\Phi} \| =1 $. Then, 
\begin{align}
& \left( \tilde{\Phi} ,  \left( \tilde{X}_{\fer , R } \HI \, 
-   \tilde{H}_{\I}    \tilde{X}_{\fer , R } 
 \right)  \Psi \right)    \notag \\
 & = \sum_{j=1}^3 \int_{\Rthree} \chiIx \left(   \Phi , 
 \left( \left( X_{\fer , R} \psidaggerx  (\mbf{x})  \alpha^j  \psi  (\mbf{x}) 
-  \tilde{\psi}^{\dagger} (\mbf{x})  \tilde{\alpha}^j  \tilde{\psi} (\mbf{x}) 
  X_{\fer , R} \right) \tens A_{j} (\mbf{x})
  \right) \Psi \right) \dx  . \notag   
\end{align}
Then we have
\begin{align}
& \left| \left( \tilde{\Phi} ,  \left( \tilde{X}_{\fer , R } \HI \, 
-   \tilde{H}_{\I}    \tilde{X}_{\fer , R } 
 \right)  \Psi \right)  \right|  \notag \\
 & \leq   \sum_{j=1}^3 \int_{\Rthree}  \left| \chiIx \right| \, 
  \left\|  X_{\fer , R} \psi^{\dagger} (\mbf{x})  \alpha^j  \psi  (\mbf{x}) 
-  \tilde{\psi}^{\dagger} (\mbf{x})  \tilde{\alpha}^j  \tilde{\psi} (\mbf{x}) 
  X_{\fer , R} \right\| \,  \left\| \left(  \1_{\D} \tens A_{j} (\mbf{x})  \right) \Psi \right\| \dx   \notag \\
& \leq   \sum_{j=1}^3 \sum_{l,l'=1}^4 |\alpha_{l , l'}^j | \int_{\Rthree}  \left| \chiIx \right| \, 
  \|  \left( X_{\fer , R} \psi_{l} (\mbf{x})^{\ast}   \psi_{l'} (\mbf{x})
-  \tilde{\psi}_{l} (\mbf{x})^{\ast}  \tilde{\psi}_{l'} (\mbf{x}) 
  X_{\fer , R} \right) \|  \left\| \left(  \1_{\D} \tens A_{j} (\mbf{x})  \right) \Psi \right\| \dx .  \notag
\end{align}
By Corollary \ref{coro32a} \textbf{(i)}, we have 
$ \|  \left( 
X_{\fer , R} \psi_{l} (\mbf{x})^{\ast}  \psi_{l'} (\mbf{x}) 
-  \tilde{\psi}_{l} (\mbf{x})^{\ast}   \tilde{\psi}_{l'} (\mbf{x}) 
  X_{\fer , R} 
\right) \| \leq  \delta_{\, \fer , R}^{3, l, l'} (\mbf{x}) $. We also see that  $ \| A_{j} (\mbf{x}) (N_{\rad} +1)^{-1/2}\| \leq 3 \sum\limits_{r=1,2}\| h_{r}^j\|  $. Then   it follows that
\begin{align}
 & \left|  \left( \tilde{\Phi} ,  \left( \tilde{X}_{\fer , R } \HI \, 
-   \tilde{H}_{\I}    \tilde{X}_{\fer , R } 
 \right)  \Psi \right)  \right|  \notag \\
&   \quad \qquad \leq     \sum_{r=1,2} \sum_{j=1}^3 \sum_{l,l'=1}^4    |\alpha_{l , l'}^j | \| h_{r}^j  \|
 \left( \int_{\Rthree} | \chiIx  |
   \delta_{\, \fer , R}^{3, l, l'} (\mbf{x}) \dx  \frac{}{} \right)    \,
\|( \1_{\D } \tens (N_{\rad} +1 )^{1/2}) \Psi \|  .  \label{9/9.2}
\end{align}
Since (\ref{9/9.2})  holds for all $\tilde{\Phi} \in \ms{F}_{\Dir} \tens  \ms{F}_{\Dirac} \tens \Frad $ with $\| \tilde{\Phi} \| =1 $,  it follows that 
\begin{equation}
\left\|  \left( \tilde{X}_{\fer , R } \HI \, 
-   \tilde{H}_{\I}    \tilde{X}_{\fer  , R } 
 \right)  \Psi   \right\|    \leq    \delta_{\, \fer , \I } (R) \,  
\|( \1_{\D} \tens (N_{\rad} +1 )^{1/2}) \Psi \|  ,
\end{equation} 
where  $\delta_{\, \fer , \I } (R)  =  3 \sum\limits_{r=1,2} \sum\limits_{j=1}^3 \sum\limits_{l,l'=1}^4 \, |\alpha_{l , l'}^j | \, \| h_{r}^j  \| \, 
  \|  \chi_{\I}  \, 
   \delta_{\, \fer , R}^{3, l, l'} \|_{L^1} $.
  We see that   $ \lim\limits_{R \to \infty}
 \delta_{\, \fer  ,\I } (R) =0 $, and hence \textbf{(ii)} follows. \\
\textbf{(iii)} 
Let $\Psi \in \ms{D}(H_m)$. We set $Q_{ l  }(\mbf{x})= \psi_{l} (\mbf{x})^{\ast}  \psi_{l} (\mbf{x}) $. Then for all $\tilde{\Phi} \in \ms{F}_{\Dir} \tens  \ms{F}_{\Dir} \tens \Frad $ with $\| \tilde{\Phi} \| =1 $, 
\begin{align}
& \left( \tilde{\Phi} ,  \left( \tilde{X}_{\fer , R } \HII \, 
-   \tilde{H}_{\II}    \tilde{X}_{\fer , R } 
 \right)  \Psi \right)    \notag \\
 & = \sum_{l, l'=1}^4\int_{\Rthree \times \Rthree} \frac{\chiIIx \chiIIy}{|\mbf{x}-\mbf{y}|}  \left(   \tilde{\Phi} , 
 \left( ( X_{\fer , R} Q_{ l }(\mbf{x})  Q_{l' } (\mbf{y} )
-   (( {Q}_{ l }(\mbf{x}) {Q}_{l'} (\mbf{y} ) \tens \1_{\D}   ) 
  X_{\fer , R} ) \tens \1_{\rad}
  \right) \Psi \right) \dx  \dy. \notag
\end{align} 
Then we have 
\begin{align}
& \left|  \left( \tilde{\Phi} ,  \left( \tilde{X}_{\fer , R } \HII \, 
-   \tilde{H}_{\II}    \tilde{X}_{\fer , R } 
 \right)  \Psi \right)   \right|  \notag \\
 & \leq    \sum_{l, l'=1}^4\int_{\Rthree \times \Rthree}
 \frac{ | \chiIIx \chiIIy | }{|\mbf{x}-\mbf{y}|}   
\left\|  \left( X_{\fer , R} Q_{ l }(\mbf{x})  Q_{l' } (\mbf{y} )
-   ( {Q}_{ l }(\mbf{x})  {Q}_{l' } (\mbf{y} ) \tens \1_{\D}   ) 
  X_{\fer , R} \right)  \Psi  \right\| 
  \dx  \dy     .  \notag 
\end{align}
From Corollary  \ref{coro32a} \textbf{(ii)}, it holds that $\left\|   X_{\fer , R} Q_{ l }(\mbf{x})  Q_{l' } (\mbf{y} )
-   ({Q}_{ l }(\mbf{x}) {Q}_{l' } (\mbf{y} ) \tens \1_{\D}   ) 
  X_{\fer , R}  \right\|  \leq \delta_{\fer , R}^{4, l, l'} (\mbf{x} , \mbf{y})$. Then we have
\begin{equation}
\left|  \left( \tilde{\Phi} ,  \left( \tilde{X}_{\fer , R } \HII \, 
-   \tilde{H}_{\II}    \tilde{X}_{\fer , R } 
 \right)  \Psi \right)   \right|   \leq \left( \sum_{l, l'=1}^4\int_{\Rthree \times \Rthree} 
\frac{ | \chiIIx \chiIIy | }{|\mbf{x}-\mbf{y}|}   \delta_{\fer , R}^{4, l, l'} (\mbf{x} , \mbf{y})   \dx  \dy   \right) \,  \left\|  \Psi  \right\|    . \notag
\end{equation}
This implies that 
\begin{equation}
\left\|   \left( \tilde{X}_{\fer , R } \HII \, 
-   \tilde{H}_{\II}    \tilde{X}_{\fer , R } 
 \right)  \Psi \right\|   \leq  \delta_{\, \fer , \II } (R)  \, \left\|  \Psi  \right\| , \notag  
\end{equation}
where $ \delta_{ \, \fer , \II } (R)  = \sum\limits_{l, l'=1}^4 \int_{\Rthree \times \Rthree} \frac{ | \chiIIx \chiIIy |}{|\mbf{x}-\mbf{y}|}   \delta_{\, \fer , R}^{4, l, l'} (\mbf{x} , \mbf{y})   \dx  \dy   $. 
We see that $\lim\limits_{R \to \infty }
\delta_{\, \fer , \II } (R) =0$, and thus the proof is obtained. $\blacksquare$\\

$\;$  \\
We define  $\tilde{Y}_{\bos , R } : \ms{F}_{\QED} \to \FDir \tens \Frad \tens \Frad $ by
\[ 
\tilde{Y}_{\bos , R } = \1_{\D} \tens  Y_{\bos , R}  .
\]

\begin{proposition} \label{9/9.b}  \normalfont 
Assume \textbf{(A.1)} - \textbf{(A.3)}.   Then it holds that for all $\Psi \in \ms{D}(H_m )$, 
\begin{align*}
& \textbf{(i)}
\left\|  \left( \tilde{Y}_{\bos , R } ( \1_{\D} \tens  \Hradm )
-  (  \1_{\D} \tens   \Hradm \tens  \1_{\rad}   +  \1_{\QED} \tens  \Hradm  )
 \tilde{Y}_{\bos , R }  \right) \Psi   \right\|  \notag \\
 & \qquad \qquad   \leq  \frac{ c_{ \, \bos }}{R}  \,  
  \left( \left\|    (\1_{\D} \tens  N_{\rad}   )   \Psi \right\|   + 
 \left\| \Psi \right\|  \right) , \\
 & \textbf{(ii)} 
 \left\|   \left( \tilde{Y}_{\bos , R } \, \HI
-   (      \HI \tens \1_{\rad}  ) \tilde{Y}_{\bos , R } 
 \right)  \Psi \right\|  \leq 
  \delta_{ \, \bos , \I } (R )  \left( 
  \|  (\1_{\D} \tens  N_{\rad}^{1/2}  )  \Psi  \| + \|  \Psi  \|  \right)  ,
\end{align*}
 where $c_{ \, \bos} \geq 0 $    and  $\delta_{ \, \bos  ,\I } (R)  \geq 0 $  satisfying   $\lim\limits_{R \to \infty } 
 \delta_{ \, \bos , \I } (R)  =0 $.
\end{proposition}
\textbf{(Proof)} 
\textbf{(i)} It   immediately  follows from Lemma \ref{lemma33a} \textbf{(i)}. \\
 \textbf{(ii)} Let $\Psi \in \ms{D}(H_m )$ and $\tilde{\Xi} \in  \FDir \tens \Frad  \tens \Frad   $ with $\| \tilde{\Xi} \| =1$. 
We see that
\begin{align}
& \left( \tilde{\Xi} ,  \left( \tilde{Y}_{\bos , R } \, \HI
-   (      \HI \tens \1_{\rad}  ) \tilde{Y}_{\rad , R } 
 \right)  \Psi \right)    \notag \\
 & = \sum_{j=1}^3 \int_{\Rthree} \chiIx \left(   \tilde{\Xi} , 
\left( \psidaggerx  \alpha^j  \psi (\mbf{x}) \right)
 \tens \left( Y_{\bos , R }  A_{j} (\mbf{x}) \, -   (   A_{j} (\mbf{x})\tens \1_{\rad}  )
 Y_{\bos , R } )
 \right)  \Psi \right) \dx  , \notag .
\end{align}
Then,
\begin{align}
& \left| \left( \tilde{\Xi} ,  \left( \tilde{Y}_{\bos , R } \, \HI
-   (      \HI \tens \1_{\rad}  ) \tilde{Y}_{\bos , R } 
 \right)  \Psi \right)  \right|    \notag \\
 & \leq  \sum_{j=1}^3 \int_{\Rthree} | \chiIx | \,   
\left\| (  \psidaggerx  \alpha^j  \psi (\mbf{x}) \tens \left( Y_{\bos , R }  A_{j}  \, -   (   A_{j} (\mbf{x})\tens \1_{\rad}  )
 Y_{\bos , R } )  \right)  \Psi  \right\| \dx  \notag \\
&  \leq \left(  \sum_{j=1}^3 \sum_{l,l'=1 }^4 |\alpha^j_{l,l'}| c_{\, \D}^{\, l} c_{\, \D}^{\, l'} \right)
\int_{\Rthree} | \chiIx |  \left\|
 \left( \1_{\D} \tens \left( Y_{\bos , R }  A_{j}  \, -   (   A_{j} (\mbf{x})\tens \1_{\rad}  )
 Y_{\bos , R } ) \right)  \right)  
\Psi  \right\| \dx.  \notag 
\end{align}
From Lemma \ref{lemma33a} \textbf{(ii)}, it holds that   
\begin{equation}
\left\| ( \1_{\D} \tens \left( Y_{\bos , R }  A_{j}  \, -   (   A_{j} (\mbf{x})\tens \1_{\rad}  )  Y_{\bos , R } )  \right)  \Psi  \right\|
\leq  \delta_{\, \bos ,R}^j (\mbf{x}) \|(\1_{\D} \tens (N_{\rad}+1 )^{1/2})  \Psi   \| , \notag
\end{equation}
where $ \delta_{\bos , R}^j (\mbf{x}) \geq 0 $  is the error term, and hence,
\begin{equation}
 \left| \left( \tilde{\Xi} ,  \left( \tilde{Y}_{\bos , R } \, \HI
-   (      \HI \tens \1_{\rad}  ) \tilde{Y}_{\bos , R } 
 \right)  \Psi \right)  \right| \leq  \delta_{\, \bos , \I} (R) \, \|(\1_{\D} \tens (N_{\rad}+1 )^{1/2})  \Psi   \| , \label{9/9.3}
\end{equation}
where $\delta_{\, \bos ,\I} (R)   = \sum\limits_{l,l'=1 }^4 |\alpha^j_{l,l'}| c_{\, \D}^{\, l} c_{\, \D}^{\, l'} \int_{\Rthree} | \chiIx |  \delta^j_{\, \bos } (\mbf{x})  dx  $.
Since (\ref{9/9.3}) holds for all $\tilde{\Xi} \in  \FDir \tens \Frad  \tens \Frad   $ with $\| \tilde{\Xi} \| =1$, we have 
\begin{equation}
 \left\|   \left( \tilde{Y}_{\bos , R } \, \HI
-   (      \HI \tens \1_{\rad}  ) \tilde{Y}_{\rad , R } 
 \right)  \Psi  \right\| \leq  \delta_{\, \bos , \I} (R) \, \|(\1_{\D} \tens (N_{\rad}+1 )^{1/2})  \Psi   \| . \notag
\end{equation}
Since $\lim\limits_{R \to \infty }\delta_{\, \bos , \I}(R)=0$, the proof is obtained. $\blacksquare$ \\

$\; $ \\
Here we introduce a new norm defined by
\[ \qquad 
 \|  \Psi \|_{ \lambda  , \, \lambda  '  } \; = \; \|(  N_{\D}^{\, \lambda /2} \tens \1_{\rad} ) \Psi  \| + \|( \1_{\D} \tens N_{\rad}^{ \, \lambda ' /2}  )\Psi  \| + \| \Psi \| , \quad \Psi \in \ms{D} (N_{\D}^{\lambda  /2 } \tens N_{\rad}^{ \lambda  ' /2}  ) .
 \]

$\;$ \\  
From Proposition \ref{9/9.a} and Proposition \ref{9/9.b}, the next corollary follows.
$\;$ \\
\begin{corollary}  \label{9/9.c} \normalfont 
 Assume \textbf{(A.1)} - \textbf{(A.3)}. Then for all $\Psi \in \ms{D}(H_m)$, 
\begin{align*}
&\textbf{(i)} \;  \left\|  \left( \tilde{X}_{\fer , R } H_{m}
-  ( \tilde{H}_{m}   +  \1_{\D} \tens  H_{\D} \tens \1_{\rad } )
 \tilde{X}_{\fer , R}   \right) \Psi  \right\| 
\leq \delta_{ \, \fer   } (R)  \| \Psi \|_{2,1 } ,  \\
&\textbf{(ii)} \; \left\|  \left( \tilde{Y}_{\bos , R } H_{m}
-  ( H_{m} \tens  \1_{\rad}   +  \1_{\D} \tens \1_{\rad} \tens  \Hradm  )
 \tilde{Y}_{\bos , R }  \right) \Psi \right\| \leq \delta_{ \,\bos    } (R)
  \| \Psi \|_{0, 2} \, .
\end{align*}  
Here $ \delta_{ \, \fer  } (R) \geq 0 $ and $ \delta_{ \, \bos } (R) \geq 0 $ are error terms which satisfy that 
$ \lim\limits_{R \to \infty} \delta_{ \, \fer  } (R)=0 $ and $ \lim\limits_{R \to \infty} \delta_{ \, \bos } (R) =0 $, respectively. \\
\end{corollary}

\begin{lemma} \label{9/9.d} \normalfont  \label{LformboundHm}
Assume \textbf{(A.1)} - \textbf{(A.3)}. Let $q_{\, \fer , R} = (j^{\, 0}_{\, \fer , R})^2$ and $q_{\, \bos , R} = (j^{\, 0}_{\, \bos , R})^2$. Then, for all  $\Psi \in 
\ms{ D} (H_m ) $ with $\| \Psi \| =1$, 
\begin{align*}
 (\Psi, H_m \Psi )     \geq & E_{0} (H_m)  + \, m  \,  +  \,   (M-m)  
 \left( \Psi,   ( \1_{\D } \tens  \Gamma_{\bos} (q_{\, \bos , R}) )  \Psi \right) \\
&    - M \left( \Psi ,  
 \left( \Gamma_{\fer} (q_{\, \fer , R}) \tens \Gamma_{\bos} (q_{\, \bos , R}) \right) \Psi \right) +  \left(  \delta_{\, \fer } (R ) \|  \Psi  \|_{2,1 } + \delta_{\, \bos} (R )  \|  \Psi \|_{0,2 }  \right) .
 \end{align*}
\end{lemma}
\textbf{(Proof)} 
Let $\Psi \in \ms{ D} (H_m ) $ with $\| \Psi \| =1$.  By Lemma Corollary \ref{9/9.c} \textbf{(ii)},
\begin{align}
 (\Psi , H_m \Psi )
& = \left(  \Psi , 
 \tilde{Y}_{\, \bos ,R}^{\ast} \tilde{Y}_{\, \bos ,R} H_m  \Psi \right) \notag \\
&  \geq ( \Psi ,  \tilde{Y}_{\, \bos ,R}^{\ast}  (  H_m  \tens  \1_{\rad})
   \tilde{Y}_{\, \bos ,R}   \Psi )  + ( \Psi , \tilde{Y}_{\, \bos ,R}^{\ast}
 ( \1_{\D} \tens \1_{\rad}  \tens \Hradm )
   \tilde{Y}_{\, \bos ,R}   \Psi )  - \delta_{\, \bos } (R)  \| \Psi \|_{0,2} . \notag   
   \end{align}
  We see that 
$ \Hradm \geq m (  \1_{\rad} - \Prad ) $ 
with $ \Prad = E_{N_{\rad}}(\{ 0\})$ where  $E_{X}(J)$ denotes the spectral projection on a Borel set $J \in \ms{B}(\mbf{R})$ for a self-adjoint operator $X$. Then  
\begin{align}
(\Psi , H_m \Psi ) &  \geq 
(   \Psi ,  \tilde{Y}_{\, \rad ,R}^{\ast}  (  H_m \tens  \1_{\rad}  )
   \tilde{Y}_{\, \rad ,R}  \Psi )  + m   \notag \\
&   \qquad \qquad \quad  - m (  \Psi , \tilde{Y}_{\, \rad ,R}^\ast (\1_{\D} \tens \1_{\rad}  \tens   \Prad )
  \tilde{Y}_{\, \rad ,R}  \Psi )  - \delta_{\, \bos } (R)  \| \Psi \|_{0,2 }   \notag \\
&  \geq (  \Psi ,   \tilde{Y}_{\, \bos ,R}^{\ast} (  H_m \tens  \1_{\rad} )
   \tilde{Y}_{\, \bos ,R}   \Psi )  + m-m (\Psi , ( \1_{\D} \tens 
 \Gammab ( q_{\, \bos, R})  \Psi ) - \delta_{\, \bos } (R)  \| \Psi \|_{0,2 }  . \label{9/10.1}
\end{align}  
 Here we used $ 
 {Y}_{\, \bos ,R}^{\ast} ( \1_{\rad} \tens   \Prad )
   {Y}_{\, \bos ,R}  =    \Gammab (q_{\, \bos, R}) $ in the last line. 
We evaluate  the first term in the right hand side of (\ref{9/10.1}). 
Let $\tilde{\tilde{X}}_{\fer ,R}= \tilde{X}_{\, \fer ,R} \tens \1_{\rad}$. 
By  Corollary  \ref{9/9.c} \textbf{(i)}, 
\begin{align}
 & (  \Psi ,  \tilde{Y}_{\, \bos ,R}^{\ast}   (  H_m \tens \1_{\rad} )
  \tilde{Y}_{\, \bos ,R}   \Psi ) \notag \\
& = \left(   \Psi , \tilde{Y}_{\, \bos ,R}^{\ast}
    ( ( \tilde{X}_{\, \fer ,R}^{\ast} \tilde{X}_{\, \fer ,R} H_m)    \tens  \1_{\rad} )  
  \tilde{Y}_{\, \bos ,R}   \Psi ) \right)  \notag \\
&\geq    \left(    \Psi , 
\tilde{Y}_{\, \rad ,R}^{\ast} 
 \tilde{\tilde{X}}^{\ast}_{\, \fer ,R} (    \tilde{H}_m \tens  \1_{\rad})
 \tilde{\tilde{X}}_{\, \fer ,R}  \tilde{Y}_{\, \bos ,R}   \Psi \right)  \notag  \\
& \qquad  \qquad \qquad  \quad + \left(  \Psi_n ,  
 \tilde{Y}_{\, \bos ,R}^{\ast} 
 \tilde{\tilde{X}}^{\ast}_{\, \fer ,R} ( \1_{\D } \tens H_{\D} \tens  \1_{\rad} )
  \tilde{\tilde{X}}_{\, \bos ,R}  \tilde{Y}_{ \bos ,R} \Psi \right) 
 - \delta_{\, \fer  } (R) \| \tilde{Y}_{  \bos ,R} \Psi \|_{2,1}^{\sim }, \label{9/9.3}
\end{align}
 where  we set 
\[
\| \tilde{ \Phi}  \|_{ \lambda , \lambda '  }^{\sim} \; = \; \|(  N_{\D}^{\lambda /2} \tens \1_{\rad}\tens \1_{\rad} )   \tilde{ \Phi} \| + \|( \1_{\D} \tens N_{\rad}^{\lambda ' /2} \tens \1_{\rad} ) \tilde{ \Phi} \| + \| \tilde{ \Phi} \| ,
\]
for $  \tilde{ \Phi} \in \ms{D} ( N_{\D}^{\lambda /2} \tens \1_{\rad}\tens \1_{\rad})  \cap \ms{D}(\1_{\D} \tens N_{\rad}^{\lambda ' /2} \tens \1_{\rad} ) $.  We see that  
\begin{align*}
\| \tilde{Y}_{  \bos ,R} \Psi  \|_{ 2,1 }^{\sim} & = 
 \|  \tilde{Y}_{  \bos ,R} ( N_{\D} \tens \1_{\rad} )  \Psi  \|
+ \|( \1_{\D} \tens N_{\rad}^{1/2} \tens \1_{\rad} ) \tilde{Y}_{  \bos ,R} \Psi \| + \| \tilde{Y}_{  \bos ,R} \Psi  \| \\
 &  \leq \|  \tilde{Y}_{  \bos ,R} ( N_{\D} \tens \1_{\rad} )  \Psi  \|  + \|\tilde{Y}_{  \bos ,R} ( \1_{\D} \tens  N_{\rad}^{1/2}  )  \Psi \|+ \| \tilde{Y}_{  \bos ,R} \Psi  \|
=  \| \Psi \|_{2,1} ,
\end{align*}
and 
$ H_{\D} \geq M (  \1_{\D}- P_{\D}) $  
with $ P_{\D} =  E_{N_{\D}}(\{ 0 \})$. Then we have 
\begin{align} 
(\ref{9/9.3})& 
\geq  E_{0} (\tilde{H}_m ) +   M   - M 
 \left(    \Psi , \tilde{Y}_{\, \bos ,R}^{\ast} \tilde{\tilde{X}}_{\, \fer ,R}^{\ast}  (\1_{\D} \tens \POmegaD \tens \1_{\rad} )    \tilde{\tilde{X}}_{\, \fer ,R}  \tilde{Y}_{\, \bos ,R}   \Psi \right)  - \delta_{\, \fer } (R) \|  \Psi \|_{ 2,1} 
  \notag \\
  & \geq E_{0} (H_m ) +   M - M ( \Psi , ( \Gammaf ( q_{ \, \fer  , R} ) \tens \1_{\rad}  ) \Psi )
  - \delta_{\, \fer , m } (R) \|  \Psi \|_{2,1} . \notag
\end{align}
Here we used $  E_{0} (\tilde{H}_m)= E_{0} (H_m)$ and  $   {X}_{\, \fer ,R}^{\ast} ( \1_{\D} \tens   P_{\D} )
   {X}_{\, \fer ,R}  =    \Gammaf (q_{\, \fer, R} ) $   
 in the last line. 
Thus we have
\begin{align}
(\Psi,  H_m \Psi )   & \geq E_{0} (H_m ) + m + M 
-  M \left( \Psi , ( \Gammaf \left( q_{ \, \fer  , R} \right) \tens \1_{\rad}  ) \Psi \right) \notag  \\
& \qquad \qquad  -m \left( \Psi , \left(  \1_{\D} \tens
 \Gammab \left( q_{ \, \bos  , R}  \right) \right)  \Psi \right)
 -\delta_{\, \bos  } (R) \|  \Psi \|_{0,2} -
 \delta_{\, \fer } (R) \|  \Psi \|_{2,1} .
\end{align}
Note that 
\begin{align}
 \1_{\D} \tens \1_{\rad}    & \geq \Gammaf \left( q_{ \, \fer  , R} \right) \tens 
\1_{\rad }+  (\1_{\D} -  \Gammaf \left( q_{ \, \fer  , R}  \right)  )   \tens  
\Gammab \left( q_{ \, \bos  , R} \right)  \notag \\
& = \Gammaf \left( q_{ \, \fer  , R} \right) \tens 
\1_{\rad }+ \1_{\D} \tens  \Gammab \left( q_{ \, \bos  , R} \right)
  -  \Gammaf \left( q_{ \, \fer  , R}  \right)     \tens  
\Gammab \left( q_{ \, \bos  , R} \right)  . \notag
\end{align}
Then we have
\begin{align*}
 (\Psi, H_m \Psi )    &  \geq E_{0} (H_m)  + \, m  \,  +  \,  
(M-m)   \left( \Psi,   (\1_{\D}   \tens \Gammab \left( q_{ \, \bos  , R}  \right) ) \Psi \right)  \\
& \quad- M \left( \Psi ,  (\Gammaf \left( q_{ \, \fer  , R}  \right) \tens \Gammab
  \left( q_{ \, \bos  , R}  \right) ) \Psi \right) -\delta_{\, \bos  } (R) \|  \Psi \|_{0,2} -
 \delta_{\, \fer } (R) \|  \Psi \|_{2,1} .
\end{align*}
Thus the proof is obtained. $\blacksquare$ \\

\begin{lemma} \label{9/9.e}\normalfont  
Assume \textbf{(A.1)} - \textbf{(A.3)}. Then
 for all $ 0< \epsilon < \frac{1}{c_{\I} |\kappaI |}$, 
\begin{equation}
\qquad  \qquad 
\| H_{0, m} \Psi \| \leq  L_{\epsilon} \| H_{m}  \Psi \|  +   R_{\epsilon} \|  \Psi \|  , \qquad  \Psi \in \ms{D}(H_m )  ,  \notag
\end{equation}
where $L_{\epsilon} = \frac{1}{1- c_{\I} |\kappaI |\epsilon} $ and  
$R_{\epsilon}= \frac{1}{1- c_{\I} |\kappaI |\epsilon}\left( c_\I |\kappaI  | \, ( \frac{1 }{2 \epsilon} +1 )  + |\kappaII | \, \|  \HII\| \right)$.
\end{lemma}
\textbf{(Proof)}  Let $\Psi \in \ms{D}(H_m)$.  Since $H_{0,m} = H_m - \kappaI \HI - \kappaII \HII $, we see that  
\begin{equation}
\|H_{0,m} \Psi  \|  \leq  \| H_m \Psi \| + \left| \kappaI \right| \, \left|  \HI  \Psi \right\|
 + \left| \kappaII \right| \, \,  \| \HII  \| \, \| \Psi \| . \notag 
\end{equation}
From (\ref{HImbound'}),  it holds that  
$ \| \HI \Psi \| \leq c_{\I } \epsilon   \| H_{0 , m } \Psi  \|  + c_{\I} ( 
\frac{1 }{2 \epsilon} +1 ) \| \Psi \| $ for all $\epsilon > 0$. Hence 
\begin{equation}
(1- c_{\, \I } |\kappaI | \epsilon  )\| H_{0,m} \Psi  \|  \leq \| H_m \Psi   \|
+ \left( c_\I |\kappaI  | \, \left( \frac{1 }{2 \epsilon} +1 \right)   +  | \kappaII | \, \| \HII \|  \,  \right)
\| \Psi \| . 
\end{equation}
 Taking  $\epsilon >0$ such that $ \epsilon < \frac{1}{c_{\, \I  }|\kappaI |}$, we obtain the proof. $\blacksquare $

$\;$ \\
Since $\| N_{\D} \Psi \| \leq  \frac{1}{M} \| H_{\D} \Psi  \|$, $\Psi \in \ms{D}(H_{\D})$, and $\| N_{\rad} \Phi \| \leq  \frac{1}{m} \| H_{\rad} \Phi  \|$, $\Phi \in \ms{D}(H_{\rad})$,
the next corollary follows from Lemma \ref{9/9.e}. \\

\begin{corollary}  \label{9/9.f}\normalfont
  Assume \textbf{(A.1)} - \textbf{(A.3)}.  Then for all $ 0< \epsilon < \frac{1}{c_{\I} |\kappaI |}$ and $ \Psi \in \ms{D}(H_m )$,
\begin{align*}
& \textbf{(i)} \; \| (N_{\D} \tens \1_{\rad} ) \Psi \| \leq  
\frac{L_{\epsilon}}{M} \|H_{m} \Psi  \|  + \frac{R_\epsilon}{M} \| \Psi \| ,     \\
& \textbf{(ii)} \;\| (\1_{\D} \tens N_{\rad} ) \Psi \| \leq 
 \frac{L_{\epsilon }}{m} \|H_{m} \Psi  \|  + \frac{R_\epsilon}{m} \| \Psi \|  . 
\end{align*}  
\end{corollary}

$\;$ \\
{\large \textbf{(Proof of Theorem \ref{Massive-Case} )}} \\
It is enough to show that $ \sigma_{\ess} (H_m )  \subset [E_{0} (H_m ) + m , \infty )$. Let
$\lambda \in \sigma_{\ess} (H_m )$. Then by the Weyl's theorem, there exists a sequence 
$\{ \Psi_n \}_{n=1}^{\infty}$ of $\ms{D}(H_m )$ such that (i) $\| \Psi_n \| =1$, $n \in \mbf{N}$, (ii) s-$\lim\limits_{ n \to \infty} (H_m -\lambda ) \Psi_n =0 $, and (iii) w-$\lim\limits_{n \to \infty} \Psi_n = 0$. 
Since $ | \lambda- (\Psi_n , H_m \Psi_n ) | \leq |  ( \Psi_n , ( H_m- \lambda  )\Psi_n |
 \leq \|  ( H_m- \lambda  )\Psi_n \| $, it holds that   
$  \lambda = \lim\limits_{n \to \infty} (\Psi_n , H_m \Psi_n ) $. Here  we show that 
 \begin{equation} 
\lim\limits_{n \to \infty} (\Psi_n , H_m \Psi_n ) \geq  E_{0} (\Hm ) + m  , \notag 
\end{equation}
 and then,   the proof is obtained.   
Let   $m \leq M$. From   Lemma \ref{LformboundHm}, 
\begin{align}
 ( \Psi_{n} ,H_m \Psi_{n} ) \geq  & E_{0} (H_m)  + m - M  (\Psi_n , (\Gammaf (q_{\, \fer ,R} ) \tens  \Gammab (q_{\, \bos , R}) ) ,\Psi_n  )  \notag \\
 &  \qquad \qquad \qquad \quad  \;\;- \delta_{\, \fer  } (R) \|  \Psi_n \|_{2,1} -\delta_{\, \bos  , m } (R) \|  \Psi_n \|_{0,2 }. \notag
 \end{align}
 Since s-$\lim\limits_{ n \to \infty} (H_m -\lambda ) \Psi_n =0 $,
we can set
\[
E_{m} = \sup_{n \in \mbf{N}} \| H_m \Psi_n  \| < \infty . 
\]
Let  $0 \leq  \lambda \leq 2 $ and $0 \leq  \lambda' \leq 2$. From Corollary \ref{9/9.f},  it is seen that  for all $ 0< \epsilon < \frac{1}{c_{\I} |\kappaI |}$,
\begin{align}
\| \Psi_{n} \|_{\lambda ,\lambda ' }
& = \|( N_{\D}^{\lambda /2} \tens \1_{\rad} ) \Psi_n \|    +  \|(  \1_{\D} \tens N_{\rad}^{\lambda ' } ) \Psi_n \| + \| \Psi_n \| \notag  \\
& \leq \|( N_{\D} \tens \1_{\rad} ) \Psi_n \|    +  \|(  \1_{\D} \tens N_{\rad} ) \Psi_n \| + 3
\| \Psi_n \| \notag \\
& \leq (\frac{1}{M} + \frac{1}{m} ) \left( L_{\epsilon }  \|H_{m} \Psi_n \|  + 2R_{\epsilon} \right) +3\| \Psi_n \|  \notag \\ 
& \leq E_{m} L_{\epsilon} (\frac{1}{M} + \frac{1}{m} ) +   2 (\frac{1}{M} + \frac{1}{m} )R_{\epsilon} +3  . \notag
\end{align}  
Then we have 
\begin{equation}
( \Psi_{n} ,H_m \Psi_{n} ) \geq   E_{0} (H_m)  + m - M  (\Psi_n , (\Gammaf (q_{\, \fer ,R} ) \tens  \Gammab (q_{\, \bos , R}) ) ,\Psi_n  )  - \delta_{m, \epsilon} (R) , \label{9/9.6}
\end{equation}
where $\delta_{m , \epsilon} (R) =  c_{\, m , \epsilon}  ( \delta_{\, \bos } (R) +  \delta_{ \, \fer } (R) ) $ with $c_{\, m , \epsilon }  = E_{m} L_{\epsilon} (\frac{1}{M} + \frac{1}{m} ) +   2 (\frac{1}{M} + \frac{1}{m} )R_{\epsilon} +3$. 
It is seen that
\begin{align}
 &\left| ( \Psi_n , ( \Gammaf (q_{\, \fer ,R} ) \tens  \Gammab (q_{\, \bos , R}) ) \Psi_n  ) \right| \notag \\
& \qquad \qquad  \leq \| (H_{0,m} +1)^{1/2} \Psi_n  \| \,  \|
(H_{0,m} +1)^{-1/2}  (\Gammaf (q_{\, \fer ,R} ) \tens  \Gammab (q_{\, \bos , R}) ) \Psi_n   \| .  \label{9/16.1}
\end{align}
From Lemma \ref{9/9.e}, we see that 
\[
\| (H_{0,m} +1)^{1/2} \Psi_n \| \leq \| H_{0,m} \Psi_n \| + \|\Psi_n  \|  \leq  L_\epsilon \| H_{m} \Psi_n \| + ( R_\epsilon +1)  \|\Psi_n  \| = E_{0} (H_m) L_{\epsilon} + R_\epsilon +1 ,
\]
 and hence,
\begin{equation}
\sup\limits_{n \in \mbf{N}} \| (H_{0,m} +1)^{1/2} \Psi_n \| \leq E_{m} L_\epsilon  +  R_\epsilon +1 .  \label{9/16.2}
\end{equation}
It holds that  
\begin{align}
   (H_{0,m} +1)^{-1/2} ( \Gammaf (q_{\, \fer ,R} ) \tens  \Gammab (q_{\, \bos , R}) ) = 
  & (H_{0,m} +1)^{-1/2} ( (H_{\D} + 1)^{1/2} \tens (\Hradm +1)^{1/2} ) \notag \\
  & \;  \times ( (H_{\D} + 1)^{-1/2}  \Gammaf (q_{ \, \fer ,R} )  ) \tens  
( (\Hradm +1)^{-1/2} \Gammab (q_{ \, \bos ,R} ) )  ) ,
    \notag
\end{align}
and hence,  $(H_{0,m} +1)^{-1/2} ( \Gammaf (q_{\, \fer ,R} ) \tens  \Gammab (q_{\, \bos , R}) )$ is compact, since $\| (H_{0,m} +1)^{-1/2} ( (H_{\D} + 1)^{1/2} \tens (\Hradm +1)^{1/2} )\| \leq 1$ and  $ \left(  (H_{\D} + 1)^{-1/2}\Gammaf (q_{ \, \fer ,R})  \right) \tens  
\left(   (\Hradm +1)^{-1/2} \Gammab (q_{ \, \bos, R}) \right)$ is compact.  Therefore it holds that 
\begin{equation}
 \lim_{n \to \infty}  \left\| (H_{0,m} +1)^{-1/2} ( \Gammaf (q_{\, \fer ,R} ) \tens  \Gammab (q_{\, \bos , R}) ) \Psi_n \right\|=0. \label{9/16.3} 
\end{equation}
From (\ref{9/16.1}) -  (\ref{9/16.3}) we have $
\lim\limits_{n \to \infty}\left| \left( \Psi_n , (\Gammaf (q_{ \, \fer ,R}) \tens  
\Gammab (q_{ \, \bos ,R}) ) ,\Psi_n  \right) \right|
 =0 $. Then by taking the limit of (\ref{9/9.6}) as $R \to \infty$, we have
$ \lim\limits_{n \to \infty} (\Psi_n , H_m  \Psi_n ) \geq  E_{0}(H_m ) + m $.
   $\blacksquare$ \\


\section{Derivative Bounds}
From Theorem \ref{Massive-Case},   $H_{m}$ has the ground state. Let  $\Psi_{m}$ be  the normalized ground state of $H_{m}$, i.e.
\begin{equation}
\qquad  \qquad H_{m} \Psi_m = E_{0}(H_m ) \Psi_m , \quad \| \Psi_m  \| = 1.  \notag
\end{equation}

\subsection{Electron-Positron Derivative Bounds}
We introduce the distribution kernel of the annihilation operator for the Dirac field. 
For all $ \Psi = \left\{  \Psi^{(n)} = {}^{t}\left( \Psi^{(n)}_1 , \cdots ,  \Psi^{(n)}_4  \right)  \right\}_{n=0}^{\infty}
\in \ms{D} ( H_{\D} )$, we set 
\[ \qquad 
C_{l}(\mbf{p})\Psi^{(n , \nu )}( \mbf{p}_{1} , \cdots ,  \mbf{p}_{n} )
= \delta_{\, l , \nu } \sqrt{n+1} \Psi^{(n+1 , \nu )}( \mbf{p} , \mbf{p}_{1} , \cdots ,  \mbf{p}_{n} ) . \quad l=1 ,\cdots 4 .
\]
Let
\[
 b_{1/2} (\mbf{p}) = C_{1} (\mbf{p}), \; \; b_{-1/2} (\mbf{p}) = C_{2} (\mbf{p}) , \;
  \;   d_{1/2} (\mbf{p}) = C_{3} (\mbf{p}), \; \;   d_{-1/2} (\mbf{p}) = C_{4} (\mbf{p})  .
 \]
 $\;$ \\
 Then it follows that for all  $\Phi \in \FDirac$ and $ \Psi \in \ms{D}( H_{\D} )$,
 \begin{align*}
\qquad \quad & ( \Phi ,  b_{s} (f)  \Psi ) \, = \,  \int_{\Rthree} f(\mbf{p})^{\ast} ( \Phi ,  b_{s} (\mbf{p})  \Psi ) d \mbf{p}  , \quad  \quad   \; f  \in \ms{D}(\omega_{\, M}) , \\
& ( \Phi ,  d_{s} (g)  \Psi ) \, = \,  \int_{\Rthree} g(\mbf{p})^{\ast} ( \Phi ,  d_{s} (\mbf{p})  \Psi ) d \mbf{p} , \quad  \quad   \;  g \in \ms{D}(\omega_{\, M})  .
 \end{align*}
The number operator for electrons and positrons are defined by  
\[ N_{\D}^{+} = d \Gammaf \left( \left( \begin{array}{cc} \1 & O \\ O& O \end{array} \right) \right) , \qquad N_{\D}^{-} = d \Gammaf \left( \left( \begin{array}{cc} O & O \\ O& \1 \end{array} \right) \right) ,
\]
respectively. It holds that for all  $\Phi , \Psi \in \ms{D}( H_{\D} )$,
\begin{align}
&(\Phi , N_{\D}^{+}  \Psi ) = \sum_{\pm1/2}\int_{\Rthree}( b_{s}(\mbf{p})  \Phi ,  b_{s}(\mbf{p}) \Psi) d \mbf{p}  , \notag \\
&(\Phi , N_{\D}^{-}  \Psi ) = \sum_{\pm1/2} \int_{\Rthree}( d_{s}(\mbf{p})  \Phi ,  d_{s}(\mbf{p}) \Psi) d \mbf{p} \notag  .
\end{align}

$\; $ \\
By the canonical anti-commutation relation, it is proven in (\cite{Ta09} ; Section III)  that 
\begin{align}
&[ \psidaggerx \alpha^j \psix  , b_{s}(f ) ] = - \sum_{l,l'=1}^4 \alpha_{l,l'}^{j}
( f , f_{s,\mbf{x}}^{l}  ) \;  \psi_{l'}(\mbf{x}) ,  \label{ori1}
\\
&[ \psidaggerx \alpha^j \psix , d_{s}(g ) ] = \sum_{l,l'=1}^4 \alpha_{l,l'}^{j}
( g ,g_{s ,\mbf{x}}^{l'}  ) \; \psi_{l}(\mbf{x})^{\ast} ,  \label{ori2}
 \end{align}
 and for  $\rho(\mbf{x})=\psidaggerx \psi (\mbf{x})$, 
\begin{align}
&[ \rho(\mbf{x}) \rho (\mbf{y}), \; b_{s}(f  ) ] 
= -\sum_{l=1}^4 \left( ( f  , f_{s, \mbf{y}}^{l}  )
\rho (\mbf{x}) \psi_{l}(\mbf{y}) 
+ ( f , f_{s, \mbf{x}}^{l}  ) \; \psi_{l}(\mbf{x})
 \rho (\mbf{y}) \right) ,   \label{ori3}  \\
&[ \rho (\mbf{x}) \rho (\mbf{y}), \; d_{s}( g ) ] 
= \sum_{l=1}^4 \left( ( g , g_{s, \mbf{y}}^{l}  )
\rho (\mbf{x}) \psi_{l}(\mbf{y}) ^{\ast}
+ (g , g_{s, \mbf{x}}^{l}  ) \;  \psi_{l}(\mbf{x})^{\ast}
 \rho (\mbf{y}) \right) . \label{ori4}
\end{align}

$\; $ \\
Let $X$ and $Y$  be operators on a Hilbert space. The weak commutator is defined by
\[
[X, \,Y ]^0 (\Phi , \Psi ) = (X^{\ast}\Phi , Y \Psi  ) - (Y^{\ast}\Phi , X \Psi  ) ,
\]
where $ \Psi \in \ms{D}(X) \cap \ms{D} ( Y) $ and  $ \Phi \in \ms{D}(X^{\ast}) \cap \ms{D} ( Y^{\ast}) $. \\

\begin{lemma} \label{9/12.a}  \normalfont 
Assume \textbf{(A.1)} -  \textbf{(A.3)}. Then it holds that for all $f \in L^2 (\Rthree)$,  
\begin{align*}
\textbf{(i)} & \; \;  [\HI , b_{s}(f) \tens \1 ]^0 (\Phi , \Psi )  \,
= \, \int_{\Rthree} f(\mbf{p})^{\ast} \left( \Phi , K_{s}^{+} (\mbf{p})   \Psi  \right)
 d \mbf{p} ,  \quad \Phi \in \ms{F}_{\QED} ,\; \Psi \in \ms{D}(H_m ) ,  \\
 \textbf{(ii)} & \; \;
[\HII , b_{s}(f)\tens \1 ]^0 (\Phi , \Psi)  \,
= \, \int_{\Rthree} f(\mbf{p})^{\ast} \left( \Phi , (S_s^{\, +} (\mbf{p}) + T_s^{+} (\mbf{p}) )    \Psi  \right) d \mbf{p} ,  \quad \Phi , \Psi \in \ms{F}_{\QED}  . 
\end{align*}
Here $ K_s^{+}(\mbf{p})$, 
$S_s^{\, +} (\mbf{p})   $  and $T_s^{+} (\mbf{p})   $   are operators which satisfy
\begin{align}
& (\Phi , K_s^{+}(\mbf{p}) \Psi ) = -\sum_{j=1}^3 \sum_{l,l'=1}^4 \alpha^{j}_{l,l'}\, 
 \int_{\Rthree} \chiIx   f_{s , \mbf{x}}^{\, l} (\mbf{p}) 
\left( \Phi , \,  (\psi_{l'}(\mbf{x}) \tens A_j (\mbf{x}))  \Psi \right) \dx , \notag \\
 & (\Phi , S^{+}_{s}(\mbf{p}) \Psi ) = -\sum_{l=1}^4  \int_{\Rthree \times \Rthree} 
\frac{\chiIIx \chiIIy}{|\mbf{x}-\mbf{y}|} f_{s , \mbf{y} }^{\, l}(\mbf{p}) 
\left( \Phi , \,  ( \rho (\mbf{x}) \psi_{l}(\mbf{y}) \tens \1 ))  \Psi \right)  \dx \dy  , 
\notag \\ 
&(\Phi , T^{+}_{s}(\mbf{p}) \Psi ) = -\sum_{l=1}^4 \int_{\Rthree \times \Rthree} 
\frac{\chiIIx \chiIIy}{|\mbf{x}-\mbf{y}|}  f_{s , \mbf{x}}^{\, l }(\mbf{p})  
\left( \Phi , \,  (\psi_{l}(\mbf{x})  \rho (\mbf{y}) \tens \1 ))  \Psi \right)  \dx \dy .
\notag
\end{align}
\end{lemma}
\textbf{(Proof)} \\
\textbf{(i)} Let $\Phi \in \ms{F}_{\QED}$ and  $ \Psi \in \ms{D}(H_m )$. By (\ref{ori1}), we have
\begin{align}
[\HI , b_{s}(f) \tens \1 ]^0 (\Phi , \Psi)  
& =\sum_{j=1}^3 \int_{\Rthree}
 \chiIx    [ \psidaggerx \alpha^j \psi (\mbf{x} ) \tens A_{j} (\mbf{x}) , b_{s}(f) \tens \1 ]^0 
( \Phi , \Psi )   \dx  \notag \\
 & = \sum_{j=1}^3\int_{\Rthree}
\chiIx   \left( \Phi , \left( [ \psidaggerx \alpha^j \psi (\mbf{x} ), b_{s}(f) ]   \tens A_{j} (\mbf{x}) \right)   \Psi \right) \dx  \notag  \\
& = -  \sum_{ j=1}^3  \sum_{ l, l'=1}^4 \alpha_{l , l'}^j \int_{\Rthree}  \chiIx
\, (f,  f_{s, \mbf{x}}^{\, l})  \left( \Phi , ( \psi_{l'} (\mbf{x} ) \tens A_{j} (\mbf{x})) \Psi \right)  \dx .  \notag 
\end{align}
Let   $\ell_{s, \mbf{p}} : \ms{F}_{\QED} \times  \ms{F}_{\QED} \to \mbf{C}  $ be  a functional defined by 
\[
\ell_{s, \mbf{p}} (\Phi ' , \Psi ') = 
 - \sum_{j=1}^3 \sum_{l, l'=1}^4 \alpha_{l , l'}^j  \int_{\Rthree}  \chiIx
\,   f_{s, \mbf{x}}^{\, l} (\mbf{p})  \left( \Phi ' , ( \psi_{l'} (\mbf{x} ) \tens A_{j} (\mbf{x})) \Psi '\right)  \dx   , 
\]
for $  \Phi ' \in \FQED , \Psi ' \in \ms{D}(H_{0, m} )$.  We see that 
\[
 \ell_{s, \mbf{p}} (\Phi ' , \Psi ' )  \leq 
c_{ \, \I , \, s, \mbf{p}}  \| \Phi ' \|  \,  \| (\1 \tens ( \Hradm +1)^{1/2}) \Psi ' \| ,
\]
where $c_{\, \I , \, s, \mbf{p}} = \sum\limits_{j=1}^3 \sum\limits_{l, l'=1}^4 | \alpha_{l , l'}^j| \, \| \chi_{\, \I}  \|_{L^1} \, |f_{s}^{\, l}(\mbf{p})|c_{\, \D}^{\, l'} c_{\rad}^{\, j} $.  Then from the  Riesz Representation theorem, we can define an operator  $ K_{s}^{\, +} (\mbf{p})$  which satisfy 
$\ell_{s, \mbf{p}} (\Phi ', \Psi ' ) = (\Phi ' , K_{s}^{\, +} (\mbf{p}) \Psi' )   $. Then it holds that 
\[
 [\HI , b_{s}(f) ]^0 (\Phi , \Psi) =  \int_{\Rthree} f(\mbf{p})^{\ast}
 \ell_{s, \mbf{p}} (\Phi , \Psi ) 
 d \mbf{p} = \int_{\Rthree} f(\mbf{p})^{\ast} \left( \Phi , K_{s}^{+} (\mbf{p})   \Psi  \right)  d \mbf{p} .  \]
\textbf{(ii)}
From (\ref{ori3}), we see that for all  $\Phi , \Psi  \in \ms{F}_{\QED}$, 
\begin{align}
[\HII , b_{s}(f) \tens \1 ]^0 (\Phi , \Psi)  & = \int_{\Rthree \times \Rthree} \frac{\chiIIx \chiIIy}{| \mbf{x}-\mbf{y} | } [ \rho (\mbf{x} ) 
\rho (\mbf{y} ) \tens  \1  , b_{s}(f) \tens \1
]^0 (\Phi ,\Psi ) \dx \dy  \notag \\
& =\int_{\Rthree \times \Rthree} \frac{\chiIIx \chiIIy}{| \mbf{x}-\mbf{y} | }
\left( \Phi , \left( [  \rho (\mbf{x} )  \rho (\mbf{y} ) , b_{s}(f)  ]  \tens \1 \right)  \Psi \right) \dx \dy \notag  \\ 
& = -\sum_{l=1}^4 \int_{\Rthree \times \Rthree} \frac{\chiIIx \chiIIy}{| \mbf{x}-\mbf{y} | } \left\{ 
( f  , f_{s, \mbf{y}}^{\, l}  ) ( \Phi , 
( \rho (\mbf{x}) \psi_{l} (\mbf{y}) \tens \1 ) \Psi ) \right. \notag \\
& \qquad \qquad \qquad \qquad   \left. + ( f , f_{s, \mbf{x}}^{\, l}  ) (\Phi ,  \, 
 ( \psi_{l} (\mbf{x})  \rho (\mbf{y}) \tens \1 )\Psi ) \right\} \dx \dy .  \notag 
\end{align}
We set  functionals  $q_{s, \mbf{p}} $ and $r_{s,\mbf{p}}  $ on $\ms{F}_{\QED} \times
 \ms{F}_{\QED}$  by 
\begin{align*}
&\quad  q_{s, \mbf{p}} (\Phi '  , \Psi ' ) = -\sum_{l=1}^4 \int_{\Rthree \times \Rthree} \frac{\chiIIx \chiIIy}{| \mbf{x}-\mbf{y} | } f_{s, \mbf{y}}^{\, l}  (\mbf{p}) ( \Phi ' , 
( \rho (\mbf{x}) \psi_{l} (\mbf{y}) \tens \1 ) \Psi ' ) \dx \dy , \quad \Phi ', \, \Psi ' \in \FQED ,  \\
&\quad   r_{s, \mbf{p}} (\Phi '', \Psi '') =  -\sum_{l=1}^4 \int_{\Rthree \times \Rthree} \frac{\chiIIx \chiIIy}{| \mbf{x}-\mbf{y} | } f_{s, \mbf{x}}^{\, l}  (\mbf{p}) ( \Phi '' , 
(\psi_{l} (\mbf{x}) \rho (\mbf{y})  \tens \1 ) \Psi '' ) \dx \dy,  \quad \Phi '' , \, \Psi '' \in \FQED .
\end{align*}
We see that 
\begin{align*}
& q_{s, \mbf{p}} (\Phi ' , \Psi ' )  \leq c_{ \, \II ,  \, s, \mbf{p}}  \| \Phi ' \| \, \| \Psi '\| ,  \notag \\
& r_{s, \mbf{p}} (\Phi '' , \Psi '' )  \leq  c_{ \, \II ,  \, s, \mbf{p}}  \| \Phi '' \| \, \| \Psi '' \| ,
\end{align*}
where $c_{ \, \II , \, s, \mbf{p}} = \sum\limits_{l, l'=1}^4 
 \left\| \frac{\chiIIx \chiIIy}{|\mbf{x}-  \mbf{y}| }  \right\|_{L^1} \,
 |f_{s}^{\, l}(\mbf{p})|(c_{\, \D}^{\, l'})^2  c_{\, \D}^{\, l} $. 
  Then from Riesz Representation theorem, we can define  operators $ S_s^{+} (\mbf{p})$ and  $ T_s^{+} (\mbf{p})$ such that $q_{s, \mbf{p}} (\Phi ' , \Psi ' ) = (\Phi ', S_{s}^{+} (\mbf{p}) \Psi ' )   $ and $r_{\mbf{p}} (\Phi '' , \Psi '' ) = (\Phi '' , T_s^{+} (\mbf{p}) \Psi '' )   $, respectively. 
Then it holds that  
\begin{align*}
[\HII , b_{s}(f) ]^0 (\Phi , \Psi)  
& =  \int_{\Rthree} \overline{f(\mbf{p})} \left( q_{s,  \mbf{p}}  (\Phi , \Psi )
+ r_{ s, \mbf{p}}  (\Phi , \Psi ) \right)
 d \mbf{p} \\ 
&=  \int_{\Rthree} \overline{f(\mbf{p})} \left( \Phi , (S_s^{+} (\mbf{p}) + T_s^{+} (\mbf{p}) )    \Psi  \right) d \mbf{p}.
\end{align*}
Thus proof is obtained.  $\blacksquare $

$\; $ \\
In a similar way to Lemma \ref{9/12.a}, the following lemma is also proven. \\

\begin{lemma} \label{9/12.b}  \normalfont 
Assume \textbf{(A.1)} -  \textbf{(A.3)}. Then it holds that for all $g \in L^2 (\Rthree)$,  
\begin{align*}
\textbf{(i)} \; & \; \;  [\HI , d_{s}(g) \tens \1 ]^0 (\Phi , \Psi )  \,
= \, \int_{\Rthree} g(\mbf{p})^{\ast} \left( \Phi , K_{s}^{\,-} (\mbf{p})   \Psi  \right)
 d \mbf{p} ,  \qquad \Phi \in \ms{F}_{\QED} ,\Psi \in \ms{D}(H_m ) ,  \\
 \textbf{(ii)} & \; \;
[\HII , d_{s}(g)\tens \1 ]^0 (\Phi , \Psi)  \,
= \, \int_{\Rthree} g(\mbf{p})^{\ast} \left( \Phi , (S_s^{\, -} (\mbf{p}) + T_s^{-} (\mbf{p}) )    \Psi  \right) d \mbf{p} ,  \qquad \Phi , \Psi \in \ms{F}_{\QED}  . 
\end{align*}
Here $ K_s^{\, -}(\mbf{p})$, 
$S_s^{\, -} (\mbf{p})   $  and $T_s^{-} (\mbf{p})   $   are operators which satisfy
\begin{align}
& (\Phi , K_s^{\, -}(\mbf{p}) \Psi ) = \sum_{j=1}^3 \sum_{l,l'=1}^4 \alpha^{j}_{l,l'}\, 
 \int_{\Rthree} \chiIx   g_{s , \mbf{x}}^{\, l'} (\mbf{p}) 
\left( \Phi , \,  (\psi_{l}(\mbf{x})^{\ast} \tens A_j (\mbf{x}))  \Psi \right) \dx , \notag  \\
 & (\Phi , S^{-}_{s}(\mbf{p}) \Psi ) = \sum_{l=1}^4  \int_{\Rthree \times \Rthree} 
\frac{\chiIIx \chiIIy}{|\mbf{x}-\mbf{y}|} g_{s , \mbf{y} }^{\, l}(\mbf{p}) 
\left( \Phi , \,  ( \rho (\mbf{x}) \psi_{l}(\mbf{y})^{\ast} \tens \1 ))  \Psi \right)  \dx \dy  , \notag \\ 
&(\Phi , T^{-}_{-}(\mbf{p}) \Psi ) = \sum_{l=1}^4 \int_{\Rthree \times \Rthree} 
\frac{\chiIIx \chiIIy}{|\mbf{x}-\mbf{y}|}  g_{s , \mbf{x}}^{\, l }(\mbf{p})  
\left( \Phi , \,  (\psi_{l}(\mbf{x})^{\ast}  \rho (\mbf{y}) \tens \1 ))  \Psi \right)  \dx \dy ,
\notag
\end{align}
$\; $
\end{lemma}

\begin{lemma} \label{9/12.c}  \normalfont 
Assume \textbf{(A.1)} - \textbf{(A.5)}. Let $\Psi \in \ms{D}(H_m )$. Then, 
$K_{s}^{\pm }(\mbf{p}) \Psi$, $S_{s}^{\pm }(\mbf{p}) \Psi$ and $T_{s}^{\pm }(\mbf{p}) \Psi$, $s= \pm 1/2$,  are strongly differentiable for all $\mbf{p} \in \Rthree \backslash O_{\D}$.
\end{lemma}
\textbf{(Proof)}   We  show that $K_{s}^{+}(\mbf{p}) \Psi$ is strongly differentiable. 
Let $\Phi \in \ms{F}_{\QED}$ with $ \| \Phi \| =1$. From \textbf{(A.4)}, $K_{s}^{+}(\mbf{p}) \Psi$  is weakly differentiable for all $\mbf{p} \in \Rthree \backslash O_{\D}$, and we have
\[ 
\partial_{p^{\nu}} (\Phi , K_{s}^{+ }(\mbf{p}) \Psi )
= -\sum_{j=1}^3 \sum_{l,l'=1}^4 \alpha^{j}_{l,l'}\, 
 \int_{\Rthree} \chiIx   \partial_{p^\nu}f_{s , \mbf{x}}^{\, l} (\mbf{p}) 
\left( \Phi , \,  (\psi_{l'}(\mbf{x}) \tens A_j (\mbf{x}))  \Psi \right) \dx ,
\]
and $ 
 |\partial_{p^{\nu}} (\Phi , K_{s}^{\pm }(\mbf{p}) \Psi )| \leq \sum\limits_{j=1}^3 \sum\limits_{l , l'=1}^4  | \alpha^{j}_{l, l'} | c_{\D}^{\, l'} c_{\rad}^{\, j} \left( \int_{\Rthree}|  \partial_{p^\nu}f_{s , \mbf{x}}^{\, l} (\mbf{p})  | \dx \, \right) \| (\1 \tens \Hradm^{1/2} ) \Psi \| $. 
Then the Riesz representation theorem shows  that there exists a vector $\Xi_{\Psi} (\mbf{p}) \in \ms{F}_{\QED}$ such that 
$( \Phi , \Xi_{\Psi} (\mbf{p}) )= \partial_{p^{\nu}} (\Phi , K_{s}^{\pm }(\mbf{p}) \Psi )$. 
Let $\mbf{e}_{\nu} = (\delta_{\nu , j})_{j=1}^3 $. It is seen that 
\begin{align}
&(\Phi ,  \frac{K_{s}^{+ }(\mbf{p} + \epsilon \mbf{e}_{\nu}) - 
 K_{s}^{+ }(\mbf{p} )}{\epsilon}  \Psi) -  
( \Phi , \Xi (\mbf{p}) ) \notag \\
& = -\sum_{j=1}^3 \sum_{l , l'=1}^4  \alpha^{j}_{l, l'} \int_{\Rthree} \chiIx
\left( \frac{f_{s , \mbf{x}}^{\, l }(\mbf{p} + \epsilon \mbf{e}_{\nu}) - 
 f_{s , \mbf{x}}^{\,l } (\mbf{p})}{\epsilon}  - \partial_{p^{\nu}} f_{s , \mbf{x}}^{l }  (\mbf{p})\right)  
(\Phi , ( \psi_{l'}(\mbf{x}) \tens A_{j} (\mbf{x}) )  \Psi ) \dx , \notag 
\end{align}
and hence, 
\begin{align}
&| (\Phi , (  \frac{K_{s}^{+ }(\mbf{p} + \epsilon \mbf{e}_{\nu}) -  K_{s}^{+ }(\mbf{p} )}{\epsilon}  \Psi - \Xi (\mbf{p}))  |  \notag \\
& \leq  \sum_{j=1}^3 \sum_{l , l'=1}^4  | \alpha^{j}_{l, l'} | c_{\D}^{\, l'} c_{\rad}^{\, j'} \left( \int_{\Rthree} | \chiIx |
\left| \frac{f_{s , \mbf{x}}^{\, l }(\mbf{p} + \epsilon \mbf{e}_{\nu}) - 
 f_{s , \mbf{x}}^{\, l } (\mbf{p}) }{\epsilon}  - \partial_{p^{\nu}} f_{s , \mbf{x}}^{\, l }  (\mbf{p})\right|  
  \dx \right)  \|    \Psi \| .  \label{9/13.1}
\end{align}
Since (\ref{9/13.1}) holds for all $\Phi \in \ms{F}_{\QED}$ with $ \| \Phi \| =1$, we have 
\begin{align}
& \left\|  \frac{K_{s}^{+ }(\mbf{p} + \epsilon \mbf{e}_{\nu}) -  K_{s}^{+ }(\mbf{p} )}{\epsilon}  \Psi - \Xi (\mbf{p}) \right\|  \notag \\
&   \leq 
\sum_{j=1}^3 \sum_{l , l'=1}^4  | \alpha^{j}_{l, l'} | c_{\D}^{\, l'} c_{\rad}^{\, j'} \left( \int_{\Rthree} | \chiIx |
\left| \frac{f_{s , \mbf{x}}^{\, l }(\mbf{p} + \epsilon \mbf{e}_{\nu}) - 
 f_{s , \mbf{x}}^{\, l } (\mbf{p}) }{\epsilon}  - \partial_{p^{\nu}} f_{s , \mbf{x}}^{\, l }  (\mbf{p})\right|  
  \dx \right)  \|    \Psi \|  \to 0 ,   \notag
\end{align}
as $\epsilon \to 0$. 
 Thus $K_{s}^{+ }(\mbf{p} ) \Psi $ is strongly differentiable. 
Similarly, it is proven that $ K_{s}^{\, -}(\mbf{p} ) \Psi$, $ S_{s}^{\pm  }(\mbf{p} ) \Psi$ and
$ T_{s}^{\pm  }(\mbf{p} ) \Psi$ are strongly differentiable  for all $\mbf{p} \in \Rthree \backslash O_{\D}$. $\blacksquare $ \\

\begin{lemma}  \label{9/12.d}  \normalfont
For all $\Phi , \Psi  \in \ms{D}(H_{\D})$, it holds that 
\begin{align*}
&\textbf{(i)} \; \; [H_{\D}  , b_{s}(f) ]^0 (\Phi , \Psi )=-  \left(  \Phi , b_{s}( 
\omega_{\, M} f )\Psi  
\right)  , \\
& \textbf{(ii)} \; \;
[H_{\D}  , d_{s}(f) ]^0 (\Phi , \Psi )=-  \left(  \Phi , d_{s}( \omega_{\, M} f )\Psi  
\right) .
\end{align*}

\end{lemma}
\textbf{(Proof)} It holds that for all $\Phi  \in \ms{F}_{\Dirac}^{\, \fin} (\ms{D} ( \omega_{\, M} )$,
\[
 [H_{\D}, b^{\dagger}_{s}(f)   ]\Phi = b^{\dagger}_{s}( \omega_{\, M} f)  \Phi . 
\]
 Let  $\Psi \in \ms{D} (H_m )$. Then
\[
(H_{\D} \Phi,    b_{s}(f)   \Psi ) -  ( b_{s}(f)   \Phi , H_{\D} \Psi )
= ( [ b^{\dagger}_{s}(f) ,  H_{\D} ] \Phi , \Psi  ) = (- b^{\dagger}_{s}( \omega_{\, M} f) \Phi ,  \Psi) ,
\]
and hence,
\begin{equation}
(H_{\D} \Phi,    b_{s}(f)   \Psi ) -  ( b_{s}(f)   \Phi , H_{\D} \Psi ) =  -(  \Phi , 
b_{s}( \omega_{\, M} f) \Psi) .  \label{9/11.1}
\end{equation}
Since $ \ms{F}_{\Dirac}^{\, \fin} (\ms{D} ( \omega_{\, M} ))$ is a core of $H_{\D}$ and 
$ b_{s}(f)$ is bounded, (\ref{9/11.1}) holds for all $\Phi \in \ms{D}(H_{\D})$. 
Hence \textbf{(i)} follows. Similarly, we can also prove \textbf{(ii)}.  $\blacksquare $ \\

\begin{proposition} \label{EP-Pullth}  \normalfont 
\textbf{(Electron-Positron Pull-Through Formula)}  \\
Assume \textbf{(A.1)} - \textbf{(A.3)}. Then that 
\begin{align*}
& \textbf{(i)} \; \; \;  (b_s (\mbf{p}) \tens \1  ) \Psi_m  = 
 (H_m- E_0 (H_m) + \omega_{M}(\mbf{p}))^{-1} \left( \kappaI K^{\,+}_{s} (\mbf{p} ) +
 \kappaII( S^{\,+}_{s} (\mbf{p} ) + T^{\,+}_{s} (\mbf{p} ) \right) \Psi_m ,
 \\ 
& \textbf{(ii)} \; \; (d_s (\mbf{p}) \tens \1  ) \Psi_m  
 =   (H_m- E_0 (H_m) + \omega_{M}(\mbf{p}))^{-1} \left( \kappaI K^{\,-}_{s} (\mbf{p} ) +
 \kappaII( S^{\,-}_{s} (\mbf{p} ) + T^{\,-}_{s} (\mbf{p} ) \right) \Psi_m ,
 \end{align*}
  for almost everywhere $\mbf{p} \in \Rthree$.
 
\end{proposition}
\textbf{(Proof)} Let $\Phi, \in \ms{D} (H_m )$.
By Lemma \ref{9/12.d} \textbf{(i)}, we have  
\begin{align} 
& [\Hm , b_{s}(f) \tens \1 ]^0 (\Phi , \Psi_m )  \notag \\
&=  -  \left(  \Phi , ( b_{s}( \omega_M f ) \tens \1 ) \Psi_m  \right) + \kappaI [\HI , b_{s}(f) \tens \1 ]^0 (\Phi , \Psi_m ) +
\kappaII [\HII , b_{s}(f) \tens \1 ]^0 (\Phi , \Psi_m ) .  \notag
\end{align}
On the other hand,  $H_m \Psi_m=E_{0}(H_m) \Psi_m$ yields that
\begin{equation}
  [\Hm , b_{s}(f) \tens \1 ]^0 (\Phi , \Psi_m ) = \left(   (  \Hm- E_{0}(\Hm )  ) \Phi , ( b_{s}(f) \tens \1 ) \Psi_m \right) . \notag
\end{equation}
Then, we have
\begin{align}
& (    (\Hm- E_{0}(\Hm ))  \Phi , ( b_{s}(f) \tens \1 ) \Psi_m  )   + ( \Phi , ( b_{s}( \omega_M f ) \tens \1 )  \Psi_m ) \notag \\
&\qquad \qquad \qquad   =  \kappaI [\HI , b_{s}(f) \tens \1 ]^0 (\Phi , \Psi_m ) +
\kappaII [\HII , b_{s}(f) \tens \1 ]^0 (\Phi , \Psi_m ) . \notag
\end{align}
By Lemma \ref{9/12.a},  it follows that 
\begin{align}
& \int_{\Rthree} f(\mbf{p})^{\ast} \left(    (\Hm- E_{0}(\Hm ) + \omega_M (\mbf{p}))
\Phi , ( b_{s}(\mbf{p}) \tens \1 ) \Psi_m  \frac{}{} \right)  d \mbf{p} \notag \\
&\quad  \qquad \quad 
= \int_{\Rthree} f(\mbf{p})^{\ast} \left( \Phi ,  \left( \kappaI K^{\,+}_{s} (\mbf{p} ) +
 \kappaII( S^{\,+}_{s} (\mbf{p} ) + T^{\,+}_{s} (\mbf{p} ) \right) \Psi_m  \right) d \mbf{p}.
 \label{9/12.3}
\end{align}
Since (\ref{9/12.3}) holds for all $f  \in L^2 ({\Rthree} )$, it follows that 
\begin{equation}
  (    (\Hm- E_{0}(\Hm ) + \omega_M (\mbf{p}))  \Phi , ( b_{s}(\mbf{p}) \tens \1 ) \Psi_m  )   
=( \Phi ,  \left( \kappaI K^{\,+}_{s} (\mbf{p} ) +
 \kappaII( S^{\,+}_{s} (\mbf{p} ) + T^{\,+}_{s} (\mbf{p} ) \right) \Psi_m ) 
    , \notag
\end{equation}
for almost everywhere $\mbf{p} \in \Rthree$. 
 This implies that   $ ( b_{s}(\mbf{p}) \tens \1 ) \Psi_m \in  \ms{D} (\Hm ) $ and
\begin{equation}
 ( \Hm- E_{0}(\Hm ) + \omega_M (\mbf{p})) ( b_{s}(\mbf{p}) \tens \1 ) \Psi_m   
= \left( \kappaI K^{\,+}_{s} (\mbf{p} ) +
 \kappaII( S^{\,+}_{s} (\mbf{p} ) + T^{\,+}_{s} (\mbf{p} ) \right) \Psi_m   . \label{9/12.7}
 \end{equation}
 From (\ref{9/12.7}), we obtain \textbf{(i)}.   Similarly,  \textbf{(ii)} is also proven.   $\blacksquare $\\

\begin{theorem} \label{EP-DB}  \normalfont (\textbf{Electron-Positron Derivative Bounds}) \\
Assume \textbf{(A.1)} - \textbf{(A.5)}. Then,  it holds that  for all $\mbf{p} \in \mbf{R}^3  \backslash O_{\D}$ and $0 < \epsilon < \frac{1}{c_{\I} |\kappaI |}$,  
\begin{align*}
&\textbf{(i)} \; \;  \left\| \partial_{p^{\nu}} ( b_{s} (\mbf{p}) \tens \1 ) \Psi_{m}  \right\| 
\leq \left( ( L_{\epsilon} E_{0}(H_{m} )  +  R_\epsilon + 1 \,)
|\kappaI |+ 2|\kappaII | \frac{}{}
\right) F_{s , +}^{\nu}(\mbf{p})  , \\
&\textbf{(ii)} \; \; \left\| \partial_{p^{\nu}} (d_{s} (\mbf{p}) \tens \1 ) \Psi_{m}  \right\| 
\leq  \left( ( L_{\epsilon} E_{0}(H_{m} )  +  R_\epsilon + 1 \,)
|\kappaI |+ 2|\kappaII | \frac{}{}
\right)  F_{s , -}^{\nu}(\mbf{p}) . 
\end{align*} 
Here $  F_{s ,\pm }^{\nu}  $ are functions satisfying $ F_{s , \pm }^{\nu} \in L^{2} (\Rthree ) $, $s= \pm  1/2$, $\nu=1 , \cdots , 3$. 
\end{theorem}
\textbf{(Proof)} 
Let    $R_{m,M}(\mbf{p}) =(H_m- E_0 (H_m) + \omega_{\, M}(\mbf{p}))^{-1}  $. From Proposition \ref{EP-Pullth} it holds that for  all $\Phi \in \FQED $ with $ \| \Phi \|=1$, 
\begin{align}
( \Phi , \partial_{p^{\nu}} ( b_{s} (\mbf{p}) \tens \1 ) \Psi_{m} ) 
& =   \kappaI   \left( \Phi, \partial_{p^{\nu}} R_{m,M}(\mbf{p})  K^{\,+}_{s} (\mbf{p} )    \Psi_m \right)  
 +  \kappaII \left( \Phi,  \partial_{p^{\nu}} R_{m,M}(\mbf{p}) 
 S^{\,+}_{s} (\mbf{p} )    \Psi_m \right) \notag \\
& \qquad \qquad  + \kappaII  \left( \Phi, \partial_{p^{\nu}} R_{m,M}(\mbf{p}) 
 T^{\,+}_{s} (\mbf{p} )    \Psi_m \right) . \label{9/13.3} 
\end{align}
Here we evaluate the three terms in  the right-hand side of  (\ref{9/13.3}) as follows. \\
(First term) We see that 
\begin{align}
&  \left( \Phi,\partial_{p^{\nu}} R_{m,M}(\mbf{p}) 
 K^{\,+}_{s} (\mbf{p} )    \Psi_m \right)  \notag \\
& =- \sum_{j=1}^{3} \sum_{l,l'=1}^{4} \alpha^{j}_{l,l'}\, \partial_{p^{\nu}} \left(   f_{s}^{\, l} (\mbf{p})
 \int_{\Rthree} \chiIx   e^{-i \mbf{p} \cdot \mbf{x}}
\left( R_{m,M}(\mbf{p}) \Phi , \,  (\psi_{l'}(\mbf{x}) \tens A_j (\mbf{x}))  \Psi_m \right) \dx \right) , \notag  \\
& = -\sum_{j=1}^{3} \sum_{l,l'=1}^{4} \alpha^{j}_{l,l'}\, \left\{  ( \partial_{p^{\nu}}  f_{s}^{\, l} (\mbf{p}))
 \int_{\Rthree} \chiIx   e^{-i \mbf{p} \cdot \mbf{x}}
\left( R_{m,M}(\mbf{p}) \Phi , \,  (\psi_{l'}(\mbf{x}) \tens A_j (\mbf{x}))  \Psi_m \right) \dx \right. , \notag  \\
& \qquad \qquad \qquad  -i  f_{s}^{\, l} (\mbf{p}) \int_{\Rthree} \chiIx   x^{\nu} \, e^{-i \mbf{p} \cdot \mbf{x}}
\left( R_{m,M}(\mbf{p}) \Phi , \,  (\psi_{l'}(\mbf{x}) \tens A_j (\mbf{x}))  \Psi_m \right) \dx 
\notag \\
& \qquad \qquad \qquad  \left.  - \frac{ f_{s}^{\, l} (\mbf{p}) p^{\nu} }{ \omega_{\, M}(\mbf{p})}
   \int_{\Rthree} \chiIx   e^{-i \mbf{p} \cdot \mbf{x}}
\left(  R_{m,M}(\mbf{p})^2 \Phi , \,  (\psi_{l'}(\mbf{x}) \tens A_j (\mbf{x}))  \Psi_m \right)
\dx \right\} \notag  .
\end{align}
Since  $ \| R_{m,M}(\mbf{p})\|  \leq \frac{1}{\omega_{\, M}(\mbf{p})} \leq  \frac{1}{M}$ and $\|\Phi \|=1$,    we have 
{\small
\begin{align*}
& \left| \int_{\Rthree} \chiIx   e^{-i \mbf{p} \cdot \mbf{x}}
\left( R_{m,M}(\mbf{p}) \Phi , \,  (\psi_{l'}(\mbf{x}) \tens A_j (\mbf{x}))  \Psi_m \right) \dx \right|  \leq  \frac{ c_{\, \D}^{\, l'}  c_{\rad}^{\, j}}{ M}
 \| \chi_{\I} \|_{L^1}   \|    ( \1 \tens (\Hradm +1)^{1 /2} ) \Psi_{m} \|,  \\
 & \left| \int_{\Rthree} \chiIx   x^{\nu} \, e^{-i \mbf{p} \cdot \mbf{x}}
\left( R_{m,M}(\mbf{p}) \Phi , \,  (\psi_{l'}(\mbf{x}) \tens A_j (\mbf{x}))  \Psi_m \right)
 \dx \right|   \leq \frac{ c_{\, \D}^{\, l'}  c_{\rad}^{\, j}}{ M} \, 
\| |\mbf{x}| \chi_{\I} \|_{L^1}  \| ( \1 \tens (\Hradm +1)^{1 /2} )   \Psi_{m} \|, \\
& \left| \int_{\Rthree} \chiIx   e^{-i \mbf{p} \cdot \mbf{x}}
\left(  R_{m,M}(\mbf{p})^2 \Phi , \,  (\psi_{l'}(\mbf{x}) \tens A_j (\mbf{x}))  \Psi_m \right)
\dx  \right|   \leq \frac{c_{\, \D}^{\, l'}  c_{\rad}^{\, j}}{ M^2} \,
\| \chi_{\I} \|_{L^1}   \| ( \1 \tens (\Hradm +1)^{1 /2} ) \Psi_{m} \|  .
\end{align*}
}
It is seen that   $ \| ( \1 \tens (\Hradm^{1/2} +1)^{1/2}) \Psi_m \| \leq 
  \|  H_{0,m} \Psi_{m} \| +  \| \Psi_m\| = \|  H_{0,m} \Psi_{m} \| +1 $, and hence,
\begin{align}
& \left| \partial_{p^{\nu}} \left( \Phi, R_{m,M}(\mbf{p}) K_{s}^{\, +} (\mbf{p})\Psi_m \right) \right|  \notag \\
& \leq   \| (1+ |\mbf{x}|) \chi_{\I} \|_{L^1} \sum_{j=1}^{3} \sum_{l,l'=1}^{4} c_{\, \D}^{\, l'}  c_{\rad}^{\, j}
 \left( 
 \frac{| \partial_{p^{\nu}}f_{s }^{\, l}(\mbf{p}) |}{ M} 
+ \frac{ | f_{s }^{\, l}(\mbf{p}) | \,   }{M} 
+  \frac{ | f_{s }^{\, l}(\mbf{p}) | \,   }{M^2}
\right)   \, \left( \|   H_{0,m}  \Psi_m \| +1 \frac{}{}  \right) .
\label{9/13.I}
 \end{align}
(Second term)   
 It is seen that
\begin{align}
&  \left( \Phi, \partial_{p^{\nu}}R_{m,M}(\mbf{p}) 
 S^{\,+}_{s} (\mbf{p} )    \Psi_m \right)  \notag \\
& =- \sum_{l=1}^4 \partial_{p^{\nu}} \left( f_{s }^{\, l}(\mbf{p}) \int_{\Rthree \times \Rthree} \frac{\chiIIx \chiIIy}{|\mbf{x}-\mbf{y}|} e^{-i \mbf{p} \cdot \mbf{y}}
\left( R_{m,M}(\mbf{p}) \Phi , \,  ( \rho (\mbf{x}) \psi_{l}(\mbf{y}) \tens \1 ))  \Psi_m \right)  \dx \dy
 \right) \notag   \\
&  =- \sum_{l=1}^4 \left\{ ( \partial_{p^{\nu}}  f_{s }^{\, l}(\mbf{p})) \int_{\Rthree \times \Rthree} \frac{\chiIIx \chiIIy}{|\mbf{x}-\mbf{y}|} e^{-i \mbf{p} \cdot \mbf{y}}
\left( R_{m,M}(\mbf{p}) \Phi , \,  ( \rho (\mbf{x}) \psi_{l}(\mbf{y}) \tens \1 ))  \Psi_m  \right)  \dx \dy  \right. \notag \\
&\qquad \quad - i  f_{s }^{\, l}(\mbf{p}) \int_{\Rthree \times \Rthree} \frac{\chiIIx \chiIIy}{|\mbf{x}-\mbf{y}|} y^{\nu}  e^{-i \mbf{p} \cdot \mbf{y}}
\left( R_{m,M}(\mbf{p}) \Phi , \,  ( \rho (\mbf{x}) \psi_{l}(\mbf{y}) \tens \1 ))  \Psi_m  \right)  \dx \dy  \notag \\
& \qquad \quad \left. - \frac{f_{s }^{\, l}(\mbf{p}) p^{\nu} }{ \omega_{\, M}(\mbf{p})} \int_{\Rthree \times \Rthree} \frac{\chiIIx \chiIIy}{|\mbf{x}-\mbf{y}|}   e^{-i \mbf{p} \cdot \mbf{y}} \left( R_{m,M}(\mbf{p})^2 \Phi , \,  ( \rho (\mbf{x}) \psi_{l}(\mbf{y}) \tens \1 ))
  \Psi_m  \right)  \dx \dy \right) . \label{9/13.9}
\end{align}
By  evaluating  the right-hand side of (\ref{9/13.9}), we have 
\begin{equation}
 \left| \partial_{p^{\nu}} \left( \Phi, R_{m,M}(\mbf{p}) 
 S^{\,+}_{s} (\mbf{p} )    \Psi_m \right)  \right|
   \leq  \| (1+ |\mbf{x}|) \chi_{\I} \|_{L^1} \sum_{l,l'=1}^{4} ( c_{\, \D}^{\, l'})^2 c_{\, \D}^{\, l}\left( 
 \frac{| \partial_{p^{\nu}}f_{s }^{\, l}(\mbf{p}) |}{ M} 
+ \frac{ | f_{s }^{\, l}(\mbf{p}) | \,   }{ M} 
+  \frac{ | f_{s }^{\, l}(\mbf{p}) | \,   }{M^2}
\right)     . \label{9/13.II}
\end{equation}
(Third term)  We see that 
\begin{align}
&  \left( \Phi, \partial_{p^{\nu}} R_{m,M}(\mbf{p}) 
 T^{\,+}_{s} (\mbf{p} )    \Psi_m \right)  \notag \\
& =-\sum_{l=1}^4 \partial_{p^{\nu}} \left( f_{s }^{\, l}(\mbf{p}) \int_{\Rthree \times \Rthree} \frac{\chiIIx \chiIIy}{|\mbf{x}-\mbf{y}|} e^{-i \mbf{p} \cdot \mbf{y}}
\left( R_{m,M}(\mbf{p}) \Phi , \,  ( \psi_{l}(\mbf{x}) \rho (\mbf{y})  \tens \1 ))  \Psi_m  \right)  \dx \dy
 \right) \notag   \\
&  =- \sum_{l=1}^4 \left\{ ( \partial_{p^{\nu}}  f_{s }^{\, l}(\mbf{p})) \int_{\Rthree \times \Rthree} \frac{\chiIIx \chiIIy}{|\mbf{x}-\mbf{y}|} e^{-i \mbf{p} \cdot \mbf{y}}
\left( R_{m,M}(\mbf{p}) \Phi , \,  (\psi_{l}(\mbf{x}) \rho (\mbf{y})  ) \tens \1 ))  \Psi_m  \right)  \dx \dy  \right. \notag \\
&\qquad \quad - i  f_{s }^{\, l}(\mbf{p}) \int_{\Rthree \times \Rthree} \frac{\chiIIx \chiIIy}{|\mbf{x}-\mbf{y}|} y^{\nu}  e^{-i \mbf{p} \cdot \mbf{y}}
\left( R_{m,M}(\mbf{p}) \Phi , \,  (\psi_{l}(\mbf{x}) \rho (\mbf{y})  ) \tens \1 ))  \Psi_m \right)  \dx \dy  \notag \\
& \qquad \quad \left. - \frac{f_{s }^{\, l}(\mbf{p}) p^{\nu} }{ \omega_{\, M}(\mbf{p})} \int_{\Rthree \times \Rthree} \frac{\chiIIx \chiIIy}{|\mbf{x}-\mbf{y}|}   e^{-i \mbf{p} \cdot \mbf{y}} \left( R_{m,M}(\mbf{p})^2 \Phi , \,  (\psi_{l}(\mbf{x}) \rho (\mbf{y})  ) \tens \1 ))  \Psi_m  \right)  \dx \dy \right) . \label{9/13.10}
\end{align}
We estimate the right-hand side of the absolute value of (\ref{9/13.10}), and then, 
\begin{equation}
 \left| \partial_{p^{\nu}} \left( \Phi, R_{m,M}(\mbf{p}) 
 T^{\,+}_{s} (\mbf{p} )    \Psi_m \right)  \right|   \leq 
  \| (1+ |\mbf{x}|) \chi_{\I} \|_{L^1} \sum_{l,l'=1}^{4}c_{\, \D}^{\, l} ( c_{\, \D}^{\, l'})^2 
 \left( 
 \frac{| \partial_{p^{\nu}}f_{s }^{\, l}(\mbf{p}) |}{M} 
+ \frac{ | f_{s }^{\, l}(\mbf{p}) | \,   }{ M} 
+  \frac{ | f_{s }^{\, l}(\mbf{p}) | \,   }{M^2}
\right)  . \label{9/13.III}
\end{equation}
From (\ref{9/13.I}), (\ref{9/13.II}) and (\ref{9/13.III}),  we have
\begin{align}
& \left| ( \Phi , \partial_{p^{\nu}} ( b_{s} (\mbf{p}) \tens \1 ) \Psi_{m} ) \right| 
 \notag \\
&  \leq    \sum_{l=1}^4 c_{+}^{\,l}  \left( 
 \frac{| \partial_{p^{\nu}}f_{s }^{\, l}(\mbf{p}) |}{ M} 
+ \frac{ | f_{s }^{\, l}(\mbf{p}) | \,   }{M} 
+  \frac{ | f_{s }^{\, l}(\mbf{p}) | \,   }{ M^2}
\right)   \left(   |\kappa_{\, \I }| \,  \| H_{0,m} \Psi_m \|  + |\kappaI | + 2 |\kappa_{\, \II }|    \frac{}{} \right) ,  \notag 
\end{align}
where $c_{+}^{\,l} =  \| (1+ |\mbf{x}|) \chi_{\I} \|_{L^1} \, \times  \max \left\{ 
\sum\limits_{j=1}^{3}\sum\limits_{l'=1}^{4} |\alpha^j_{l,l'}| c_{\, \D}^{\, l'} c_{\rad}^{\, j} , \;   \sum\limits_{l'=1}^{4}  ( c_{\, \D}^{\, l'})^2  c_{\, \D}^{\, l}   \right\}$. 
By the definition of   $f_{s }^{\, l}(\mbf{p}) = \frac{\chi_{\D}(\mbf{p}) \,  u_{s}^{\,l}(\mbf{p})}{
\sqrt{(2 \pi )^3  } }$, we have
\begin{equation}
\left| ( \Phi , \partial_{p^{\nu}} ( b_{s} (\mbf{p}) \tens \1 ) \Psi_{m} ) \right| 
  \leq   F_{s ,+}^{\nu}(\mbf{p})
   \, \left(   |\kappa_{\, \I }| \, \| H_{0,m} \Psi_m \| + |\kappaI |+ 2 |\kappa_{\, \II }|    \frac{}{} \right) , \label{9/13.11}
\end{equation}
where
\[
F_{s ,+}^{\nu}(\mbf{p}) = \frac{1}{\sqrt{(2 \pi )^3}} \sum_{l=1}^4 c_{+}^{\, l} \left(
  \frac{| \partial_{p^{\nu}} \chi_{\D}(\mbf{p}) \,   |}{ M} 
+ \frac{ | \chi_{\D}(\mbf{p}) \partial_{p^{\nu}} u_{s}^{\,l}(\mbf{p}) | \,   }{M} + \frac{ |\chi_{\D}(\mbf{p})  | \,   }{ M}
+  \frac{ |\chi_{\D}(\mbf{p})  | \,   }{ M^2}
\right)  . 
\]
We see that    (\ref{9/13.11}) holds  for all $\Phi \in \ms{F}_{\QED}$ with 
$\| \Phi \| =1$, and this implies that 
\begin{equation}
\left\|  \partial_{p^{\nu}} ( b_{s} (\mbf{p}) \tens \1 ) \Psi_{m} ) \right\| 
  \leq    F_{s ,+}^{\nu}(\mbf{p})
   \, \left(   |\kappa_{\, \I }| \|  H_{0,m}  \Psi_m \| +| \kappaI | + 2 |\kappa_{\, \II }|   \frac{}{} \right)  . \notag
\end{equation}
From Lemma \ref{9/9.e}, it holds that for all $0 < \epsilon < \frac{1}{c_{\I} |\kappaI |}$, 
\[
   \|  H_{0,m} \Psi_{m} \|   \leq 
 \, L_{\epsilon} \| H_{m} \Psi_m \| +  R_{\epsilon}  \| \Psi_m \| 
=  \, L_{\epsilon} E_{0}(H_m)  + R_{\epsilon}    . 
\]  
Thus  \textbf{(i)} is obtained. Similarly, \textbf{(ii)} is also proven in a same way as \textbf{(i)}.  $\blacksquare $\\

\subsection{Photon Derivative Bound}
In a similar to the Dirac field, we introduce the distribution kernel of the annihilation operator for the radiation field.
For all $ \Psi = \left\{  \Psi^{(n)} = \left( \Psi^{(n)}_1 ,   \Psi^{(n)}_2  \right)  \right\}_{n=0}^{\infty}
\in \ms{D} ( H_{\rad , m } )$, we define $a_{r}(\mbf{k})$, by
\[
a_{r}(\mbf{k})\Psi^{(n)}_{\varrho} ( \mbf{k}_{1} , \cdots ,  \mbf{k}_{n} )
= \delta_{\, r , \varrho  } \sqrt{n+1} \Psi^{(n+1 )}_{\varrho }( \mbf{k} , \mbf{k}_{1} , \cdots ,  \mbf{k}_{n} ) , \qquad \varrho = 1,2. 
\]
It holds that 
\begin{equation}
\qquad \qquad 
(\Phi ,  a_{r}(h)  \Psi )=
\int_{\Rthree}h(\mbf{k})^{\ast}(\Phi ,  a_{r}(\mbf{k}) \Psi) d \mbf{k}, 
\quad \Phi \in \ms{F}_{\rad} , \; \Psi \in  \ms{D}(H_{\rad , m})  .
\end{equation}

\begin{lemma}  \label{9/13.c}\normalfont
Assume (\textbf{A.2}). Then  for all $\Phi , \Psi  \in \ms{D}(H_{\rad ,m })$,
\begin{align*}
& \textbf{(i)} \; \;
[H_{\rad , m }  \, , a_{r} (h)]^0 (\Phi , \Psi )=  \left(  \Phi , a_{r}( \omega_m h )\Psi
\right) , \\
&\textbf{(ii)} \; \; [A_{j}(\mbf{x} )  , a_{r}(h) ]^0 (\Phi , \Psi )=
-  (h, h_{r , \mbf{x}}^j  )\left(  \Phi , \Psi \right)  .
 \end{align*}
\end{lemma}
\textbf{(Proof)}
It holds that for all $\Phi  \in \ms{F}_{\rad}^{\, \fin} (\ms{D} ( \omega_{\, m} ))$,
\begin{align}
& [ H_{\rad , m}, a^{\dagger}_{r}(h)   ]\Phi =  -  a^{\dagger}_{r}(\omega_{\,m }h)\Phi ,
\label{9/14.1}  \\
&[ A_{j} (\mbf{x}), a^{\dagger}_{r}(h)   ]\Phi =  ( h_{r , \mbf{x}}^j ,h  ) \Phi .\label{9/14.2}
\end{align}
In a similar  way to
Lemma \ref{9/12.d},  we can prove \textbf{(i)}  by (\ref{9/14.1}) and \textbf{(ii)} by (\ref{9/14.2}). $\blacksquare $ \\

\begin{lemma}  \label{9/13.d} \normalfont 
Assume \textbf{(A.1)} - \textbf{(A.3)}. Then  \\
\textbf{(i)} it holds that  for all $\Phi , \Psi \in \ms{D}(H_m )$,  
\begin{equation}
 [\HI ,  \1 \tens a_{r} ( h )]^0 (\Phi , \Psi)  \,
= \, \int_{\Rthree} h(\mbf{k})^{\ast} \left( \Phi , Q_{r} (\mbf{k})   \Psi  \right)
 d \mbf{k} . \notag 
\end{equation}
Here  $ Q_r (\mbf{k})$  is  an operator which satisfy  
 \[
(\Phi , Q_{r} (\mbf{k}) \Psi ) = - \sum_{j=1}^3  \, 
 \int_{\Rthree} \chiIx   h_{r, \mbf{x}}^{\, j} (\mbf{k})
\left( \Phi , \,  (\psidaggerx \alpha^j  \psix  \tens \1 )  \Psi \right) \dx 
\]
with $ \| Q_{r} (\mbf{k}) \| \leq \|\chi_{\I} \|_{L^1}  \sum\limits_{j=1}^3 \sum\limits_{l,l'=1}^4 |h_{r}^j (\mbf{k})| \, |\alpha^{j}_{l,l'}| \, |c_{\, \D}^{\, l}| \, |c_{\, \D}^{\, l'}|$. \\
\textbf{(ii)} Additionally  assume \textbf{(A.4)} and \textbf{(A.6)}.
Then,  $Q_{r} (\mbf{k}) \Psi$ is strongly differential for all $\mbf{k} \in \Rthree \backslash O_{\rad}$. 
\end{lemma}
\textbf{(Proof)} \textbf{(i)} Let $\Phi \in \ms{D}(H_m)$
From Lemma \ref{9/13.c},
\begin{align} 
[\HI , \1 \tens a_{r}(h ) ]^0 (\Phi , \Psi ) &=  
\sum_{j=1}^3 \int_{\Rthree} \chiIx [( \psidaggerx \alpha^j \psi  (\mbf{x}) \tens A_{j}(\mbf{x}), \1 \tens a_{r}(h)]^0( \Phi , \Psi ) \dx
 \notag \\
&= \sum_{j=1}^3 \int_{\Rthree}  \chiIx [ \1 \tens  A_{j}(\mbf{x}) , \1 \tens a_{r}(h)  ]^0 
(  \Phi ,( \psidaggerx \alpha^j \psi  (\mbf{x}) \tens \1 )  \Psi   )   \dx \notag \\
& =- \sum_{j=1}^3 \int_{\Rthree}  \chiIx( h, h_{r ,\mbf{x}}^j  )(  \Phi ,( \psidaggerx \alpha^j \psi  (\mbf{x}) \tens \1 )  \Psi   )   \dx .
\end{align}
We define $\ell_{r, \mbf{k}}:\FQED \tens \FQED \to \mbf{C}$  by
\[
\ell_{r, \mbf{k}} (\Phi ' , \Psi ') = - \sum_{j=1}^3  h_{r}^j  (\mbf{k}) \int_{\Rthree} \chiIx e^{-i \mbf{k} \cdot \mbf{x}}  \,  (  ( \psidaggerx \alpha^j \psi  (\mbf{x})   \tens \1 ) \Phi ', \Psi  ' ) \dx
\]
We see that $|\ell_{r, \mbf{k}} (\Phi ' , \Psi ' ) | \leq 
\|\chi_{\I} \|_{L^1}  \sum\limits_{j=1}^3 \sum\limits_{l,l'=1}^4 |h_{r}^j (\mbf{k})| \, |\alpha^{j}_{l,l'}| \, |c_{\, \D}^{\, l}| \, |c_{\, \D}^{\, l'}|\, \| \Phi ' \| \, \| \Psi' \| $. By Riesz representation theorem, we can define  an operator $Q_{r} (\mbf{k})$ such that $\ell_{r, \mbf{k}} (\Phi ' , \Psi ')  = (\Phi ', Q_{r}(\mbf{k}) \Psi ')$. Then we have
\begin{equation}
[\HI , \1 \tens a_{r}(f) ]^0 (\Phi , \Psi )
= \int_{\Rthree} h (\mbf{k})^{\ast} \ell_{r, \mbf{k}} (\Phi , \Psi ) \dk
=   \int_{\Rthree} h (\mbf{k})^{\ast} \left(  \Phi ,Q_{r} (\mbf{k}) \Psi    \right)   \dk   .
\notag 
\end{equation}
Then \textbf{(i)} is obtained. \\
\textbf{(ii)} The strong differentiability of $ Q_{r} (\mbf{k}) \Psi $ is proven by  \textbf{(A.4)} and \textbf{(A.6)} in a similar way to Lemma \ref{9/12.c}, and the proof is omitted.  $\blacksquare $. \\

\begin{proposition} \label{PF-P}  \normalfont 
(\text{Photon Pull-Through Formula})  \\
Assume \textbf{(A.1)} - \textbf{(A.3)}. 
Then it holds that for almost everywhere $\mbf{k} \in \Rthree$, 
\begin{equation} 
(\1 \tens a_{r} (\mbf{k})) \Psi_m =\kappaI (H_m- E_0 (H_m) + \omega_{m}(\mbf{k}))^{-1} 
Q_{r} (\mbf{k}) \Psi_m  .
\end{equation}
\end{proposition} 
\textbf{(Proof)} Let $\Phi \in \ms{D}(H_m) $. 
By Lemma \ref{9/13.c} \textbf{(i)}, 
\begin{equation}
[\Hm , \1 \tens a_{r}(h ) ]^0 (\Phi , \Psi_m )= - \left( \Phi , ( \1 \tens a_{r} ( \omega_m h)) \Psi_m \right)
 +   \kappaI \, [\HI , \1 \tens a_{r}(h) ]^0 (\Phi , \Psi_m )  . \notag
 \end{equation}
 It also holds that 
\begin{equation}
[\Hm , \1 \tens a_{r}(h) ]^0 (\Phi , \Psi_m ) =
 \left( (H_{m}-E_{0}(\Hm ) ) \Phi , (\1 \tens a_{r}(h)) \Psi_m \right) . \notag
\end{equation}
Then we have
\begin{equation}
\left( (H_{m}-E_{0}(\Hm ) ) \Phi , (\1 \tens a_{r}(h)) \Psi_m \right) 
+ \left( \Phi, (\1 \tens a_{r}(\omega_m h)) \Psi_m \right) = \kappaI 
[\HI , \1 \tens a_{r}(h )]( \Phi, \Psi_{m} ).  \notag
\end{equation}
By Lemma \ref{9/13.d}, 
\begin{equation}
\int_{\Rthree} h (\mbf{k})^{\ast}
\left(  (H_{m}-E_{0}(\Hm )+\omega_{m}(\mbf{k}) ) \Phi , (\1 \tens a_{r}(\mbf{k})) \Psi_m    \right) \dk 
=  \kappaI   \int_{\Rthree} h (\mbf{k})^{\ast} \left(  \Phi ,Q_{r} (\mbf{k}) \Psi_m    \right)   \dk . \label{9/13.14}
\end{equation}
Note that  (\ref{9/13.14}) holds for all $h \in L^2 (\Rthree )$. Then  we have
\begin{equation}
\left(  (H_{m}-E_{0}(\Hm )+\omega_{m}(\mbf{k}) ) \Phi , (\1 \tens a_{r}(\mbf{k})) \Psi_m    \right) =    \left(  \Phi ,\,  \kappaI Q_r (\mbf{k}) \Psi_m    \right) ,  \label{9/13.15}
\end{equation}
for almost everywhere $ \mbf{k} \in \Rthree $.  In addition, (\ref{9/13.15}) yields that 
$ (\1 \tens a_{r}(\mbf{k})) \Psi_m \in \ms{D} (H_m )$ and 
\begin{equation}
  (H_{m}-E_{0}(\Hm )+\omega_{m}(\mbf{k}) ) (\1 \tens a_{r}(\mbf{k})) \Psi_m  =   \kappaI  Q_r (\mbf{k}) \Psi_m    . \notag 
\end{equation}
Thus the proof is obtained. $\blacksquare $

\begin{theorem} \label{P-DB} 
 \normalfont (\textbf{Photon Derivative Bounds}) \\
Assume \textbf{(A.1)}-\textbf{(A.4)} and \textbf{(A.6)}. Then it holds that  for all $\mbf{k} \in \Rthree \backslash O_{\rad}$, 
\begin{equation}
  \left\| \partial_{k^\nu} (\1 \tens a_{r} (\mbf{k})) \Psi_m  \right\|     \leq |\kappaI |  F_{r}^{\, \nu} (\mbf{k}) \notag 
\end{equation}
where $  F_{r}^{\, \nu} $ is a function which satisfy $ F_{r}^{\, \nu} \in L^2 (\Rthree ) $. 
\end{theorem}
\textbf{(Proof)} $\; $ \\
 Let $ R_{m}(\mbf{k}) = (H_m- E_0 (H_m) + \omega_{\, m}(\mbf{k}))^{-1} $. From Proposition \ref{PF-P}, it holds that  $\1 \tens a_{r} (\mbf{k}) \Psi_m =  R_{m}(\mbf{k}) 
  Q_{r} (\mbf{k})   \Psi_m   $. Then for all $\Phi \in \ms{F}_{\QED}$, 
\begin{align}
& = ( \Phi ,  \partial_{k^\nu } (\1 \tens a_{r} (\mbf{k})) \Psi_m  )  \notag \\
& =- \kappaI \sum_{j=1}^3  \, \partial_{k^{\nu}} \left(   h_{r}^{\, j} (\mbf{k})
 \int_{\Rthree} \chiIx   e^{-i \mbf{k} \cdot \mbf{x}}
\left( R_{m}(\mbf{k}) \Phi , \,  ( \psidaggerx \alpha^j  \psi(\mbf{x})
  \tens  \1 )  \Psi_m  \right) \dx \right) \notag  \\
& =- \kappaI \sum_{j=1}^3  \, \left\{  ( \partial_{k^{\nu}}  h_{r}^{\, j} (\mbf{k}))
 \int_{\Rthree} \chiIx   e^{-i \mbf{k} \cdot \mbf{x}}
\left( R_{m}(\mbf{k}) \Phi , \,  ( \psidaggerx \alpha^j  \psi(\mbf{x}) 
\tens \1 )  \Psi_m \right) \dx \right. \notag  \\
& \qquad \qquad \qquad  -i  h_{r}^{\, j} (\mbf{k}) \int_{\Rthree} \chiIx   x^{\nu} \, e^{-i \mbf{k} \cdot \mbf{x}}
\left( R_{m}(\mbf{k}) \Phi , \,  ( \psidaggerx \alpha^j  \psi(\mbf{x})
 \tens \1 )  \Psi_m  \right) \dx 
\notag \\
& \qquad \qquad \qquad  \left.  - \frac{ h_{r}^{\, j} (\mbf{k}) k^{\nu} }{\omega_{\, m}(\mbf{k})}
   \int_{\Rthree} \chiIx   e^{-i \mbf{k} \cdot \mbf{x}}
\left(  R_{m}(\mbf{k})^2 \Phi , \,  ( \psidaggerx \alpha^j  \psi(\mbf{x})
 \tens \1 )  \Psi_m \right)
\dx \right\}  . \label{9/13.16}
\end{align}
By estimating the absolute value of the right-hand side of  (\ref{9/13.16}), we have
\begin{align*}
& \left| \partial_{k^{\nu}} \left( \Phi, R_{m}(\mbf{k})Q (\mbf{k})\Psi_m \right) \right| 
\notag \\
& \leq  \| (1+ |\mbf{x}| ) \chi_{\I} \|_{L^1}|\kappaI |
\sum_{j=1}^3 \sum_{l,l'=1}^4 \, | \alpha^j_{l,l'} | \, | c_{\, \D}^{\,l}| \, |  c_{\, \D}^{\,l'}| \, 
 \left( 
 \frac{| \partial_{k^{\nu}} h_{r }^{\, j}(\mbf{k}) |}{\omega_{\, m}(\mbf{k})} 
+ \frac{ | h_{r }^{\, j}(\mbf{k}) | \,   }{\omega_{\, m}(\mbf{k})} 
+  \frac{ | h_{r }^{\, j}(\mbf{k}) | \,   }{\omega_{\, m}(\mbf{k})^2}
\right) . \notag
 \end{align*}
From the definition of  $h_{r}^{\, j}(\mbf{k})=\frac{\chi_{\rad}(\mbf{k}) 
e_{r}^{\,j}(\mbf{k}))}{\sqrt{2 (2 \pi )^3 \omega (\mbf{k}) }}$, we have 
 \begin{equation}
  \partial_{k^\nu} h_{r}^{\, j}(\mbf{k}) = \frac{1}{\sqrt{2 (2 \pi )^3 }} \left( 
\frac{ ( \partial_{k^\nu} \chi_{\rad}(\mbf{k}) )  e_{r}^{\,j}(\mbf{k}) }{\omega(\mbf{k})^{1/2}}
+ \frac{ \chi_{\rad}(\mbf{k}) \partial_{k^\nu} e_{r}^{\,j}(\mbf{k})}{\omega (\mbf{k})^{1/2}}
- \frac{1}{2} \frac{  \chi_{\rad}(\mbf{k}) k^{\nu}}{ \omega (\mbf{k})^{5/2}} \right) . \notag
 \end{equation}
Hence, it holds that 
\begin{equation}
\left| ( \Phi ,  \partial_{k^\nu } (\1 \tens a_{r} (\mbf{k})) \Psi_{m} ) \right|
  \leq    | \kappaI | \, F_{r}^{\nu}(\mbf{k}) ,  \label{9/13.17}
\end{equation}
where
 \begin{align*}
  F_{r}^{\nu}(\mbf{k}) =\frac{\| (1+ |\mbf{x}| ) \chi_{\I} \|_{L^1}}{\sqrt{2 (2 \pi )^3 }}   & \sum_{j=1}^3 \sum_{l,l'=1}^4 \, | \left\{  \alpha^j_{l,l'} | \, | c_{\, \D}^{\,l}| \, |  c_{\, \D}^{\,l'}| \, \frac{}{} \right.\notag \\
  &   \times \left.
\left( \frac{| \partial_{k^\nu} \chi_{\rad}(\mbf{k}) |+| \chi_{\rad}(\mbf{k}) \partial_{k^\nu} e_{r}^{\,j}(\mbf{k})| 
 + | \chi_{\rad}(\mbf{k}) | }{  \omega (\mbf{k})^{3/2}}
+ \frac{3}{2} \frac{ | \chi_{\rad }(\mbf{k}) |}{ \omega (\mbf{k})^{5/2}}  \right) \right\} . \notag 
 \end{align*}
 Since (\ref{9/13.17}) holds for all $\Phi \in \FQED$, we have
 \[
\| \partial_{k^{\nu}} ( \1  \tens a_{r}(\mbf{k}) ) \Psi_{m}  \| \leq  | \kappaI | F_{r}^{\nu}(\mbf{k}) .
 \]
 The condition \textbf{(A.6)}  yields that  $ F_{r}^{\nu} \in L^2 (\Rthree) $, and hence the proof is obtained. $\blacksquare $


\section{Proof of Theorem \ref{Main-Theorem} }
Let  $\{ \Psi_m\}_{m>0}$  be the sequence of the  normalized ground state of $H_m$, $m>0$. Then there exists a subsequence of $\{ \Psi_{m_{j}} \}_{j=1}^{\infty}$ with $m_{j+1} < m_{j} $, $j \in \mbf{N}$, such that the weak limit $\Psi_{0} := $w-$\lim\limits_{j \to \infty } \Psi_{m_{j}} $ exists.

\begin{lemma}    \label{9/16.a} \normalfont
Suppose \textbf{(A.1)} - \textbf{(A.3)}. Then, \\
$\qquad $ \textbf{(i)} $\ms{D}_{0}$ is a common  core of $H_{\QED}$ and $H_{m}$, $m>0$, and 
 $H_{m}$ strongly converges to $H_{\QED}$ on $\ms{D}_{0}$ \\
$\qquad $ \textbf{(ii)} $  \lim\limits_{m \to \infty} E_{0}(H_m) = E_{0}(H_{\QED} ) $.
\end{lemma}
\textbf{(i)} Since $\ms{D}_{0}$ is a core of $H_{0,m}$,  $\ms{D}_{0}$ is also a core of $H_{m}$. It is directly proven that $ \lim\limits_{m \to 0} H_{m} \Psi = H_{\QED} \Psi $ for all $\Psi \in \ms{D}_0$. \\
\textbf{(ii)} We see that   $(\Psi , H_{m} \Psi )  \geq ( \Psi, H_{\QED} \Psi ) \geq {E_{0}(H_{\QED})}$, for all $\Psi \in \ms{D}_{0}$. Hence $\inf\limits_{m>0}E_{0}(H_{m}) \geq E_{0}(H_m )$. From \textbf{(i)}, it follows that    $H_{m}$ converges to $H_{\QED}$ as $m \to 0$ in the strong resolvent sense, and this yields that  $\limsup\limits_{m \to 0}E_{0}(H_{m}) \leq E_{0}(\HQED )$. Hence \textbf{(ii)} follows. $\blacksquare $\\

$\;$ \\
From  Lemma \ref{9/16.a} \textbf{(ii)}, we can set 
\[
 E_{\infty} = \sup\limits_{j\in \mbf{N}} |E_{0}(H_{m_j })|  \; < \infty . \\
\]

\begin{lemma} \normalfont  \label{9/14.a}  \textbf{(Number Operator Bounds)} \\
Suppose \textbf{(A.1)} - \textbf{(A.6)}. Then, for all  $
 0 < \epsilon < \frac{1}{c_\I | \kappaI |} ,$
\begin{align*}
 \quad \mbf{(i)} \; \; & 
\sup_{j\in \mbf{N} } \| (N_{\D}^{1/2} \tens \1 ) \Psi_{m_j } \|
 \leq \left( \frac{L_{\epsilon}}{M}E_{\infty} + \frac{R_{\epsilon} }{M}\right)^{1/2} , \notag \\
\mbf{(ii)} \; \;&  \sup_{j\in \mbf{N} } \| (\1 \tens N_{\rad}^{1/2} ) \Psi_{m_j } \|
 \leq c_0 |\kappaI|  \, \left\|  \frac{ \chi_{\rad}}{\omega^{3/2}} \right\|   , \notag 
\end{align*}
where $c_{0} = \sqrt{\frac{11}{2(2\pi)^3}} \sum\limits_{j=1}^3 \sum\limits_{l,l'=1}^4 |\alpha^{j}_{l,l'}| c_{\D}^{\,l} c_{\D}^{\,l'} $. 
\end{lemma}
\textbf{(Proof)} 
\textbf{(i)} We see that $ \|( N_{\D}^{1/2} \tens \1 ) \Psi_{m} \|^2  
 = (\Psi_m , ( N_{\D} \tens \1 ) \Psi_{m} )\leq \|( N_{\D} \tens \1 ) \Psi_{m} \|$, and Corollary \ref{9/9.f} yields that for all $0 < \epsilon < \frac{1}{c_\I | \kappaI | }$,
\begin{equation}
  \|( N_{\D} \tens \1 ) \Psi_{m} \|
 \leq \frac{L_{\epsilon}}{M} \| H_m \Psi_m  \| + \frac{ R_{\epsilon} }{M}\|\Psi_m  \| = 
 \frac{L_{\epsilon}}{M}E_{0}(H_m ) + \frac{R_{\epsilon}}{M} .
 \notag 
\end{equation}
Hence \textbf{(i)}  follows. \\
\textbf{(ii)} From the photon pull-through formula in Proposition \ref{PF-P}, it follows that
\begin{align}
 (\Psi_m , ( \1 \tens N_{\rad}) \Psi_m ) 
 & = \sum_{r=1,2} \int_{\Rthree} \| (\1 \tens a_{r}(\mbf{k})) \Psi_m \|^2 d \mbf{k} \notag \\
 & = |\kappaI |^2 \sum_{r=1,2} \int_{\Rthree}\| (H_m -E_{0}(H_m ) + \omega_{m}(\mbf{k})) Q_{r}(\mbf{k}) \Psi_m \|^2 d \mbf{k} \notag \\
 & \leq |\kappaI|^2 \frac{11}{{2(2\pi)^3}} \sum_{r=1,2} \sum_{j=1}^3 \sum_{l,l'=1}^4 |\alpha^{j}_{l,l'}|^2 ( c_{\D}^{\,l} c_{\D}^{\,l'} )^2 \left( \int_{\Rthree}   \frac{| \chi_{\rad}|^2}{|\mbf{k}|^3}  d \mbf{k} \right) . \label{9/15.1}
\end{align}
From (\ref{9/15.1}),  we obtain \textbf{(ii)}. $\blacksquare $ \\

\begin{proposition}  \label{9/15.a} \normalfont 
Assume \textbf{(A.1)}-\textbf{(A.6)}. Let $ F \in C_0^{\, \infty} (\mbf{R}^3)$ which satisfy   
$0 \leq F \leq 1$ and $F(\mbf{x})=1$ for  $|\mbf{x}| \leq 1$, 
and set  $ F_{ R}(\mbf{x}) = F (\frac{  \mbf{x} }{R} ) $. Let   $\hat{\mbf{p}} = - i \nabla_{\mbf{p}}  $ and  $\hat{\mbf{k}} = - i \nabla_{\mbf{k}}  $. Then 
for all  $0 < \epsilon < \frac{1}{c_{\I} \kappaI}$, $R \geq 1$ and $R' \geq 1$,
\begin{align*}
&\textbf{(i)} \; \;  \,   \sup_{j\in \mbf{N} } \|  (  \1 - 
 \Gammaf ( F_{ R} (\hat{\mbf{p}}) ) \tens \1 )  \Psi_{m_j } \|)  \leq  
\frac{c_{1,\epsilon  }}{\sqrt{R}} , \\
&\textbf{(ii)} \; \;  \,    \sup_{  j\in \mbf{N}} \|  ( \1 \tens ( \1 -  \Gammab ( 
F_{R'}(\hat{\mbf{k}}) ))   \Psi_{m_j } \| )  \leq   \frac{c_2}{\sqrt{R'}}  , \notag 
\end{align*}
where 
\begin{equation}
c_{1, \epsilon } =   \left( 4 \frac{L_{\epsilon} E_{\infty} +  R_{\epsilon}}{M}\right)^{1/4}
\left(   \left( \frac{L_{\epsilon} E_{\infty} +  R_{\epsilon}}{M}\right)^{1/2}
+  ( L_{\epsilon} E_{\infty} + R_{\epsilon} +1 ) |\kappaI| + 2|\kappaII  |  \sum_{s=\pm 1/2} \sum_{\nu =1}^3  \sum_{\tau= \pm } \|F_{s ,\tau}^{\nu} \| 
\right)^{1/2}  \notag 
\end{equation}
  and 
\begin{equation}
c_2 =    |\kappaI|^{1/2} \left(  c_0  \left\| \frac{\chi_{\rad}}{\omega^{3/2}} \right\| \,\right)^{1/2}  \, \left(
c_0  \left\| \frac{\chi_{\rad}}{\omega^{3/2}} \right\|
+ \, \sum_{r=1,2} \sum_{\nu =1}^3 \| F_{r}^{\nu} \|_{L^2}\right)^{1/2} . \notag 
\end{equation}
\end{proposition}
\textbf{(Proof)}
It follows that  $ (\1 -  \Gammaf (F_{ R} (\hat{\mbf{p}}) )^2 \leq  \1 - \Gammaf (F_{ R} (\hat{\mbf{p}}) ) \leq 
 \sqzf{1- F_{ R} (\hat{\mbf{p}})}  $, and then,
\begin{align}
 \|  ( ( \1 -  \Gammaf (F_{ R} (\hat{\mbf{p}})  ) \tens \1 )  \Psi_{m} \|^2 
&\leq ( \Psi_m ,\left(    \dGammaf ( 1- F_{ R} (\hat{\mbf{p}})  ) \tens \1 \right) \Psi_{m} ) \notag \\
& = \sum_{s= \pm 1/2}
\left(  \int_{\Rthree} \left( (b_s(\mbf{p}) \tens \1 )\Psi_{m}   , 
(1- F_{ R} (\hat{\mbf{p}})  )
(b_s(\mbf{p}) \tens \1 ) \Psi_m \right)  d \mbf{p}  \right. \notag \\
& \qquad \quad  \left. +  \int_{\Rthree} \left( (d_s(\mbf{p}) \tens \1 )\Psi_{m}   ,
 (1-    F_{ R} (\hat{\mbf{p}}) )(d_s(\mbf{p}) \tens \1 ) \Psi_{m } \right)  d \mbf{p}  \right) . \label{9/15.2}
\end{align}
We evaluate the two terms in the right-hand side of (\ref{9/15.2}). 
The   first term is  estimated as  
\begin{align}
&\left| \int_{\Rthree} \left( (b_s(\mbf{p}) \tens \1 )\Psi_{m}   ,
 (1- F_{ R} (\hat{\mbf{p}})  )(b_s(\mbf{p}) \tens \1 ) \Psi_m \right)  d \mbf{p} \right| \notag \\
& \leq \left( \int_{\Rthree} \left\| (b_s(\mbf{p}) \tens \1 )\Psi_m  \right\|^2 d \mbf{p}  
 \right)^{1/2} \times \left( \int_{\Rthree} \left\| (1- F_{ R} (\hat{\mbf{p}})  )(b_s(\mbf{p}) \tens \1  ) \Psi_m \right\|^2  d \mbf{p} \right)^{1/2} \notag \\
 & =  \| ( N_{\D}^{+} \tens \1 )^{1/2} \Psi_m \| \times  \left( \int_{\Rthree}
 \left\| (1-F_{ R} (\hat{\mbf{p}})  )(b_s(\mbf{p}) \tens \1  ) \Psi_m \right\|^2  d \mbf{p} \right)^{1/2}. \notag
\end{align}
It is seen that 
\begin{align}
 & \int_{\Rthree} \| (1- F_{ R} (\hat{\mbf{p}}) )(b_s(\mbf{p}) \tens \1 ) \Psi_m  \|^2  d \mbf{p} 
\notag \\
&  \leq 4  \int_{\Rthree} \| (1- F_{ R} (\hat{\mbf{p}}) ) \, \frac{1}{1+ \hat{\mbf{p}}^2}  (b_s(\mbf{p}) \tens \1 ) \Psi_m  \|^2  d \mbf{p}   \notag \\
& \qquad \qquad \qquad  \qquad    + 4 \sum_{\nu=1}^3
 \int_{\Rthree} \| (1-F_{ R} (\hat{\mbf{p}})  ) \, \frac{(\hat{p}^\nu )^2}{1+ \hat{\mbf{p}}^2} 
 \, (b_s(\mbf{p}) \tens \1 ) \Psi_m  \|^2  d \mbf{p}  . \notag
\end{align}
Note that for all $\mbf{p} \in \Rthree$,
\[
\sup_{\mbf{p} \in \Rthree } \left|
\left(  1- F_{ R} (\mbf{p})\right) \, \frac{1}{ 1+ \mbf{p}^2} \right| \leq  \frac{1}{R^2} \; , \quad  \; \;  \; \; \sup_{\mbf{p} \in \Rthree } \left|
 \left( 1- F_{ R} (\mbf{p})\right) \, \frac{p^{\nu}}{ 1+ \mbf{p}^2} \right| \leq  \frac{1}{R} .
\]
Then by the electron derivative bounds in Theorem \ref{EP-DB} \textbf{(i)} and the spectral decomposition theorem, we have
\begin{align}
& \int_{\Rthree} \left\| (1- F_{ R} (\hat{\mbf{p}}))(b_s(\mbf{p}) \tens \1 ) \Psi_m \right\|^2  d \mbf{p} \ \notag \\
& \leq  \frac{4}{R^4} 
  \int_{\Rthree} \left\| (b_s(\mbf{p}) \tens \1 ) \Psi_m \right\|^2  d \mbf{p}
+  \frac{4}{R^2} \sum_{\nu =1}^3  \int_{\Rthree} \left\| \partial_{p^{\nu}}(b_s(\mbf{p}) \tens \1 ) \Psi_m \right\|^2  d \mbf{p}   \notag \\
& \leq  \frac{4}{R^4} 
  \| ( N_{\D}^{+} \tens \1 )^{1/2} \Psi_m \|^2 +   \frac{ c_m( \epsilon )^2}{R^2} \sum_{\nu =1}^3
   \int_{\Rthree}|F_{s, +}^{\nu}(\mbf{p})|^2 d \mbf{p} , \notag 
  \end{align}
where $c_m ( \epsilon ) =  2 ( L_{\epsilon} E_{0} (H_m ) + R_{\epsilon} +1) |\kappaI| + 4|\kappaII  |$. 
Therefore,
\begin{align}
& \left| \int_{\Rthree} (b_s(\mbf{p}) \tens \1 )\Psi_m  , (1-  F_{ R} (\hat{\mbf{p}}) )(b_s(\mbf{p}) \tens \1 ) \Psi_m  ) d \mbf{p} \right| \notag \\
& \leq \| ( N_{\D}^{+} \tens \1 )^{1/2} \Psi_m \| \times \left(
\frac{2}{R^2} \| ( N_{\D}^+ \tens \1 )^{1/2} \Psi_m \| +  \frac{c_m ( \epsilon )}{R} \sum_{\nu =1}^3 \|F_{s ,+}^{\nu} \|_{L^2} \right) . \label{9/15.3}
\end{align}
 In a same way as the first term,  we can estimate the second term in the right-hand side of (\ref{9/15.2})  by the positron derivative bounds in  Theorem \ref{EP-DB} \textbf{(ii)}, and  then,
\begin{align}
& \left| \int_{\Rthree} (d_s(\mbf{p}) \tens \1 )\Psi_m  , (1-   F_{ R} (\hat{\mbf{p}}) )(d_s(\mbf{p}) \tens \1 ) \Psi_m  ) d \mbf{p}  \right| \notag \\
& \leq \| ( N_{\D}^{-} \tens \1 )^{1/2} \Psi_m \| \times \left(
\frac{2}{R^2} \| ( N_{\D}^- \tens \1 )^{1/2} \Psi_m \| +  \frac{c_m ( \epsilon )}{R} \sum_{\nu =1}^3  \|F_{s , -}^{\nu} \|_{L^2} \right) . \label{9/15.4}
\end{align}
From (\ref{9/15.3}) and  (\ref{9/15.4}), we have for all $R>1$, \\
\begin{align}
& \|  ( ( \1 -  \Gammaf (F_{ R} (\hat{\mbf{p}}) )) \tens \1 )  \Psi_{m} \|^2  \notag  \\
&\leq \frac{1}{R}\sum_{\tau = \pm} \| ( N_{\D}^{\tau} \tens \1 )^{1/2} \Psi_m \|  \left( 
2 \| ( N_{\D}^{\tau} \tens \1 )^{1/2} \Psi_m \| +  c_m ( \epsilon )
\sum_{s=\pm 1/2} \sum_{\nu =1}^3  \|F_{s ,\tau}^{\nu} \|_{L^2} \right) . \label{9/15.7}
\end{align}
 From Lemma \ref{9/14.a} \textbf{(i)},
 \[
\sup_{j \in \mbf{N}} \| ( N_{\D}^{\pm} \tens \1 )^{1/2} \Psi_{m_j} \| \leq 
\sup_{j \in \mbf{N}} \| ( N_{\D} \tens \1 )^{1/2} \Psi_{m_j} \|  \leq 
  \left( \frac{L_{\epsilon}}{M} E_{\infty} +\frac{R_{\epsilon}}{M} \right)^{1/2}  ,
\]
and we see that 
\[
\sup_{j \in \mbf{N}} c_{m_{j}} ( \epsilon ) 
= \sup_{j \in \mbf{N}} \left(2 ( L_{\epsilon} E_{0}(H_{m_j}) + R_{\epsilon} +1 ) |\kappaI| + 4|\kappaII  |  \right)
\; \leq  \;
 2 ( L_{\epsilon} E_{\infty} + R_{\epsilon} +1 ) |\kappaI| + 4|\kappaII  |.
\] 
Hence \textbf{(i)} follows.  \\
 \textbf{(ii)}
In a similar way to the proof of \textbf{(i)}, it follows that   $ (\1 -  \Gammab (  F_{R'}(\hat{\mbf{k}})))^2 \leq  \1 - \Gammab (F_{R'}(\hat{\mbf{k}})) \leq 
 \sqzb{1-F_{R'}(\hat{\mbf{k}}) }  $, and hence,
\begin{align}
 \|   ( \1 -  \Gammab ( \1 \tens  F_{R'}(\hat{\mbf{k}}))  )  \Psi_{m} \|^2 
&\leq ( \Psi_m ,\left(    \1 \tens \dGammab ( 1- F_{R'}(\hat{\mbf{k}}) )   \right) \Psi_{m} ) \notag \\
& = \sum_{r= 1,2}
  \int_{\Rthree} \left(  ( \1 \tens a_{r}(\mbf{k}) )\Psi_{m}   , 
(1- F_{R'}(\hat{\mbf{k}}) )
(\1 \tens a_{r}(\mbf{k}) ) \Psi_m \right)  d \mbf{k}  .
 \label{9/15.5}
\end{align}
We see that 
\begin{align}
&\left| \int_{\Rthree} \left(  ( \1 \tens a_{r}(\mbf{k}) )\Psi_{m}   , 
(1- F_{R'}(\hat{\mbf{k}}) )
(\1 \tens a_{r}(\mbf{k}) ) \Psi_m \right)  d \mbf{k}  \right| \notag \\
& \leq \left( \int_{\Rthree} \left\| ( \1 \tens a_{r}(\mbf{k})) \Psi_m  \right\|^2 d \mbf{p}  
 \right)^{1/2} \times \left( \int_{\Rthree} \left\| (1- F_{R'}(\hat{\mbf{k}}) )
(  \1 \tens a_{r}(\mbf{k})) \Psi_m \right\|^2  d \mbf{p} \right)^{1/2} \notag \\
 & =  \| (  \1 \tens N_{\rad}^{1/2} ) \Psi_m \| \times  \left( \int_{\Rthree} \left\| (1-F_{\bos , R'} )( \1 \tens a_{r}(\mbf{k})   ) \Psi_m \right\|^2  d \mbf{k} \right)^{1/2} . \notag
\end{align}
By  the photon derivative bounds  in Theorem \ref{P-DB} and the spectral decomposition theorem, 
\begin{align}
 & \int_{\Rthree} \| (1-F_{R'}(\hat{\mbf{k}}) )(  \1 \tens a_{r}(\mbf{k})) \Psi_m  \|^2  d \mbf{k}
 \notag \\
 & \leq 4 \int_{\Rthree} \| (1- F_{R'}(\hat{\mbf{k}}) ) \, \frac{1}{1+ \hat{\mbf{k}}^2} 
 (\1 \tens a_{r}(\mbf{k})) \Psi_m  \|^2  d \mbf{k} \notag \\
& \qquad \qquad + 4 \sum_{\nu =1}^3 \int_{\Rthree} \|
 (1- F_{R'}(\hat{\mbf{k}}) ) \, \frac{(\hat{k}^\nu )^2}{1+ \hat{\mbf{k}}^2} 
 \, (\1 \tens a_{r}(\mbf{k}) ) \Psi_m  \|^2  d \mbf{k} \notag  \\
 &  \leq \frac{4}{R'^4} 
  \int_{\Rthree} \left\| (\1 \tens a_{r}(\mbf{k}) ) \Psi_m \right\|^2  d \mbf{k}
+ \frac{4}{R'^2}  \sum_{\nu=1}^3\int_{\Rthree}  \left\| \partial_{k^{\nu}}( \1 \tens a_{r}(\mbf{k}) ) \Psi_m \right\|^2  d \mbf{k} \notag \\
& \leq \frac{4}{R'^4}  \| (  \1 \tens N_{\rad}^{1/2} ) \Psi_m \|^2 + \frac{ 4 |\kappaI |^2}{R'^2}\sum_{\nu =1}^3
   \int_{\Rthree}|F_{r}^{\nu}(\mbf{k})|^2 d \mbf{k} . \notag 
\end{align}
Then we have
\begin{align}
& \left| \int_{\Rthree} \left(  ( \1 \tens a_{r}(\mbf{k}) )\Psi_{m}   , 
(1- F_{R'}(\hat{\mbf{k}}) )
(\1 \tens a_{r}(\mbf{k}) ) \Psi_m \right)  d \mbf{k}  \right| \notag \\
& \leq  \| (  \1 \tens N_{\rad}^{1/2} ) \Psi_m \| \times \left(  \frac{4}{{R'}^4} \| (  \1 \tens N_{\rad}^{1/2} ) \Psi_m \|^2
+ \frac{ 4|\kappaI |^2 }{{R'}^2} \sum_{\nu =1}^3
   \| F_{r}^{\nu} \|_{L^2}^2 \right)^{1/2} \notag , 
\end{align}
and hence, for all $R'>1$, 
\begin{align}
&  \|  ( ( \1 -  \Gammab ( \1 \tens F_{R'}(\hat{\mbf{k}} ))  )  \Psi_{m} \|^2   \notag \\
&\leq  \frac{2}{R'}   \| (  \1 \tens N_{\rad}^{1/2} ) \Psi_m \| \times \left(  \| (  \1 \tens N_{\rad}^{1/2} ) \Psi_m \|
+|\kappaI |\, \sum_{r=1,2} \sum_{\nu =1}^3 \| F_{r}^{\nu} \|_{L^2}\right) .
\end{align}
From Lemma \ref{9/14.a} (\textbf{ii}), it holds that  $ \sup\limits_{j \in \mbf{N}} \| (  \1 \tens N_{\rad}^{1/2} ) \Psi_{m_j} \| < c_{0} |\kappaI|
\left\| \frac{\chi_{\rad}}{\omega^{3/2}} , \right\| $. Therefore  the proof is obtained. $\blacksquare$ \\

$\;$ \\
{\large\textbf{(Proof of Theorem \ref{Main-Theorem})}}\\
From Proposition \ref{9/16.a} and  a general theorem (\cite{AH97} ; Lemmma 4.9), it is enough to show that   w-$ \lim\limits_{j \to \infty} \Psi_{m_j} \neq 0$. 
We see that 
\begin{align}
 \1_{\D} \tens \1_{\rad } & =   ( \1_{\D } -\Gammaf (F_{R}(\hat{\mbf{p}} ) ) \tens \1_{\rad}   +   \Gammaf ( F_{R}(\hat{\mbf{p}} ) ) \tens ( \1_{\rad} -  \Gammab ( F_{R'}(\hat{\mbf{k}} ) ))     \notag \\
& \quad + \Gammaf ( F_{R}(\hat{\mbf{p}} ) ) \tens ( F_{R'}(\hat{\mbf{k}} ) ) E_{N_{\rad}}([0, n] )) + \Gammaf ( F_{R}(\hat{\mbf{p}}  )) \tens ( \Gammab ( F_{R'}(\hat{\mbf{k}} )  ) E_{N_{\rad}}([ n+1, \infty) )) . \notag 
\end{align}
Then by Proposition \ref{9/15.a}, we have for all $0 < \epsilon < \frac{1}{c_{\I} |\kappaI |}$, $R>1$ and $R'> 1 $, 
\begin{align}
& \left\| \left( \Gammaf ( F_{R}(\hat{\mbf{p}}  )) \tens ( \Gammab (  F_{R'}(\hat{\mbf{k}} ) ) E_{N_{\rad}}([0, n ] ) )\right) \Psi_{m_j}  \right\| 
\notag \\
& \geq 1-  \left(  \frac{}{} \| (( \1_{\D} -\Gammaf ( F_{R}(\hat{\mbf{p}}  ))) \tens \1_{\rad} ) \Psi_{m_j}  \|  
  \right. \notag \\
& \qquad \qquad \qquad  \left. + \| ( \1_{\D} \tens ( \1_{\rad } -  
\Gammab ( F_{R'}(\hat{\mbf{k}} ) ) )  \Psi_{m_j}  \|
+ \| ( \1_{\D} \tens  E_{N_{\rad}}([n+1, \infty) )) \Psi_{m_j}  \| \frac{}{}
 \right) . \notag \\
 & \geq 1-  \left( \frac{c_{1, \epsilon }}{\sqrt{R}} + \frac{c_2}{\sqrt{R'}}  + \| ( \1_{\D} \tens  E_{N_{\rad}}([n+1, \infty) )) \Psi_{m_j}  \|  \right) , \notag
\end{align}
It is seen that
\[
 \sqrt{n+1 }\|  ( \1_{\D} \tens  E_{N_{\rad}}([n+1, \infty) )) \Psi_{m_j}  \| \leq
 \| ( \1_{\D } \tens  N_{\rad}^{1/2} ) \Psi_{m_j}  \|  \leq c_{0} |\kappaI | \left\| \frac{\chi_{\rad}}{\omega^{3/2}}\right\| . 
\] 
Then from Lemma \ref{9/14.a} \textbf{(ii)}, we have
\begin{equation}
 \sup\limits_{j \geq 1}\|  ( \1_{\D} \tens  E_{N_{\rad}}([n+1, \infty) )) \Psi_{m_j}  \| \leq \frac{c_3 }{(n+1)^{1/2}}. \notag 
\end{equation}
where $c_{3}= c_{0} |\kappaI | \left\| \frac{\chi_{\rad}}{\omega^{3/2}}\right\| $. 
Then it follows that  
\begin{equation}
\| \Gammaf ( F_{R}(\hat{\mbf{p}} ) \tens ( \Gammab ( F_{R'}(\hat{\mbf{k}} ) E_{N_{\rad}}([0, n ] ))\Psi_{m_j}  \|
\geq 1 - \left( \frac{c_{1 , \epsilon}}{R} + \frac{c_2}{R'}  + \frac{c_3}{(n+1)^{1/2}} \right) .
\label{9/15.9}
\end{equation} 
We also see that
\begin{align}
& \| \Gammaf (  F_{R}(\hat{\mbf{p}} )) \tens (   F_{R'}(\hat{\mbf{k}} )) E_{N_{\rad}}([0, n ] ))\Psi_{m_j}  \|^2 
\notag \\ 
& = ((H_0 +1) E_{N_{\rad}}([0, n ] ) \Psi_{m_j} ,(H_0 +1)^{-1} 
( \Gammaf (F_{R}(\hat{\mbf{p}} )^2) \tens ( \Gammab (  F_{R'}(\hat{\mbf{k}} )^2 ) E_{N_{\rad}}([0, n ] )) ) \Psi_{m_j}  ) \notag \\
& \leq \| (H_0 +1)\Psi_{m_j} \| \, \| (H_0 +1)^{-1} \, ( \Gammaf ( F_{R}(\hat{\mbf{p}} )^2 ) \tens (
 \Gammab (  F_{R'}(\hat{\mbf{k}} )^2 ) E_{N_{\rad}}([0, n ] )))  \, \Psi_{m_j}  \| . \notag
\end{align}
We see that  $\| (H_0 +1)\Psi_{m_j} \| \leq \| H_0 \Psi_{m_j} \| +1 \leq   \| H_{0,m_{j}} \Psi_{m_j} \| +1 $ and  Lemma \ref{9/9.e} yields that 
 \[
 \| H_{0,m_{j}} \Psi_{m_j} \|
\leq \frac{L_{\epsilon } }{M}  \| H_{m_j}  \Psi_{m_j}  \| + \frac{R_{\epsilon }}{M} 
\leq \frac{L_{\epsilon}}{M} E_{\infty} + \frac{R_{\epsilon}}{M}  .
\]
Then we have
\begin{align}
&\| \Gammaf (  F_{R}(\hat{\mbf{p}} )) \tens (   F_{R'}(\hat{\mbf{k}} )) E_{N_{\rad}}([0, n ] ))\Psi_{m_j}  \| \notag   \\
& \qquad \qquad \qquad \leq  c_{4 ,\epsilon}\| (H_0 +1)^{-1} \,
\left( ( \Gammaf ( F_{R}(\hat{\mbf{p}} ) ) \tens (  \Gammab (  F_{R'}(\hat{\mbf{k}} )^2 ) E_{N_{\rad}}([0, n ] ))) \right) \, \Psi_{m_j}  \|^{1/2}, \label{9/15.10}
\end{align}
where $c_{4 , \epsilon }= (\frac{L_{\epsilon}E_{\infty} + R_{\epsilon} }{M} +1)^{1/2}$.  
 From (\ref{9/15.9}) and (\ref{9/15.10})
\begin{equation}
\|(H_0 +1)^{-1}\Gammaf (  F_{R}(\hat{\mbf{p}} )^2) \tens ( \Gammab (  F_{R'}(\hat{\mbf{k}})^2  E_{N_{\rad}}([0, n ] ))\Psi_{m_j}   \| \geq  \frac{1}{c_{4, \epsilon }}^2 \left( 1 - \left( \frac{c_{1,\epsilon}}{\sqrt{R}} + \frac{c_2}{\sqrt{R'}}  + \frac{c_3}{(n+1)^{1/2}} \right) \right)^2 \notag
\end{equation}
Since $(H_0 +1)^{-1} \left( \Gammaf ( F_{R}(\hat{\mbf{p}} )^2) \tens ( \Gammab (   F_{R'}(\hat{\mbf{k}})^2) E_{N_{\rad}}([0, n ] ))) \right) $ is compact, we have 
\begin{equation}
 \|(H_0 +1)^{-1}\Gammaf (  F_{R}(\hat{\mbf{p}})^2 ) \tens ( \Gammab (  F_{R'}(\hat{\mbf{k}})^2 ) E_{N_{\rad}}([0, n ] ))\Psi_{0}   \| \geq  \frac{1}{c_{4, \epsilon }^2}\left( 1 - \left( \frac{c_{1,\epsilon}}{\sqrt{R}} + \frac{c_2}{\sqrt{R'}}  + \frac{c_3}{(n+1)^{1/2}} \right) \right)^2  ,
 \label{9/15.20}
\end{equation}
where we set $\Psi_{0} =$ w-$ \lim\limits_{j \to \infty} \Psi_{m_j} $.
Then for sufficiently large $R>0$, $R'>0 $ and $n >0$, the right-hand side of (\ref{9/15.20}) is greater than zero, and hence $  \Psi_{0}  \ne 0 $.  \\
$\;$ \\
\textbf{(Multiplicity)} 
Assume dim ker $(H -E_{0}(H_{\QED})) = \infty $. Let $ \Psi_{l} $, $l \in \mbf{N}$, be the  ground states. Let $\ms{M} $ be  the closure of the  linear hull of  $\{ \Psi_l  \}_{l=0}^{\infty}$. Then $\ms{H}_{\QED}$ is decomposed as $\ms{H}_{\QED} = \ms{M} \oplus \ms{M}^{\bot} $. Let 
$\{ \Phi_l \}_{l=0}^{\infty}$ be a complete orthogonal  system of  $\ms{M}^{\bot}$. 
We can set a complete orthonormal system $\{ \Xi_l \}_{l=0}^{\infty}$ of $\ms{H}_{\QED}$ by 
$  \Xi_{2 l-1 } = \Psi_{l} $ and $  \Xi_{2 l} = \Phi_{l} $ for all $l \in \mbf{N}$. 
Since  $\{ \Xi_l \}_{l=0}^{\infty}$  is a complete orthonormal system
, w-$\lim\limits_{l \to \infty}\Xi_l =0$. 
 On the other hand, $\Xi_{2 l-1 }$ is ground state for all $l \in \mbf{N}$, and hence $H_{\QED}\Xi_{2 l-1 }=E_{0}(\HQED) \Xi_{2 l-1 } $. In a same argument  of the proof of the existence of the ground state, we have 
\begin{equation}
 \|(H_0 +1)^{-1}\Gammaf (  F_{R}(\hat{\mbf{p}})^2 ) \tens ( \Gammab (  
F_{R'}(\hat{\mbf{k}})^2 ) E_{N_{\rad}}([0, n ] ))\Xi_{2l-1}   \| \geq  \frac{1}{\tilde{c}_{4, \epsilon }^2}\left( 1 - \left( \frac{ \tilde{c}_{1,\epsilon}}{\sqrt{R}} + \frac{c_2}{\sqrt{R'}}
  + \frac{c_3}{(n+1)^{1/2}} \right) \right)^2 ,
 \label{9/15.15}
\end{equation}
where  $\tilde{c}_{1, \epsilon} $ and    $\tilde{c}_{4, \epsilon} $ are  the constants $ c_{1 , \epsilon}$ and $ c_{1 , \epsilon}$ replacing   $E_{\infty}$  with $E_{0}(\HQED)$. 
Then by taking  sufficiently  large $R>0$, $R'>0 $ and $n >0$, we have w-$\lim\limits_{l \to \infty}\Xi_{2l-1} \ne 0$,  but this is contradict to  w-$\lim\limits_{l \to \infty}\Xi_l = 0$. Hence  dim ker $(H_{\QED} -E_{0}(H_{\QED})) < \infty $. $\blacksquare $\\

$\;$ \\
$\;$ \\
{\large \textbf{[Concluding remarks]}} \\
\textbf{(1) The case of Massless Dirac field} \\
It is not realistic model, but we can consider the system of a massless Dirac field coupled to the radiation field. In such a case,  by replacing (\textbf{A.5}) with  similar  conditions  to (\textbf{A.6}), we can also prove  the existence of the ground state  in a same ways as $\HQED$.    
$\; $  \\
 \textbf{(2) Infrared divergent problem} \\
For  some  systems of particles coupled to massless Bose fields,   the existence of the ground states without  infrared regularity conditions was obtained (refer to e.g.,  Bach-Fr\"{o}hlich-Sigal \cite{BFS99}, Griesemer-Lieb-Loss \cite{GLL01} and Hasler-Herbst \cite{HaHe11}), and  non-existence of the ground states for  other  other systems  was  also investigated (see e.g., Arai-Hirokawa-Hiroshima \cite{AHH99}).   
To prove the existence or non-existence  of the ground state   of $\HQED$ without infrared regularity conditions is left for future study.


\begin{thebibliography}{99}
\small{ 
\bibitem{Am04}
 Z. Ammari, Scattering theory for a class of fermionic Pauli-Fierz models, \textit{J. Funct. Anal.} \textbf{208}  (2004)
 302-359.
 \bibitem{Arai}
A. Arai, \textit{Fock spaces and quantum fields}, Nippon-Hyoronsha,   (in Japanese),  2000.  
 \bibitem{AH97}
A. Arai and M. Hirokawa, On the existence and uniqueness of ground states of a generalized spin-boson model, \textit{J. Funct. Anal.} \textbf{151}  (1997)  455-503. 
\bibitem{AHH99}
 A. Arai, M. Hirokawa and H. Hiroshima, On the absence of eigenvectors of Hamiltonians in a class of massless quantum field models without infrared cutoff, \textit{J. Funct. Anal.} \textbf{168}  (1999), 470-497. 
   \bibitem{BFS98}
 V. Bach, J. Fr\"{o}hlich and I. M. Sigal,
 Quantum electrodynamics of confined nonrelativistic particles. \textit{Adv. Math.} \textbf{137}  (1998) 299-395.
  \bibitem{BFS99} 
   V. Bach, J. Fr\"{o}hlich and I. M. Sigal,  Spectral analysis for systems of atoms and molecules coupled to the quantized radiation field, \textit{Comm. Math. Phys.}  \textbf{207} 
  (1999) 249-290. 
    \bibitem{BDG04} J. -M. Barbaroux, M. Dimassi, and J. -C. Guillot, Quantum electrodynamics of relativistic bound states with cutoffs, \textit{J.Hyper. Differ. Equa.} \textbf{1} (2004) 271-314.
    \bibitem{BFG14}
 J. -M. Barbaroux, J. Faupin and J. -C. Guillot
 Spectral theory near thresholds for weak interactions with massive particles.
 arXiv:1401.0649.  
  \bibitem{BD} 
  J. D. Bjorken and S. D. Drell, \textit{Relativistic quantum fields},
  McGraw-Hill, 1965.
 \bibitem{DeGe99}
J. Derezi\'{n}ski and C. G\'{e}rard, Asymptotic completeness
in quantum field theory. Massive Pauli-Fierz Hamiltonian, \textit{Rev. Math. Phys.} \textbf{11}  (1999) 383-450.   
  \bibitem{DiGu03}
M. Dimassi and J. -C. Guillot, The quantum electrodynamics of relativistic bound states with cutoffs.I, \textit{Appl. Math. Lett.} \textbf{16}   (2003) 551-555.
\bibitem{FuUs14}
S. Futakuchi and K. Usui, Time-ordered exponential on the complex plane and Gell-Mann - Low formula as a mathematical theorem, arXiv:1407.1690. 
 \bibitem{FGS02}
J. Fr\"ohlich, M. Griesemer and B. Schlein, 
Asymptotic completeness for Rayleigh scattering, \textit{Ann. H. Poincar\'{e}} \textbf{3}  (2002) 107-170. 
  \bibitem{Ge00}
C. G\'{e}rard,  On the existence of ground states for massless Pauli-Fierz Hamiltonians,
 \textit{Ann. H. Poincar\'e,} \textbf{1}   (2000) 443-459.
 \bibitem{GeEr}
C. G\'{e}rard, 
 A remark on the paper : On the existence of ground states for massless Pauli-Fierz
Hamiltonians, (preprint). 
   \bibitem{GLL01}
M. Griesemer, E. Lieb and M. Loss, Ground states in non-relativistic quantum electrodynamics,  \textit{Invent. Math.} \textbf{145}   (2001) 557-595.
\bibitem{Gu15}
J. -C. Guillot, 
Weak interactions in a background of a uniform magnetic field. A mathematical model for the  the inverse $\beta$ decay. I, arxiv : 1502.05524.
\bibitem{HaHe11}
D.  Hasler and I. Herbst, Ground states in the spin boson model, \textit{Ann. Henri Poincar\'e} \textbf{12} (2011) 621-677. 
\bibitem{Hida11} 
T. Hidaka, Existence of a ground state for the Nelson model with a singular perturbation.
\textit{J. Math. Phys.} \textbf{52} (2011) 022102.
  \bibitem{LiLo03}
E. Lieb and M. Loss, Existence of atoms and molecules in non-relativistic quantum electrodynamics,  \textit{Adv. Theor. Math. Phys.} \textbf{7} (2003) 667-710.
    \bibitem{Sp98}
H. Spohn, Ground state of quantum particle coupled to a scalar boson field, \textit{Lett. Math. Phys.} \textbf{44} (1998)   9-16.
\bibitem{Ta09}
 T. Takaesu, On the spectral analysis of quantum electrodynamics with spatial cutoffs. I, \textit{J. Math. Phys.} \textbf{50}  (2009) 06230.
 \bibitem{Ta11}
 T. Takaesu, Ground states of Yukawa models with cutoffs,
 \textit{Inf. Dim. Anal. Quantum Prob.  Related Topics}, \textbf{14} (2011) 225-235.
 \bibitem{Tha}
 B. Thaller, \textit{The Dirac equation}, Springer, 1992.
  } 
\end{thebibliography}
\end{document}